%% file: main.tex
\documentclass[pdflatex,sn-mathphys-num]{submission/sn-jnl}
\usepackage[T1]{fontenc}
\usepackage{lmodern}
\usepackage[utf8]{inputenc}
\usepackage{graphicx}
\usepackage{amsmath,amssymb,amsfonts}
\usepackage{amsthm}
\usepackage{booktabs}
\usepackage{array}
\usepackage{tabularx}
\usepackage{threeparttable}
\usepackage{longtable}
\usepackage{pdflscape}
\usepackage{placeins}
\usepackage{tikz}
\usepackage{pgfplots}
\usepackage[most]{tcolorbox}
\usepackage{setspace}
\usepackage{xcolor}
\usepackage{url}
\usepackage[protrusion=true,expansion=false]{microtype}
\usetikzlibrary{arrows.meta,positioning,fit,calc,shapes.geometric,backgrounds,shadows.blur,shadings}
\usepgfplotslibrary{groupplots}
\pgfplotsset{compat=1.18}
\input{visual_style}

\newcommand{\DeFi}{DeFi}

\newcommand{\ind}{\mathbf{1}}
\makeatletter
\newcommand{\suspendrefereelines}{\@ifundefined{lineno@off}{}{\lineno@off}}
\makeatother
\newcommand{\resumerefereelines}{\ifdefined\linenoon\linenoon\fi}

\theoremstyle{plain}
\newtheorem{theorem}{Theorem}
\newtheorem{lemma}{Lemma}
\newtheorem{proposition}{Proposition}
\theoremstyle{definition}

\AtBeginDocument{%
  \hypersetup{%
    pdftitle={Is Decentralized Finance Actually Decentralized? An Interdisciplinary Framework Integrating Network Theory, Agent-Based Simulation, and Longitudinal Evidence from Aave, GHO Issuance, and Cross-Chain Expansion},
    pdfauthor={Anonymous Authors},
    pdfkeywords={decentralized finance, Aave, GHO, stablecoins, multidimensional decentralization, network theory, agent-based simulation, infrastructure resilience}%
  }%
}

\begin{document}

\title[Is Decentralized Finance Actually Decentralized?]{Is Decentralized Finance Actually Decentralized? An Interdisciplinary Framework Integrating Network Theory, Agent-Based Simulation, and Longitudinal Evidence from Aave, GHO Issuance, and Cross-Chain Expansion}




\title[Is Decentralized Finance Actually Decentralized?]
{Is Decentralized Finance Actually Decentralized?\\
An Interdisciplinary Framework Integrating Network Theory,
Agent-Based Simulation, and Longitudinal Evidence from Aave,
GHO Issuance, and Cross-Chain Expansion}


%

\author[1,2]{\fnm{Ziqiao} \sur{Ao}\textsuperscript{$\dagger$}}

\author[3]{\fnm{Lin William} \sur{Cong}\textsuperscript{$\dagger$}}

\author[1]{\fnm{Gergely} \sur{Horvath}\textsuperscript{$\dagger$}\textsuperscript{*}}

\author[1]{\fnm{Luyao} \sur{Zhang}\textsuperscript{$\dagger$}\textsuperscript{*}\textsuperscript{$\ddagger$}}



\affil[1]{
  \orgname{Duke Kunshan University},
  \orgaddress{
    \city{Suzhou},
    \country{China}
  }
}

\affil[2]{
  \orgname{Microsoft Corporation},
  \orgaddress{
    \city{Seattle},
    \country{USA}
  }
}

\affil[3]{
  \orgname{Nanyang Technological University},
  \orgaddress{
    \country{Singapore}
  }
  }

\abstract{Decentralized finance (DeFi) can broaden access while leaving activity, network position, and infrastructure concentrated. Yet economics, finance, computer science, and network science largely examine these dimensions separately. Aave is a decentralized lending protocol whose public records cover repeated financial interactions and its GHO stablecoin. This makes Aave informative for asking whether stablecoin issuance and cross-chain expansion broaden participation, distribute activity more evenly, disperse network positions, and reduce infrastructure dependence. We integrate network theory, theorem-consistent agent-based simulation, and longitudinal analysis of 1,956,216 Aave V3 Pool events. Aave was already multichain before GHO launched; the sequence studied is GHO issuance on Ethereum (15 July 2023) and GHO's first cross-chain expansion to the existing Arbitrum market (2 July 2024). Excluding each activation week, mean weekly active position-holder addresses increased by 91.0\% around Ethereum issuance and 1.7\% around Arbitrum expansion, while activity concentration fell by 31.5\% on Ethereum but rose by 58.1\% on Arbitrum. On a common 2024 calendar, the Arbitrum--Gnosis DiD-style change is +1.9833 for log participation and -0.01842 for position-holder-event HHI; Gnosis movement, differential pre-trends, and anticipation-window sensitivity limit causal interpretation. Rule-based simulations recover the analytical equilibrium, demonstrate why aggregate growth can coexist with lower, unchanged, or higher concentration, and show why chain dispersion does not identify route or shared-component resilience. Role-aware empirical analysis further shows that network-structure conclusions vary by protocol action and scale. Intellectually, the framework explains why four dimensions of decentralization can diverge. Practically, it helps researchers, protocol designers, governance communities, and policymakers assess stablecoin growth without equating adoption with decentralization.}





\begingroup
\renewcommand{\thefootnote}{\fnsymbol{footnote}}

\footnotetext[2]{%
Authors are listed in alphabetical order by surname. Z.A. started research as an academic advisee and Signature Work mentee of L.Z and as a Summer Research Scholar (SRS) mentee of G.H. at Duke Kunshan University and is now at Microsoft Corporation.
}

\footnotetext[1]{%
Corresponding authors: Luyao Zhang
(\texttt{lz183@duke.edu}),and
Gergely Horvath (\texttt{gh117@duke.edu}),  Social Science Division and Digital
Innovation Research Center, Duke Kunshan University. Correspondence address:
8 Duke Avenue, Kunshan, Suzhou, Jiangsu 215316, China.
}
\footnotetext[3]{%
\textbf{Acknowledgment}: We have benefited from intellectual conversations at the 29th Annual Global Finance Conference, featuring Nobel Laureate Prof.~Robert Engle as keynote speaker, held in Braga, Portugal, June 20--22, 2022; the Crypto Economics Security Conference (CESC), hosted by the Berkeley Center for Responsible Decentralized Intelligence (DCI) at the University of California, Berkeley, CA, USA, October 31--November 1, 2022; and the 12th Portuguese Financial Network Conference, held in Madeira, Portugal, in July 2023. We thank Prof.~Claudio J. Tessone, Prof.~William Goetzmann, and
Prof.~Avanidhar Subrahmanyam for their insightful comments. L.Z. acknowledges support from the National Natural Science Foundation of China through the project ``Trust Mechanism Design on Blockchain: An Interdisciplinary Approach of Game Theory,
Reinforcement Learning, and Human--AI Interactions'' (Grant No.~12201266).

}

\endgroup

\maketitle

\input{sections/01_introduction}
\input{sections/02_background}
\input{sections/03_model_simulation}
\input{sections/04_data_open_science}
\input{sections/05_causal_inference}
\input{sections/06_conclusion}

\appendix
\section{Appendix guide and evidence architecture}\label{app:guide}
This opening module explains why the supplementary material is ordered by the evidentiary path from provenance to measurement, model validation, robustness, identification, and prospective extensions. It also provides the status vocabulary used throughout the appendix and links each module to the corresponding main-text claim.
\input{appendix/appendix_guide}
\input{appendix/literature_coding_audit}

\section{Data provenance, event chronology, and sample construction}\label{app:data-provenance}
This module expands the event clocks, source grades, acquisition checks, sample coverage, and Ethereum extraction summarized in Sections~\ref{sec:background} and~\ref{sec:data}. Its purpose is to show how the auditable records become the analysis panels before any structural or comparative interpretation is attempted.
\subsection{Source documentation and acquisition checks}\label{app:source-audit}
\input{appendix/source_audit_tables}
\input{appendix/real_v2_ethereum_audit}

\section{Measurement, network construction, and the address--actor boundary}\label{app:measurement-construction}
This module supports the four-dimensional measurement framework in Sections~\ref{sec:data} and~\ref{sec:causal}. It classifies each measure by research function, evidence status, and presentation coverage, then documents role decoding and the limits of translating addresses into economic actors.
\input{appendix/measurement_glossary_and_claim_map}

\input{appendix/real_v3_entity_audit}
\input{appendix/real_v4_partial_identification}
\section{Formal model and simulation validation}\label{app:model-validation}
This module supplies the derivations, protocol translation, executable update procedure, and robustness checks underlying Section~\ref{sec:model}. The analytical and simulated objects remain mechanism evidence rather than fitted observations of Aave participants.
\input{appendix/model_translation_tables}

\input{appendix/formal_model_details}
\input{appendix/simulation_protocol}
\input{appendix/simulation_sensitivity}

\section{Longitudinal and structural-network robustness}\label{app:structural-robustness}
This module extends the longitudinal membership, topology, community, coreness, and infrastructure-core diagnostics used to qualify the chain-relative patterns in Section~\ref{sec:causal}. It preserves the distinction between complete-graph estimands and display-only backbones.
\input{appendix/longitudinal_network_robustness}
\input{appendix/be_core_infrastructure_evidence}

\section{Comparative design and identification diagnostics}\label{app:comparative-diagnostics}
This module reports the common-calendar construction, pre-path, anticipation, placebo, horizon, and controlled-ITS diagnostics behind the Arbitrum--Gnosis comparison in Section~\ref{sec:causal}. These checks bound interpretation; they do not convert the one-comparison-chain design into an identified treatment effect.
\input{appendix/arbitrum_gnosis_benchmark_robustness}

\section{Evidence status, reproducibility, and future extensions}\label{app:future-evidence}
This final module consolidates what is observed, simulated, bounded, diagnostic, pending, or unavailable; specifies independent reconciliation steps; and develops the cross-disciplinary agenda introduced in Section~\ref{subsec:future-systemic}. Prospective designs are explicitly separated from results established by the current release.
\input{appendix/evidence_status_summary}
\input{appendix/subgraph_validation_protocol}

\input{appendix/future_causal_infrastructure_extensions}

\backmatter


\section*{Declarations}




\bmhead{Data availability}

The datasets supporting this study, together with their metadata, provenance
records, data dictionary, checksums, and Responsible AI documentation, are
available under the Creative Commons Attribution 4.0 International license at
\url{https://huggingface.co/datasets/zlysunshine/aave-bns-data}.

\bmhead{Code availability}

The source code and reproducibility pipeline used for data processing,
network construction, statistical analysis, and generation of the reported
results are available under the MIT license at
\url{https://github.com/sunshineluyao/aave-bns-code}.

\bmhead{Interactive demonstrations}

Two complementary interactive demonstrations accompany the data and code
releases. The research playground provides an interactive presentation of the
study's theoretical mechanisms and simulations:
\url{https://huggingface.co/spaces/zlysunshine/aave-bns-playground}.
The empirical data explorer provides interactive access to the multidimensional
protocol-network evidence and associated visual analyses:
\url{https://huggingface.co/spaces/zlysunshine/aave-bns-demo-HF}.
These demonstrations are intended for exploration and communication; the
versioned data and code repositories above constitute the authoritative
reproducibility resources.


\bibliography{references}

\end{document}

%% file: sections/01_introduction.tex
\section{Introduction}\label{sec:introduction}

Decentralized finance (\DeFi{}) seeks to provide financial services through public blockchains and smart contracts rather than conventional centralized intermediaries \cite{schar2021defi,werner2022sokdefi,makarov2022cryptocurrencies}. However, visible transactions and open code do not establish that a financial system is decentralized. A protocol may attract more participants while activity becomes concentrated among a smaller set of addresses. It may operate on more blockchains while relying on one bridge, router, or messaging service. It may distribute routine activity broadly while governance or economic control remains concentrated. The relevant question is therefore not whether a protocol is decentralized in the abstract, but \emph{which dimensions are dispersed, how they change, and what evidence can support each conclusion}.

We study this problem in Aave, an open-source lending protocol in which participants supply assets and borrow against collateral through smart contracts. Aave was already operating across multiple blockchains before its protocol-native GHO stablecoin launched. The two institutional changes examined here are therefore not Aave's entry into Ethereum and Arbitrum, but GHO issuance on Ethereum and GHO's subsequent expansion to an existing Aave market on Arbitrum. Aave's Pool contracts record supplying, withdrawing, borrowing, repaying, and liquidating in a consistent event format. Those records reveal blockchain addresses and protocol roles, but they do not by themselves reveal the people or organizations controlling the addresses, the infrastructure paths supporting an action, or the effect of a protocol change.

Economics, finance, computer science, and network science each illuminate part of this distinction. Economics studies strategic interaction and concentration; finance studies liquidity, intermediation, and dependence; computer science makes execution records auditable; and network science measures position, temporal change, and resilience. We integrate these perspectives through a \emph{Protocol-Network Decentralization Analysis} framework. It represents decentralization at time $t$ as
\begin{equation}
\boldsymbol{\Delta}_t=\left(\mathcal P_t,\mathcal A_t,\mathcal S_t,\mathcal I_t\right),
\label{eq:decentralization-vector}
\end{equation}
where $\mathcal P_t$ denotes participation, $\mathcal A_t$ activity distribution, $\mathcal S_t$ network structure, and $\mathcal I_t$ infrastructure dependence. The dimensions are reported separately rather than collapsed into one score. Figure~\ref{fig:literature-landscape} shows how complementary insights from economics, finance, computer science, and network science inform these dimensions and connects each dimension to the measures implemented in the theoretical, simulation, and empirical components. Economic-actor decentralization is an additional interpretive layer: addresses are observable, whereas common control across addresses requires independent evidence.

\input{tables/tab01_dimensions}

The analysis addresses four questions aligned with the four dimensions. First, around GHO issuance on Ethereum and GHO's first cross-chain expansion to Arbitrum, how did address-level participation change? Second, did Pool-event-frequency concentration move with participation or diverge within each chain? Third, how did network structure vary across protocol roles, actions, chains, and scale adjustments? Fourth, what additional actor, infrastructure-route, and counterfactual evidence is required to evaluate economic control, infrastructure dependence, and causal effects?

We integrate network theory, theorem-consistent agent-based simulation, and longitudinal evidence. The theory explains how heterogeneous direct benefits propagate through an economic-interaction network, why aggregate growth need not make activity more evenly distributed, and why venue dispersion does not identify infrastructure resilience. The Mesa--NetworkX simulations use 40 rule-based actors to recover the analytical equilibrium and separately test network amplification, growth--concentration divergence, and route/component dependence across deterministic seeds and updating regimes. They are synthetic mechanism checks, not calibrated Aave estimates. The empirical analysis then constructs weekly networks from 118,806 Ethereum and 1,837,410 Arbitrum Aave V3 Pool events. Figure~\ref{fig:interdisciplinary-framework} shows how the case, methods, outcomes, and evidence boundary fit together.

\input{figures/fig01_interdisciplinary_framework}

The results show why the dimensions must remain separate. Excluding the week of each reserve activation, the mean weekly number of addresses with changing lending or borrowing positions rises by 91.0\% on Ethereum and 1.7\% on Arbitrum. Over the same chain-relative windows, concentration of observed Pool-event activity falls by 31.5\% on Ethereum but rises by 58.1\% on Arbitrum. On the common 2024 calendar, Arbitrum's participation gap relative to Gnosis widens while its HHI increases less; most of the comparative movement occurs on Gnosis and the fitted level and slope changes are sensitive to the anticipation exclusion. These are descriptive and comparative longitudinal estimates, not identified effects of GHO. The analysis further shows that conclusions about network structure depend on the protocol role being measured and on adjustment for network size.

The paper makes four contributions. First, conceptually, it represents decentralization as participation, activity distribution, network structure, and infrastructure dependence rather than as a unitary verdict. Second, theoretically and computationally, it builds on the established fixed-network game and validates three mechanisms with analytical-equilibrium-checked agent-based simulations: network amplification, growth--concentration divergence, and dispersion--resilience divergence. Third, methodologically, it converts Pool events into typed weekly address--role networks with explicit roles, dates, and scale adjustments. Fourth, empirically, it combines chain-specific longitudinal evidence with a reproducible Arbitrum--Gnosis DiD-style comparison, controlled interrupted-time-series estimates, dynamic comparative contrasts, and explicit identification diagnostics. Together, these contributions make multidimensional decentralization measurable without assigning stronger meaning to the data than they can support.

The broader implication is constructive. Rather than seek a single verdict on whether \DeFi{} is decentralized, researchers and practitioners can ask which dimension matters for a given financial function, what observable pattern a proposed mechanism predicts, and what evidence would justify a stronger conclusion. The remainder develops the institutional setting and research gap (Section~\ref{sec:background}), theory and simulation (Section~\ref{sec:model}), measurement (Section~\ref{sec:data}), longitudinal comparative evidence and identification boundaries (Section~\ref{sec:causal}), and implications (Section~\ref{sec:conclusion}).

%% file: tables/tab01_dimensions.tex
\begin{table}[!b]
\caption{Four dimensions of effective decentralization and their evidence status.}
\label{tab:dimensions}
\centering
\AVTableSetup
\setlength{\tabcolsep}{2.5pt}
\renewcommand{\arraystretch}{1.30}
\begin{tabularx}{\textwidth}{@{}P{24mm}P{28mm}P{32mm}Y@{}}
\toprule
\AVTableHeader
\textbf{Dimension} & \textbf{Indicators} & \textbf{Interpretive question} & \textbf{Evidence in this study} \\
\midrule
\AVInlineIcon[AVBlue]{users}\textbf{Participation} & Active addresses; entry, exit, and retention & How broadly is access and continued use distributed? & \AVObservedTag\ Address-level breadth. \AVBoundedTag\ People and firms are not identified. \\
\addlinespace[1.5pt]
\AVInlineIcon[AVViolet]{chart}\textbf{Activity distribution} & Event shares; HHI; top-address shares & How evenly is observed protocol activity shared? & \AVObservedTag\ Address-level event frequency. \AVUnavailableTag\ Causal effect is not identified. \\
\addlinespace[1.5pt]
\AVInlineIcon[AVOrange]{network}\textbf{Network structure} & Role-specific connections; centrality; core membership & Which addresses occupy central or persistent positions? & \AVObservedTag\ Address-role networks. \AVBoundedTag\ Economic control is not inferred. \\
\addlinespace[1.5pt]
\AVInlineIcon[AVTeal]{bridge}\textbf{Infrastructure dependence} & Route shares; bridge or facilitator reliance; loss after route failure & Can activity continue if a common service becomes unavailable? & \AVDiagnosticTag\ Protocol-adjacent addresses. \AVUnavailableTag\ Verified route events and failure losses. \\
\bottomrule
\end{tabularx}
\begin{tablenotes}[flushleft]
\AVTableNoteFont
\item \emph{Notes:} The four dimensions are reported separately. Addresses are observable protocol identifiers, not verified people or firms. Any secondary composite index must retain transparent components, directions, and weights.
\end{tablenotes}
\end{table}

%% file: figures/fig01_interdisciplinary_framework.tex
\begin{figure}[!t]
\centering
\suspendrefereelines
\begin{tikzpicture}[x=1cm,y=1cm,font=\AVFigureFont,line cap=round,line join=round]
\path[use as bounding box] (0,-1.82) rectangle (12.85,7.22);
\tikzset{
  top card/.style={rounded corners=2.2mm,draw=AVRule,line width=.65pt,fill=white},
  dim card/.style={rounded corners=2.0mm,line width=.68pt,align=left},
  flow/.style={-{Latex[length=1.35mm,width=1.02mm]},line width=.74pt,draw=AVBlue,
    shorten >=.6pt,shorten <=.6pt},
  micro/.style={rounded corners=1.3mm,draw=AVRule,line width=.52pt,fill=AVLight,
    font=\AVFigureTinyFont,align=center,inner xsep=1.5mm,inner ysep=1.15mm},
  dot/.style={circle,draw=AVBlue,line width=.65pt,fill=white,minimum size=3.4mm,inner sep=0pt}
}

\node[anchor=west,font=\AVFigureTitleFont,text=AVInk,text width=124mm,align=left]
  at (.18,7.03) {Aave through a multidimensional decentralization framework};

\draw[top card] (.12,4.92) rectangle (3.47,6.66);
\node[anchor=west,font=\AVFigureSmallFont\bfseries,text=AVBlue]
  at (.34,6.41) {Aave lending market};
\node[dot] (supplier) at (.67,5.72) {};
\node[anchor=north,font=\AVFigureTinyFont,text=AVInk] at (.67,5.47) {supplier};
\node[rounded corners=1.8mm,draw=AVBlue,line width=.72pt,fill=AVMist,
  minimum width=12mm,minimum height=8mm,font=\AVFigureSmallFont\bfseries,text=AVInk]
  (aave) at (1.74,5.72) {Aave};
\node[dot] (borrower) at (2.81,5.72) {};
\node[anchor=north,font=\AVFigureTinyFont,text=AVInk] at (2.81,5.47) {borrower};
\draw[flow] (supplier.east)--(aave.west);
\draw[flow] (aave.east)--(borrower.west);

\draw[top card] (3.79,4.92) rectangle (7.12,6.66);
\node[anchor=west,font=\AVFigureSmallFont\bfseries,text=AVViolet]
  at (4.01,6.41) {Protocol changes};
\node[circle,draw=AVTeal,line width=.72pt,fill=AVTeal!7,minimum size=8.0mm,
  font=\AVFigureSmallFont\bfseries,text=AVTeal] (gho) at (4.48,5.72) {G};
\node[anchor=north,font=\AVFigureTinyFont,text=AVInk] at (4.48,5.27) {GHO};
\node[micro,draw=AVBlue!70,fill=AVMist] (eth) at (5.64,5.88) {Ethereum};
\node[micro,draw=AVViolet!70,fill=AVLilac] (arb) at (6.43,5.45) {Arbitrum};
\draw[flow,draw=AVTeal] (gho.east)--(eth.west);
\draw[flow,draw=AVViolet] (eth.south east)--(arb.north);

\draw[top card] (7.44,4.92) rectangle (12.73,6.66);
\node[anchor=west,font=\AVFigureSmallFont\bfseries,text=AVTeal]
  at (7.66,6.41) {Integrated framework};
\node[micro,draw=AVBlue!70,fill=AVMist,text width=11mm,inner xsep=.7mm] (theory) at (8.22,5.70)
  {network\\theory};
\node[micro,draw=AVViolet!70,fill=AVLilac,text width=14mm,inner xsep=.7mm] (simulation) at (9.76,5.70)
  {agent-based\\simulation};
\node[micro,draw=AVTeal!75,fill=AVTeal!6,text width=18mm,inner xsep=.7mm] (evidence) at (11.55,5.70)
  {longitudinal\\comparative evidence};
\draw[flow,draw=AVBlue] (theory.east)--(simulation.west);
\draw[flow,draw=AVTeal] (simulation.east)--(evidence.west);

\draw[flow] (3.51,5.78)--(3.75,5.78);
\draw[flow,draw=AVViolet] (7.16,5.78)--(7.40,5.78);

\node[anchor=west,font=\AVFigureSmallFont\bfseries,text=AVInk] at (.16,4.69)
  {Four distinct outcomes};

\draw[dim card,draw=AVBlue!72,fill=AVMist] (.12,3.24) rectangle (6.31,4.51);
\pic[av/icon,draw=AVBlue,scale=.40] at (.51,4.14) {av/users};
\node[anchor=west,font=\AVFigureSmallFont\bfseries,text=AVBlue]
  at (.82,4.14) {Participation};
\node[anchor=west,font=\AVFigureTinyFont,text=AVInk] at (.34,3.75)
  {How many addresses take part?};
\node[anchor=west,font=\AVFigureTinyFont\bfseries,text=AVInk] at (.34,3.46)
  {ETH +91.0\%\quad ARB +1.7\%};

\draw[dim card,draw=AVViolet!72,fill=AVLilac] (6.54,3.24) rectangle (12.73,4.51);
\pic[av/icon,draw=AVViolet,scale=.40] at (6.93,4.14) {av/chart};
\node[anchor=west,font=\AVFigureSmallFont\bfseries,text=AVViolet]
  at (7.24,4.14) {Activity distribution};
\node[anchor=west,font=\AVFigureTinyFont,text=AVInk] at (6.76,3.75)
  {How evenly is observed activity shared?};
\node[anchor=west,font=\AVFigureTinyFont\bfseries,text=AVInk] at (6.76,3.46)
  {ETH $-31.5\%$\quad ARB +58.1\%};

\draw[dim card,draw=AVOrange!80,fill=AVWarm] (.12,1.73) rectangle (6.31,3.00);
\pic[av/icon,draw=AVOrange,scale=.40] at (.51,2.63) {av/network};
\node[anchor=west,font=\AVFigureSmallFont\bfseries,text=AVOrange]
  at (.82,2.63) {Structural position};
\node[anchor=west,font=\AVFigureTinyFont,text=AVInk] at (.34,2.24)
  {Which addresses occupy central roles?};
\node[anchor=west,font=\AVFigureTinyFont\bfseries,text=AVInk] at (.34,1.95)
  {role- and scale-aware evidence};

\draw[dim card,draw=AVTeal!75,fill=AVTeal!6] (6.54,1.73) rectangle (12.73,3.00);
\pic[av/icon,draw=AVTeal,scale=.40] at (6.93,2.63) {av/bridge};
\node[anchor=west,font=\AVFigureSmallFont\bfseries,text=AVTeal]
  at (7.24,2.63) {Infrastructure dependence};
\node[anchor=west,font=\AVFigureTinyFont,text=AVInk] at (6.76,2.24)
  {How much relies on common routes?};
\node[anchor=west,font=\AVFigureTinyFont\bfseries,text=AVInk] at (6.76,1.95)
  {additional route data needed};

\draw[rounded corners=1.8mm,draw=AVRule,line width=.60pt,fill=AVLight]
  (.12,-.25) rectangle (12.73,1.43);
\node[anchor=west,font=\AVFigureSmallFont\bfseries,text=AVInk]
  at (.36,1.15) {Current evidence boundary};
\node[anchor=north west,font=\AVFigureTinyFont,text=AVInk,text width=119mm,align=left]
  at (.36,.84) {Participation and activity concentration are observed at the address level. Structural evidence is role- and scale-dependent. Economic-actor control and route resilience require additional evidence.};

\draw[rounded corners=1.8mm,draw=AVOrange,dashed,line width=.62pt,fill=white]
  (.12,-1.64) rectangle (12.73,-.39);
\node[anchor=west,font=\AVFigureSmallFont\bfseries,text=AVOrange]
  at (.36,-.65) {Future extensions --- additional evidence required};
\node[anchor=west,font=\AVFigureTinyFont,text=AVInk]
  at (.36,-1.04) {Endogenous entry and links \(\cdot\) route dependence \(\cdot\) actor resolution};
\node[anchor=west,font=\AVFigureTinyFont,text=AVInk]
  at (.36,-1.34) {Dynamic multilayer networks \(\cdot\) AI-mediated decisions};
\end{tikzpicture}
\resumerefereelines
\caption{Interdisciplinary framework and evidence boundary. Aave was already multichain before GHO launched; the institutional sequence analyzed here is GHO issuance on Ethereum followed by GHO expansion to the existing Aave market on Arbitrum. The analysis integrates network theory, theorem-consistent agent-based simulation, and longitudinal comparative protocol-event evidence while reporting participation, activity distribution, structural position, and infrastructure dependence separately. Reported percentages compare pre- and post-activation weekly means after excluding week zero. They are descriptive, addresses are not verified economic actors, and route resilience is not measured. The lower dashed band marks prospective extensions rather than current evidence.}
\label{fig:interdisciplinary-framework}
\end{figure}

%% file: sections/02_background.tex
\section{Institutional setting and research gap}\label{sec:background}

\subsection{Aave's multichain footprint and GHO's cross-chain expansion}\label{subsec:institutional}

Aave is an open-source, non-custodial liquidity protocol in which suppliers provide assets to smart-contract markets and borrowers obtain overcollateralized loans. Its institutional importance here lies in three interacting layers: public blockchains record settlement; protocol markets organize supplying, borrowing, repayment, and liquidation; and governance controls deployments, risk parameters, and approved infrastructure. Decentralization can therefore refer to access, activity, network position, governance, or technical dependence rather than one property of the codebase.

Aave's protocol footprint and GHO's asset footprint have different histories. Aave V3 first deployed on six networks in March 2022, with Arbitrum among the initial deployments, and reached Ethereum in January 2023 \cite{aave2026v3history,aave2026arbitrum}. Thus, the July 2024 event studied below is not Aave's arrival on Arbitrum; it is GHO's arrival in an Aave market that had already operated there for more than two years. Current official records list Aave deployments on Ethereum, Polygon, Avalanche, Arbitrum, Optimism, Base, BNB Chain, Scroll, Metis, Gnosis, ZKsync Era, Linea, Sonic, Celo, Soneium, Plasma, Fantom, and Harmony \cite{aave2026deployments}. These deployments define a broader comparative population, not a single common treatment date.

GHO extends this ecosystem vertically by adding a protocol-native, overcollateralized stablecoin. Governance approved its launch on Ethereum with the Aave V3 market and a FlashMinter as initial facilitators; GHO launched on 15 July 2023 \cite{aave2023gholaunch,aave2023stateq2}. Facilitators can mint and burn within governance-defined capacity limits, while borrower interest flows to the Aave DAO \cite{aavegho2026,aavefacilitators2026}. This design can deepen activity and attract participants, but it can also concentrate issuance capacity, liquidity, or routing. Stablecoin growth and decentralization are therefore distinct outcomes.

Cross-chain GHO extends the same asset horizontally. Ethereum remains the issuance anchor while approved facilitators and Chainlink's Cross-Chain Interoperability Protocol support canonical representations on destination chains \cite{aave2024crosschainlaunch,aavebridge2026}. Arbitrum was the first destination and GHO launched there on 2 July 2024, with bridge capacity subsequently expanded \cite{aave2024ghoarbitrumdate,aave2024arbitrumcapacity}. Cross-chain access can broaden venue choice while increasing common dependence on a limited route set. This tension motivates the distinction between chain dispersion and infrastructure resilience.

The GHO footprint continued to expand after the two events analyzed in this paper. By July 2025, GHO was available on Base, Lens, and Avalanche in addition to Ethereum and Arbitrum; by August 2025, a Gnosis market had also been deployed \cite{aave2025gholaterchains,aave2025ghognosis}. Later deployments are verified candidates for future longitudinal study, not observations in the present Ethereum--Arbitrum panels. This distinction is especially important for Gnosis: the paper's same-calendar 2024 Gnosis series is used only to test comparison suitability, whereas the later GHO deployment creates a separate prospective study cohort.

These changes provide verifiable clocks, not automatic experiments. Governance discussion, anticipated demand, market conditions, and chain-specific trends can precede implementation. We use the clocks to organize transparent longitudinal windows and reserve causal interpretation for designs with credible counterfactual support. Figure~\ref{fig:institutional-timeline} distinguishes Aave's pre-existing multichain development from the two GHO events examined here, identifies the weekly observation windows, and connects the comparisons to the four separately reported dimensions of decentralization.

\input{figures/fig02_institutional_timeline}

\subsection{What existing literatures explain---and leave disconnected}\label{subsec:literature}

Economics explains how strategic interaction and network position shape aggregate behavior. Fixed-network games characterize equilibrium amplification and key-player logic \cite{jackson1996strategic,galeotti2010network,ballester2006keyplayer,bramoulle2014strategic}; graphon games extend this logic to approximations and interventions in large heterogeneous networks \cite{parise2023graphon}. The present model retains a finite, observed network and uses the large-network literature as an extension rather than as an implemented estimator.

Finance focuses on intermediation, liquidity, concentration, and the allocation of rents. Research on lending protocols studies interest-rate and liquidity mechanisms \cite{gudgeon2020loanable}, while broader \DeFi{} research shows that open architecture can coexist with concentrated governance, operational dependencies, or new intermediaries \cite{schar2021defi,aramonte2021defi,makarov2022cryptocurrencies,jensen2021governance,werner2022sokdefi}. Recent empirical synthesis documents the growing use of transaction-level data while emphasizing persistent measurement and identification challenges \cite{leon2026data}. Recent decentralized-exchange studies further show that execution infrastructure can redirect arbitrage rents and that priority-fee bidding contains information about trader behavior \cite{capponi2025liquidity,capponi2026price}. These market-microstructure results motivate attention to infrastructure and role-specific behavior, but they do not establish corresponding effects in Aave lending markets.

Computer-science research makes protocol execution observable while exposing technical dependencies. Work on blockchain interoperability distinguishes architectures and trust assumptions \cite{belchior2021survey}; systematizations of \DeFi{} emphasize composability, smart-contract execution, and protocol risks \cite{schar2021defi,werner2022sokdefi}, while transaction-level composition research shows how nominally separate protocols become interdependent through nested financial and software relations \cite{kitzler2023disentangling}. Record linkage adds a separate warning: repeated addresses are not verified people or organizations without an evidence-based entity model \cite{binette2022entity}. Thus public execution records improve measurement without resolving common control or route dependence by themselves.

Network science supplies measures of centrality, coreness, communities, temporal change, and multilayer organization \cite{borgatti2000models,rombach2017core,kojaku2018core,holme2012temporal,kivela2014multilayer}. Recent methods distinguish nodewise from layerwise connectivity and detect core--periphery structure jointly across nodes and layers \cite{presigny2024duality,bergermann2025multilayercore}. They sharpen the future multilayer agenda, but the present study does not implement those estimators because verified interlayer route edges are unavailable. Financial-network research provides the economic complement: dependencies can support matching while transmitting shocks through central institutions \cite{battiston2012debtrank,elliott2014financial,acemoglu2015systemic,demange2018threat}.

Causal-inference research explains why public event dates do not identify effects by themselves. Serial dependence and few comparison units complicate uncertainty \cite{bertrand2004trust,conley2011inference,newey1987simple}; segmented and controlled interrupted-time-series designs separate level and slope changes \cite{wagner2002segmented,bernal2017interrupted,bernal2018controls}; and modern difference-in-differences work shows why staggered adoption, heterogeneous responses, limited comparison support, anticipation, and divergent pre-trends require designs beyond simple before--after contrasts \cite{callaway2021difference,goodmanbacon2021difference,sun2021estimating,roth2022pretest,rambachan2023credible,roth2023trending,borusyak2024revisiting}. These methods discipline the present comparison and specify requirements for future multi-cohort designs; they do not make the current one-comparison-chain coefficient causal.

We also adopt a narrow reporting lesson from recent applied work: show outcome paths and treatment timing, expose heterogeneity and comparison movement, and keep the identifying design distinct from the substantive mechanism \cite{azar2024minimum,brynjolfsson2025generative,colmer2025pricing}. These studies are visual and reporting exemplars only. They are not identification authority for the Aave setting and do not supply a counterfactual here.

Figure~\ref{fig:literature-landscape} makes the integration chronological. It traces selected foundations, separates DeFi and interoperability evidence from identification methods, and then shows their hand-off to the implemented protocol-event, typed-network, simulation, and comparative objects. The diagram is an intellectual lineage rather than a citation count. The earlier fixed 18-work method--domain--evidence coding remains available as provenance data in \texttt{generated/data/}; because it is neither exhaustive nor a measure of intellectual weight, its Sankey is no longer rendered or numbered in the manuscript.

\input{figures/fig02b_literature_landscape}

\subsection{Research gap and positioning}

The missing link is an integrated framework that moves from economic mechanism to an observable protocol network while preserving the limits of each source of evidence. The component literatures establish strategic amplification, financial intermediation, auditable execution, structural measurement, and identification requirements, but they authorize different claims from different evidence objects. This paper's contribution is therefore integrative rather than a claim to invent each component: it connects mechanism, simulation, role-aware temporal networks, scale-adjusted structure, and longitudinal evidence within one four-dimensional framework.

%% file: figures/fig02_institutional_timeline.tex
\begin{figure}[!tbp]
\centering
\suspendrefereelines
\begin{tikzpicture}[x=1cm,y=1cm,font=\AVFigureFont,line cap=round,line join=round]
\path[use as bounding box] (0,0) rectangle (12.85,8.20);
\tikzset{
  panel/.style={rounded corners=2mm,draw=AVRule,line width=.55pt,fill=AVLight},
  event/.style={rounded corners=1.4mm,draw=AVBlue!75,line width=.62pt,fill=white},
  future/.style={rounded corners=1.4mm,draw=AVOrange!75,line width=.62pt,dashed,fill=AVWarm},
  title/.style={anchor=west,font=\AVFigureTitleFont,text=AVInk,inner sep=0pt},
  small/.style={font=\AVFigureSmallFont,text=AVInk,align=center,inner sep=0pt},
  tiny/.style={font=\AVFigureTinyFont,text=AVInk,align=center,inner sep=0pt},
  arrow/.style={-{Latex[length=1.6mm,width=1.05mm]},line width=.72pt,draw=AVBlue}
}
\node[title] at (.14,7.97) {Protocol evolution, GHO expansion, and longitudinal comparison design};

\draw[panel] (.12,5.55) rectangle (12.73,7.67);
\node[title] at (.34,7.43) {A. Aave was already multichain};
\pic[av/icon,draw=AVBlue,scale=.40] at (.60,6.50) {av/network};

\draw[event] (1.02,5.78) rectangle (4.05,7.12);
\node[tiny] at (2.535,6.89) {March 2022};
\node[small,font=\AVFigureSmallFont\bfseries] at (2.535,6.54) {Aave V3 launches};
\node[tiny,text width=2.65cm] at (2.535,6.15) {six networks, including Arbitrum};

\draw[event] (4.48,5.78) rectangle (7.65,7.12);
\node[tiny] at (6.065,6.89) {January 2023};
\node[tiny,font=\AVFigureTinyFont\bfseries] at (6.065,6.52) {Ethereum\\deployment};
\node[tiny] at (6.065,6.06) {before GHO issuance};

\draw[event,fill=AVMist] (7.92,5.78) rectangle (12.43,7.12);
\node[small,font=\AVFigureSmallFont\bfseries] at (10.175,6.70) {Broader Aave footprint};
\node[tiny,text width=3.95cm] at (10.175,6.18)
  {contextual markets; no common treatment date};

\draw[panel] (.12,2.90) rectangle (12.73,5.37);
\node[title] at (.34,5.13) {B. Focal GHO events and weekly observation windows};
\draw[draw=AVInk,line width=.68pt] (.70,3.85)--(9.95,3.85);
\draw[draw=AVBlue!45,line width=5.2pt] (1.00,3.85)--(2.45,3.85);
\draw[draw=AVTeal!45,line width=5.2pt] (2.82,3.85)--(4.27,3.85);
\draw[draw=AVBlue!45,line width=5.2pt] (5.42,3.85)--(6.87,3.85);
\draw[draw=AVTeal!45,line width=5.2pt] (7.24,3.85)--(8.69,3.85);
\node[tiny,text=AVBlue] at (1.72,3.48) {pre};
\node[tiny,text=AVTeal] at (3.55,3.48) {post};
\node[tiny,text=AVBlue] at (6.14,3.48) {pre};
\node[tiny,text=AVTeal] at (7.96,3.48) {post};
\node[tiny,text width=2.45cm] at (2.635,4.48) {15 July 2023\\\textbf{GHO issuance}\\Ethereum};
\node[tiny,text width=2.55cm] at (7.055,4.48) {2 July 2024\\\textbf{GHO expansion}\\Arbitrum};
\node[tiny,text=AVViolet] at (2.635,3.12) {activation week excluded};
\node[tiny,text=AVViolet] at (7.055,3.12) {activation week excluded};

\draw[future] (10.20,3.17) rectangle (12.43,4.78);
\node[tiny,font=\AVFigureTinyFont\bfseries] at (11.315,4.56) {Gnosis 2024};
\node[tiny,text width=1.90cm] at (11.315,3.93) {comparison diagnostic};

\draw[panel] (.12,.12) rectangle (12.73,2.70);
\node[title] at (.34,2.46) {C. Four questions reported separately};
\foreach \x/\col/\icon/\head/\body in {
  1.64/AVBlue/users/{Participation\\$\mathcal P_t$}/{Broader\\participation?},
  4.69/AVViolet/chart/{Activity\\$\mathcal A_t$}/{Lower activity\\concentration?},
  7.74/AVOrange/network/{Structure\\$\mathcal S_t$}/{Changed role\\structure?},
  10.79/AVTeal/bridge/{Dependence\\$\mathcal I_t$}/{Route data\\needed}
}{
  \draw[rounded corners=1.5mm,draw=\col!75,line width=.58pt,fill=white]
    (\x-1.36,.28) rectangle (\x+1.36,2.20);
  \pic[av/icon,draw=\col,scale=.34] at (\x-.95,1.42) {av/\icon};
  \node[tiny,font=\AVFigureTinyFont\bfseries,text=\col,text width=2.00cm] at (\x+.25,1.55) {\head};
  \node[tiny,text width=1.98cm] at (\x+.25,.73) {\body};
}
\end{tikzpicture}
\resumerefereelines
\vspace{2pt}
\caption{\textbf{Protocol evolution, GHO expansion, and longitudinal comparison design.}
Panel A distinguishes Aave's pre-existing multichain development from the GHO events examined here. Panel B shows GHO issuance on Ethereum on 15 July 2023 and its first cross-chain expansion to Arbitrum on 2 July 2024, with chain-specific pre- and post-event windows and the activation week excluded. The Gnosis series is retained as a comparison-suitability diagnostic. Panel C connects the design to four separately reported dimensions. Contract-based measures proxy part of infrastructure dependence; route concentration and resilience require additional route data. Later GHO destinations remain future-study cohorts.}
\label{fig:institutional-timeline}
\end{figure}

%% file: figures/fig02b_literature_landscape.tex
\begin{figure}[!p]
\centering
\suspendrefereelines
\resizebox{0.93\textwidth}{!}{%
\begin{tikzpicture}[x=1cm,y=1cm,font=\AVFigureFont,line cap=round,line join=round]
\path[use as bounding box] (0,0) rectangle (12.85,18.45);
\tikzset{
  lane rule/.style={draw=AVRule,line width=.42pt},
  axis/.style={draw=AVNavy,line width=.72pt},
  tick/.style={draw=AVNavy,line width=.54pt},
  code/.style={font=\AVFigureTinyFont\bfseries,inner sep=.25mm,align=center},
  key/.style={anchor=west,font=\AVFigureTinyFont,text=AVInk,inner sep=0pt},
  source/.style={font=\AVFigureTinyFont,text=AVInk,align=center,inner sep=0pt},
  hand flow/.style={-{Latex[length=1.22mm,width=.88mm]},line width=.62pt},
  boundary/.style={draw=AVNavy,line width=.68pt}
}

\node[anchor=west,font=\AVFigureTitleFont,text=AVInk] at (.14,18.25)
  {Chronological interdisciplinary landscape and evidence hand-off};
\node[anchor=west,font=\AVFigureTinyFont,text=AVInk] at (.14,17.94)
  {Selective intellectual lineage; position denotes publication chronology, not importance or evidentiary weight.};

\fill[AVMist] (.14,17.23) rectangle (7.33,17.60);
\node[anchor=west,font=\AVFigureSmallFont\bfseries,text=AVBlue] at (.27,17.41)
  {A\quad Foundations: mechanisms, structure, dynamics, and simulation};

\node[anchor=east,font=\AVFigureTinyFont,text=AVInk] at (2.39,16.72) {strategic games};
\node[anchor=east,font=\AVFigureTinyFont,text=AVInk] at (2.39,16.30) {core--periphery};
\node[anchor=east,font=\AVFigureTinyFont,text=AVInk] at (2.39,15.88) {financial systems};
\node[anchor=east,font=\AVFigureTinyFont,text=AVInk] at (2.39,15.46) {temporal / multilayer};
\node[anchor=east,font=\AVFigureTinyFont,text=AVInk] at (2.39,15.04) {agent simulation};

\node[code,text=AVBlue] at (2.60,16.72) {F1};
\node[code,text=AVBlue] at (5.95,16.72) {F2};
\node[code,text=AVBlue] at (7.30,16.72) {F3};
\node[code,text=AVBlue] at (8.65,16.72) {F4};
\node[code,text=AVViolet] at (3.95,16.30) {F5};
\node[code,text=AVViolet] at (8.65,16.30) {F6};
\node[code,text=AVViolet] at (10.00,16.30) {F7};
\node[code,text=AVViolet] at (12.35,16.30) {F8};
\node[code,text=AVOrange] at (7.92,15.88) {F9};
\node[code,text=AVOrange] at (8.52,15.88) {F10};
\node[code,text=AVOrange] at (9.15,15.88) {F11};
\node[code,text=AVOrange] at (10.00,15.88) {F12};
\node[code,text=AVTeal] at (7.97,15.46) {F13};
\node[code,text=AVTeal] at (8.65,15.46) {F14};
\node[code,text=AVTeal] at (12.02,15.46) {F15\hspace{.8mm}F16};
\node[code,text=AVBlue] at (5.95,15.04) {F17};
\node[code,text=AVBlue] at (10.67,15.04) {F18};
\node[code,text=AVBlue] at (12.02,15.04) {F19};

\draw[axis] (2.58,14.66)--(12.42,14.66);
\foreach \x/\yr in {2.60/1996,3.95/2000,5.95/2006,7.30/2010,8.65/2014,10.00/2018,12.35/2025}{
  \draw[tick] (\x,14.66)--(\x,14.57);
  \node[anchor=north,font=\AVFigureTinyFont,text=AVInk,inner sep=0pt] at (\x,14.44) {\yr};
}
\node[anchor=west,font=\AVFigureTinyFont\bfseries,text=AVInk] at (2.58,14.10)
  {Node index (author; publication year)};

\node[key] at (2.58,13.74) {\textcolor{AVBlue}{\textbf{F1}}\quad Jackson--Wolinsky (1996)};
\node[key] at (2.58,13.44) {\textcolor{AVBlue}{\textbf{F2}}\quad Ballester et al. (2006)};
\node[key] at (2.58,13.14) {\textcolor{AVBlue}{\textbf{F3}}\quad Galeotti et al. (2010)};
\node[key] at (2.58,12.84) {\textcolor{AVBlue}{\textbf{F4}}\quad Bramoull\'e et al. (2014)};
\node[key] at (2.58,12.54) {\textcolor{AVViolet}{\textbf{F5}}\quad Borgatti--Everett (2000)};
\node[key] at (2.58,12.24) {\textcolor{AVViolet}{\textbf{F6}}\quad Rombach et al. (2014)};
\node[key] at (2.58,11.94) {\textcolor{AVViolet}{\textbf{F7}}\quad Kojaku--Masuda (2018)};
\node[key] at (2.58,11.64) {\textcolor{AVViolet}{\textbf{F8}}\quad Bergermann et al. (2025)};
\node[key] at (2.58,11.34) {\textcolor{AVOrange}{\textbf{F9}}\quad Battiston et al. (2012)};
\node[key] at (2.58,11.04) {\textcolor{AVOrange}{\textbf{F10}}\quad Elliott et al. (2014)};

\node[key] at (7.55,13.74) {\textcolor{AVOrange}{\textbf{F11}}\quad Acemoglu et al. (2015)};
\node[key] at (7.55,13.44) {\textcolor{AVOrange}{\textbf{F12}}\quad Demange (2018)};
\node[key] at (7.55,13.14) {\textcolor{AVTeal}{\textbf{F13}}\quad Holme--Saram\"aki (2012)};
\node[key] at (7.55,12.84) {\textcolor{AVTeal}{\textbf{F14}}\quad Kivel\"a et al. (2014)};
\node[key] at (7.55,12.54) {\textcolor{AVTeal}{\textbf{F15}}\quad Presigny et al. (2024)};
\node[key] at (7.55,12.24) {\textcolor{AVTeal}{\textbf{F16}}\quad Clemente et al. (2024)};
\node[key] at (7.55,11.94) {\textcolor{AVBlue}{\textbf{F17}}\quad Tesfatsion--Judd (2006)};
\node[key] at (7.55,11.64) {\textcolor{AVBlue}{\textbf{F18}}\quad Grimm et al. (2020)};
\node[key] at (7.55,11.34) {\textcolor{AVBlue}{\textbf{F19}}\quad Li et al. (2024)};

\draw[AVRule,line width=.62pt] (.14,10.70)--(12.71,10.70);

\fill[AVLilac] (.14,9.98) rectangle (8.52,10.35);
\node[anchor=west,font=\AVFigureSmallFont\bfseries,text=AVViolet] at (.27,10.16)
  {B\quad DeFi, interoperability, empirical measurement, and identification};

\node[anchor=east,font=\AVFigureTinyFont,text=AVInk] at (2.39,9.47) {DeFi / interoperability};
\node[anchor=east,font=\AVFigureTinyFont,text=AVInk] at (2.39,9.05) {measurement / synthesis};
\node[anchor=east,font=\AVFigureTinyFont,text=AVInk] at (2.39,8.63) {ITS / serial dep. / few units};
\node[anchor=east,font=\AVFigureTinyFont,text=AVInk] at (2.39,8.21) {comparative designs};

\node[code,text=AVTeal] at (7.44,9.47) {D1};
\node[code,text=AVTeal] at (8.55,9.47) {D2\hspace{.7mm}D3\hspace{.7mm}D4};
\node[code,text=AVTeal] at (9.34,9.05) {M1};
\node[code,text=AVTeal] at (10.32,9.05) {M2\hspace{.7mm}M3};
\node[code,text=AVTeal] at (12.33,9.05) {M4};
\node[code,text=AVOrange] at (2.60,8.63) {I1};
\node[code,text=AVOrange] at (3.65,8.63) {I2};
\node[code,text=AVOrange] at (4.45,8.63) {I3};
\node[code,text=AVOrange] at (5.35,8.63) {I4};
\node[code,text=AVOrange] at (6.24,8.63) {I5};
\node[code,text=AVOrange] at (8.35,8.21) {C1};
\node[code,text=AVOrange] at (9.34,8.21) {C2};
\node[code,text=AVOrange] at (10.32,8.21) {C3};
\node[code,text=AVOrange] at (11.34,8.21) {C4};

\draw[axis] (2.58,7.82)--(6.76,7.82);
\draw[axis] (7.12,7.82)--(12.42,7.82);
\draw[axis] (6.82,7.72)--(6.94,7.92);
\draw[axis] (7.00,7.72)--(7.12,7.92);
\foreach \x/\yr in {2.60/1987,3.65/2002,4.45/2004,5.35/2011,6.24/{2017/18},7.44/2020,8.35/2021,9.34/2022,10.32/2023,11.34/2024,12.33/2026}{
  \draw[tick] (\x,7.82)--(\x,7.73);
  \node[anchor=north,font=\AVFigureTinyFont,text=AVInk,inner sep=0pt] at (\x,7.60) {\yr};
}
\node[anchor=west,font=\AVFigureTinyFont\bfseries,text=AVInk] at (2.58,7.27)
  {Node index (author; publication year)};

\node[key] at (2.58,6.92) {\textcolor{AVTeal}{\textbf{D1}}\quad Gudgeon et al. (2020)};
\node[key] at (2.58,6.64) {\textcolor{AVTeal}{\textbf{D2}}\quad Sch\"ar (2021)};
\node[key] at (2.58,6.36) {\textcolor{AVTeal}{\textbf{D3}}\quad Aramonte et al. (2021)};
\node[key] at (2.58,6.08) {\textcolor{AVTeal}{\textbf{D4}}\quad Belchior et al. (2021)};
\node[key] at (2.58,5.80) {\textcolor{AVTeal}{\textbf{M1}}\quad Werner et al. (2022)};
\node[key] at (2.58,5.52) {\textcolor{AVTeal}{\textbf{M2}}\quad Kitzler et al. (2023)};
\node[key] at (2.58,5.24) {\textcolor{AVTeal}{\textbf{M3}}\quad Lyons--Viswanath-Natraj (2023)};
\node[key] at (2.58,4.96) {\textcolor{AVTeal}{\textbf{M4}}\quad Castillo Le\'on--Lehar (2026)};

\node[key] at (8.18,6.92) {\textcolor{AVOrange}{\textbf{I1}}\quad Newey--West (1987)};
\node[key] at (8.18,6.64) {\textcolor{AVOrange}{\textbf{I2}}\quad Wagner et al. (2002)};
\node[key] at (8.18,6.36) {\textcolor{AVOrange}{\textbf{I3}}\quad Bertrand et al. (2004)};
\node[key] at (8.18,6.08) {\textcolor{AVOrange}{\textbf{I4}}\quad Conley--Taber (2011)};
\node[key] at (8.18,5.80) {\textcolor{AVOrange}{\textbf{I5}}\quad Bernal et al. (2017/18)};
\node[key] at (8.18,5.52) {\textcolor{AVOrange}{\textbf{C2}}\quad Roth (2022)};
\node[key] at (8.18,5.24) {\textcolor{AVOrange}{\textbf{C4}}\quad Borusyak et al. (2024)};

\node[key] at (2.58,4.63) {\textcolor{AVOrange}{\textbf{C1}}\quad Callaway--Sant'Anna; Goodman-Bacon; Sun--Abraham (2021)};
\node[key] at (2.58,4.33) {\textcolor{AVOrange}{\textbf{C3}}\quad Rambachan--Roth; Roth et al. (2023)};

\draw[AVRule,line width=.62pt] (.14,4.08)--(12.71,4.08);

\fill[AVTeal!8] (.14,3.60) rectangle (5.86,3.97);
\node[anchor=west,font=\AVFigureSmallFont\bfseries,text=AVTeal] at (.27,3.82)
  {C\quad Contribution and evidence hand-off};

\node[source,text width=17mm] at (1.10,3.16) {network\\mechanisms};
\node[source,text width=17mm] at (3.10,3.16) {core / structural\\networks};
\node[source,text width=17mm] at (5.10,3.16) {temporal /\\multilayer};
\node[source,text width=17mm] at (7.10,3.16) {DeFi / stablecoin\\institutions};
\node[source,text width=17mm] at (9.10,3.16) {empirical\\identification};
\node[source,text width=17mm] at (11.10,3.16) {entity /\\measurement limits};

\draw[hand flow,draw=AVBlue] (1.10,2.78)--(1.10,2.70);
\draw[hand flow,draw=AVViolet] (3.10,2.78)--(3.10,2.70);
\draw[hand flow,draw=AVTeal] (5.10,2.78)--(5.10,2.70);
\draw[hand flow,draw=AVViolet] (7.10,2.78)--(7.10,2.70);
\draw[hand flow,draw=AVOrange] (9.10,2.78)--(9.10,2.70);
\draw[hand flow,draw=AVOrange] (11.10,2.78)--(11.10,2.70);
\draw[AVTeal,line width=.74pt] (1.10,2.66)--(11.10,2.66);

\node[anchor=west,font=\AVFigureSmallFont\bfseries,text=AVTeal] at (1.10,2.43)
  {This study: evidence-gated protocol-network analysis};

\draw[AVBlue,line width=1.25pt] (.78,1.95)--(1.26,1.95);
\draw[AVBlue,-{Latex[length=1.12mm,width=.82mm]},line width=.64pt] (1.02,2.12)--(1.02,1.78);
\node[source,text width=42mm,anchor=west,align=left] at (1.35,1.95) {protocol-event semantics};

\draw[AVTeal,line width=1.25pt] (6.65,1.95)--(7.13,1.95);
\draw[AVTeal,-{Latex[length=1.12mm,width=.82mm]},line width=.64pt] (6.89,2.12)--(6.89,1.78);
\node[source,text width=48mm,anchor=west,align=left] at (7.22,1.95) {typed address-role networks};

\draw[AVViolet,line width=1.25pt] (.78,1.37)--(1.26,1.37);
\draw[AVViolet,-{Latex[length=1.12mm,width=.82mm]},line width=.64pt] (1.02,1.54)--(1.02,1.20);
\node[source,text width=50mm,anchor=west,align=left] at (1.35,1.37) {network theory + theorem-consistent\\simulation};

\draw[AVOrange,line width=1.25pt] (6.65,1.37)--(7.13,1.37);
\draw[AVOrange,-{Latex[length=1.12mm,width=.82mm]},line width=.64pt] (6.89,1.54)--(6.89,1.20);
\node[source,text width=48mm,anchor=west,align=left] at (7.22,1.37) {four dimensions + common-calendar\\comparison};

\draw[AVOrange,dashed,line width=.72pt] (.68,.90)--(12.30,.90);
\node[anchor=west,font=\AVFigureTinyFont\bfseries,text=AVOrange] at (.78,.62)
  {Additional evidence required};
\node[anchor=west,font=\AVFigureTinyFont,text=AVInk,text width=116mm,align=center]
  at (.64,.27) {Verified economic-actor links \(\cdot\) route/dependency maps \(\cdot\) multiple credible comparison cohorts \(\cdot\) endogenous and dynamic networks.};

\end{tikzpicture}
}
\resumerefereelines
\caption{\textbf{Chronological interdisciplinary literature landscape and contribution boundary.}
Panel A traces selective foundations; Panel B separates DeFi and interoperability evidence from identification methods; Panel C hands those streams to the implemented contribution and a dashed future-evidence boundary. Position encodes chronology within a stream. Arrows denote author synthesis, not citation flows, influence, counts, or causal support. No item below the boundary is current evidence.}
\label{fig:literature-landscape}
\end{figure}

%% file: sections/03_model_simulation.tex
\section{Network theory and theorem-consistent agent-based simulation}\label{sec:model}

This section separates three evidence objects. The analytical model establishes conditional implications on a fixed economic-interaction network. The agent-based simulations test whether transparent update dynamics recover those implications and expose their adjustment paths, following the broader agent-based computational-economics tradition \cite{tesfatsion2006handbook}. The longitudinal analysis in Sections~\ref{sec:data}--\ref{sec:causal} measures address-level outcomes around protocol events. Simulation is therefore a bridge from mechanism to observable questions, not a substitute for empirical evidence.

\subsection{Economic-interaction model}\label{subsec:model}

Let $N=\{1,\ldots,n\}$ be potential actors, $x_i\geq0$ actor $i$'s activity, and $G=(g_{ij})$ a nonnegative, possibly directed economic-interaction matrix. The entry $g_{ij}$ records how strongly actor $j$'s activity enters actor $i$'s incentive, so the corresponding influence edge is $j\rightarrow i$. This theoretical matrix is not automatically identical to the empirical action-side-to-position-holder network constructed from Pool events. Table~\ref{tab:model-notation} summarizes the objects used below.

\input{tables/tab_model_notation}

Under protocol condition $p$, $b_i(p)$ is actor $i$'s direct net benefit and $\beta\geq0$ is the strength of strategic complementarity. Conditional on $G$ and $p$, the canonical linear--quadratic payoff is
\begin{equation}
u_i(x_i,x_{-i};p)=b_i(p)x_i-\frac12x_i^2
+\beta x_i\sum_j g_{ij}x_j.
\label{eq:payoff-main}
\end{equation}
Accordingly,
\begin{equation}
\frac{\partial x_i}{\partial\left(\sum_jg_{ij}x_j\right)}=\beta
\label{eq:beta-interpretation}
\end{equation}
on an interior best response. Thus $\beta$ measures how strongly an actor responds to weighted activity by economically connected actors, while $G$ determines where that response can propagate. In an Aave analogy, additional liquidity from Bob changes Alice's modeled borrowing or participation little when $\beta$ is low and more strongly when $\beta$ is high; Alice's response can then affect other connected actors. The parameter is not an interest rate, probability, number of links, estimated GHO effect, or blockchain-specific constant.

Because raw $\beta$ depends on the scale of $G$, define normalized network-feedback strength as
\begin{equation}
\kappa=\beta\rho(G),
\label{eq:kappa}
\end{equation}
where $\rho(G)$ is the spectral radius. The simulations normalize $G$ so that $\rho(G)=1$, making $\kappa=\beta$. Protocol events enter as heterogeneous direct-benefit changes, $b^1=b^0+\Delta b$. GHO issuance on Ethereum can represent access, borrowing, liquidity, arbitrage, repayment, or liquidation opportunities; expansion to Arbitrum can additionally represent destination availability, lower transaction costs, integrations, liquidity, and infrastructure exposure. These analogies motivate $\Delta b_i$ but do not identify its historical value or imply that every actor benefits.

\begin{lemma}[Canonical optimal response]\label{lem:optimal-response}
For fixed $x_{-i}$, Equation~\eqref{eq:payoff-main} is strictly concave in $x_i$ and has the unique best response
\begin{equation}
x_i=\max\!\left\{0,\ b_i(p)+\beta\sum_jg_{ij}x_j\right\}.
\label{eq:best-response}
\end{equation}
On an interior active set, the stacked response system is $x=b(p)+\beta Gx$.
\end{lemma}

We use a short-run benchmark: $G$ is predetermined during a comparison, the same active set remains interior, and $\beta\rho(G)<1$. This isolates the intensive margin $x_i^0\rightarrow x_i^1$. Entry, exit, and structural change $G_0\rightarrow G_1$ remain outside the present equilibrium result. Appendix~\ref{app:formal-model} gives the full proofs, richer benefit function, and boundary cases.

\subsection{Three analytical results}\label{subsec:theory-results}

\subsubsection{Network amplification}

\begin{lemma}[Walk expansion]\label{lem:walk-expansion}
Under the maintained assumptions,
\begin{equation}
(I-\beta G)^{-1}=I+\beta G+\beta^2G^2+\cdots,
\label{eq:walk-expansion}
\end{equation}
and every term is elementwise nonnegative. The $(i,j)$ entry of $G^k$ aggregates length-$k$ directed walks from actor $j$ to actor $i$.
\end{lemma}

The equilibrium is
\begin{equation}
x^{*}(G,p)=(I-\beta G)^{-1}b(p).
\label{eq:equilibrium}
\end{equation}

\begin{theorem}[Unique equilibrium and network amplification]\label{thm:equilibrium-multiplier}
Fix $G$ and an interior active set. If $G$ is nonnegative, $b(p)>0$, and $\beta\rho(G)<1$, the game has the unique equilibrium in Equation~\eqref{eq:equilibrium}. For any scalar change $t$ with $\partial b/\partial t\geq0$,
\begin{equation}
\frac{\partial x^{*}}{\partial t}=(I-\beta G)^{-1}\frac{\partial b}{\partial t}\geq0.
\label{eq:network-multiplier-main}
\end{equation}
A direct gain can therefore increase activity among actors reached through positive-weight network paths.
\end{theorem}

\paragraph{Meaning, analogy, and boundary.}
Mathematically, the inverse multiplier propagates a direct-benefit change along weighted walks. In network terms, feedback strength and path structure jointly determine amplification. In an Aave analogy, an opportunity first available to a supplier, borrower, liquidator, or intermediary can affect connected roles as their incentives respond. A three-actor line gives the intuition: Alice responds directly, Bob responds to Alice, and Chen responds through Bob. The result does not establish that the empirical Aave network equals $G$, that links are exogenous, or that GHO caused an observed change. Existence, uniqueness, and the multiplier are inherited from canonical network games \cite{ballester2006keyplayer,bramoulle2014strategic,jackson2015games}; the paper-specific use is to connect heterogeneous protocol benefits to separately measured decentralization outcomes.

\subsubsection{Growth and concentration}

Let $X=\sum_i x_i^{*}$, $q_i=x_i^{*}/X$, and
\begin{equation}
H_x=\sum_iq_i^2.
\label{eq:activity-hhi}
\end{equation}

\begin{lemma}[How an actor's share changes]\label{lem:share-change}
Along a differentiable change $t$ that preserves the active set, let $m_i=\partial x_i^{*}/\partial t$ and $M=\sum_i m_i$. Then
\begin{equation}
\frac{\partial q_i}{\partial t}=\frac{m_i-q_iM}{X}.
\label{eq:share-change}
\end{equation}
Actor $i$ gains share precisely when its marginal gain exceeds its current share of the aggregate gain.
\end{lemma}

\begin{theorem}[Growth--concentration divergence]\label{thm:growth-concentration}
Under the conditions of Lemma~\ref{lem:share-change},
\begin{equation}
\frac{\partial H_x}{\partial t}=\frac{2}{X}\sum_i(q_i-H_x)m_i.
\label{eq:hhi-sign-main}
\end{equation}
The sign of concentration change is therefore independent of the growth condition $\partial X/\partial t=\sum_i m_i>0$.
\end{theorem}

\paragraph{Meaning, analogy, and boundary.}
Mathematically, HHI falls when marginal activity is sufficiently tilted toward actors below the concentration-weighted share benchmark, remains unchanged under proportional growth, and rises when gains favor actors already above that benchmark. In network terms, amplification says how a response spreads; the recipients determine whether shares converge or diverge. In an Aave analogy, the same aggregate expansion can reflect smaller holders catching up, all holders growing proportionally, or large incumbents pulling ahead. The theorem does not identify which allocation occurred historically, translate addresses into independent economic actors, or assign a causal mechanism to a measured HHI change.

\subsubsection{Chain dispersion and infrastructure resilience}

If removing route $k$ reduces feasible activity from $X$ to $X^{-k}$, define route-removal loss
\begin{equation}
L_k=(X-X^{-k})/X.
\label{eq:removal-loss}
\end{equation}
Let $c_v$ be the share of activity recorded on chain $v$, so chain concentration is $H_c=\sum_vc_v^2$, and let $r_k$ be the share made immediately unreachable by failure of route $k$ before behavioral substitution.

\begin{lemma}[Location shares do not identify route exposure]\label{lem:location-route}
For destination share $s>0$ and $R\geq1$ routes, any nonnegative allocation satisfying $\sum_kr_k=s$ has the same chain shares, while
\begin{equation}
\frac{s}{R}\leq\max_k r_k\leq s.
\label{eq:route-exposure-bounds}
\end{equation}
\end{lemma}

\begin{theorem}[Chain dispersion does not imply route resilience]\label{thm:dispersion-resilience}
Suppose activity initially lies on one chain and share $s\in(0,1/2]$ moves to a destination chain. Chain HHI falls by $2s(1-s)$. If all destination activity depends on one non-substitutable route, direct single-route exposure remains $s$. If that exposure is divided equally across $R$ independent, non-overlapping routes, the largest direct single-route exposure is $s/R$.
\end{theorem}

\paragraph{Meaning, analogy, and boundary.}
Mathematically, venue shares leave route allocation underidentified. In network terms, where activity appears and which dependencies keep it reachable are different layers. In the focal analogy, Ethereum and Arbitrum can have balanced activity shares while remote activity still relies on one designated interoperability stack; several nominal routes can also share one critical component. The theorem does not estimate a historical route-removal loss, establish independence among internal components, or model adaptive rerouting. It characterizes the data needed to test infrastructure resilience rather than inferring resilience from chain count.

\subsection{Dynamic agent-based validation of the theoretical mechanisms}\label{subsec:simulation}

The simulations implement Equation~\eqref{eq:best-response} with transparent rule-based agents in Mesa 3 and directed networks in NetworkX; NumPy/SciPy supplies an independent analytical benchmark \cite{terhoeven2025mesa}. They illustrate mechanisms implied by the analytical results. They are not calibrated estimates of historical Aave behavior and do not constitute causal evidence about GHO issuance or cross-chain expansion.

All experiments use 40 actors. For stable configurations, each run stops when $\lVert x^{(\ell+1)}-x^{(\ell)}\rVert_2<10^{-9}$ or at its disclosed horizon (1,500 iterations in the general protocol and 700 for Experiment 2) and is checked against Equation~\eqref{eq:equilibrium}. The primary network uses deterministic seed 20260819; robustness exercises use five seeds across sparse, core--periphery, and modular networks and simultaneous, fixed sequential, and randomized asynchronous activation. The named feedback conditions are $\kappa\in\{0,0.25,0.50,0.75,0.95\}$. The $\kappa=1.05$ assumption-violation test is reported separately and is not pooled with theorem-consistent cases. Appendix~\ref{app:simulation-protocol} provides the complete protocol, and Appendix~\ref{app:simulation-sensitivity} reports validation and sensitivity checks.

\begin{figure}[p]
\centering
\suspendrefereelines
\includegraphics[width=\textwidth,height=.78\textheight,keepaspectratio]{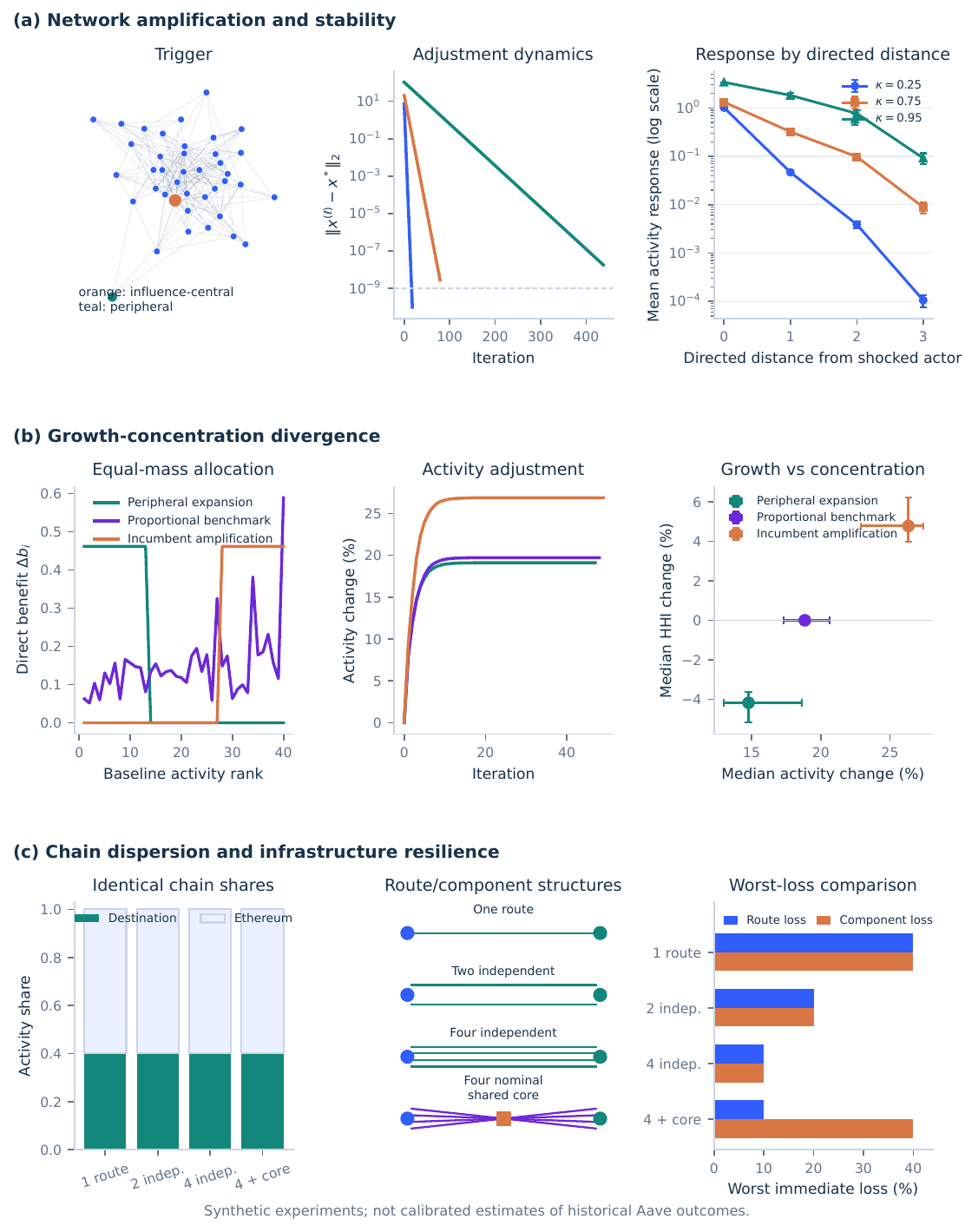}
\resumerefereelines
\caption{Theorem-consistent agent-based simulations. (a) A direct benefit propagates through a directed economic-interaction network. The middle panel shows distance to equilibrium; the right panel reports mean activity response by directed distance for $\kappa=0.25,0.75,0.95$ on a logarithmic response scale, with error bars equal to plus or minus one standard error across actors at each distance. Distance zero contains the single shocked actor. Stronger stable feedback increases amplification, persistence across network distance, and convergence time. (b) Equal-mass benefit allocations make total activity grow while activity HHI falls, remains unchanged, or rises according to the recipients of marginal activity. Points show deterministic seeds and intervals span the 5th--95th percentiles. (c) Identical 60/40 chain shares imply chain HHI of 0.52 in every synthetic configuration, while worst immediate loss depends on independent routes and shared components. All structures and outcomes are synthetic and are not calibrated estimates of historical Aave behavior or deployed-route failures.}
\label{fig:three-theory-simulations}
\label{fig:simulation-outcomes}
\end{figure}

Figure~\ref{fig:three-theory-simulations} places the three matched experiments on one page. Row (a) applies equal-mass shocks to influence-central, peripheral, and broad actor sets, traces update iterations, and compares responses over directed distance at $\kappa=0.25,0.75,0.95$. Row (b) holds shock mass fixed while allocating benefits toward peripheral actors, in proportion to baseline activity, or toward incumbents; seed-level points and 5th--95th simulation intervals remain categorical. Row (c) fixes activity shares at 60\% on the issuance chain and 40\% on a synthetic destination while varying independent routes and shared components.

\input{generated/simulation/simulation_results_macros}

The first experiment converges in \SimIterWeak{}, \SimIterStrong{}, and \SimIterNear{} simultaneous iterations at $\kappa=0.25,0.75,0.95$; the maximum stable analytical-equilibrium error is \SimMaxAnalyticError{}. The $\kappa=1.05$ boundary test does not converge. At directed distances $0,1,2,3$, mean activity responses are $(1.015,0.0467,0.00391,0.000106)$ for $\kappa=0.25$, $(1.326,0.321,0.0980,0.00887)$ for $\kappa=0.75$, and $(3.399,1.822,0.768,0.0918)$ for $\kappa=0.95$. The logarithmic axis keeps the weak-feedback profile visible. The separation among the curves shows that stronger complementarity raises not only the direct response but also the magnitude retained after additional network steps; the error bars describe within-network heterogeneity among actors at the same distance rather than historical sampling uncertainty.

The second experiment uses ten deterministic network seeds. Median activity and HHI changes are \SimPeripheralActivity{} and \SimPeripheralHHI{} under peripheral expansion, \SimProportionalActivity{} and approximately zero under proportional growth, and \SimIncumbentActivity{} and \SimIncumbentHHI{} under incumbent amplification. These are different benefit allocations, not different values of $\beta$. In the third experiment, chain HHI remains \SimChainHHI{} while worst route loss falls from 40\% to 10\% as one route becomes four independent routes; four nominal routes sharing one critical component retain a 40\% component-loss exposure.

\input{generated/simulation/simulation_results_table}

\subsection{From mechanisms to observable evidence}

Table~\ref{tab:theory-evidence-bridge} maps each theoretical result to its simulation output and empirical counterpart. The first experiment asks whether a protocol-related benefit is followed by activity beyond its most directly exposed actors, but direct exposure and the theoretical economic links are only partially observed. The second establishes why aggregate growth is insufficient to infer decentralization: holder and action-side HHI must be reported alongside participation. The third shows why dispersion across blockchains cannot establish resilience without verified information on route independence, substitutability, and shared components.

\input{tables/tab_theory_evidence_bridge}

The empirical analysis therefore reports participation, activity distribution, structural position, and infrastructure proxies separately and treats route-level resilience as theoretically characterized but presently unmeasured. Synthetic and empirical percentage changes are not compared as if they shared units, actors, or counterfactuals. Before results are interpreted, Section~\ref{sec:causal} states all joint activity--concentration directions; theoretical labels are applied only as descriptive classifications after the measured outcomes are presented.

\FloatBarrier

%% file: tables/tab_model_notation.tex
\begin{table}[t]
\centering
\caption{Core notation and evidence interpretation.}
\label{tab:model-notation}
\AVTableSetup
\begin{tabularx}{\textwidth}{P{0.10\textwidth}P{0.30\textwidth}P{0.28\textwidth}Y}
\toprule
Object & Mathematical role & Reader-facing interpretation & Evidence boundary \\
\midrule
$x_i$ & Nonnegative action of actor $i$ & Stylized protocol-activity intensity & Not a transaction count, token amount, or identified wallet \\
$b_i(p)$ & Direct net benefit under condition $p$ & Actor-specific access, cost, liquidity, or opportunity & Not an estimated GHO treatment effect \\
$g_{ij}$ & Weight with which $j$ enters $i$'s incentive & Directed economic influence $j\rightarrow i$ & Not automatically the observed Pool-event network \\
$\beta$ & Marginal response to weighted connected activity & Strategic-complementarity strength & Not an interest rate, probability, link count, or chain constant \\
$\kappa=\beta\rho(G)$ & Scale-normalized feedback strength & Comparable amplification condition across networks & Model parameter; not a decentralization score \\
$H_x$ & Sum of squared actor activity shares & Activity concentration & Address/entity interpretation requires separate evidence \\
$H_c$ & Sum of squared chain activity shares & Venue dispersion & Does not identify route independence or resilience \\
$L_k$ & Activity unavailable after route removal & Synthetic route/component stress loss & Not empirically estimated in this study \\
\bottomrule
\end{tabularx}
\end{table}

%% file: generated/simulation/simulation_results_macros.tex
\newcommand{\SimIterWeak}{18}
\newcommand{\SimIterStrong}{79}
\newcommand{\SimIterNear}{438}
\newcommand{\SimMaxAnalyticError}{1.90e-08}
\newcommand{\SimChainHHI}{0.52}
\newcommand{\SimPeripheralActivity}{+14.78\%}
\newcommand{\SimPeripheralHHI}{-4.17\%}
\newcommand{\SimProportionalActivity}{+18.85\%}
\newcommand{\SimProportionalHHI}{-0.00\%}
\newcommand{\SimIncumbentActivity}{+26.34\%}
\newcommand{\SimIncumbentHHI}{+4.77\%}

%% file: generated/simulation/simulation_results_table.tex
\begin{table}[t]
\centering
\caption{Theorem-to-simulation checks and evidence boundaries.}
\label{tab:theory-simulation-results}
\AVTableSetup
\begin{tabularx}{\textwidth}{P{0.18\textwidth}P{0.27\textwidth}P{0.27\textwidth}Y}
\toprule
Analytical result & Synthetic experiment & Verified numerical pattern & Evidence boundary \\
\midrule
Network amplification & Central benefit shock at $\kappa=0.25,0.75,0.95$ & 18, 79, and 438 iterations; maximum stable analytical error 1.90e-08 & Fixed synthetic network; no participant entry \\
Growth--concentration divergence & Equal-mass peripheral, proportional, and incumbent allocations & Peripheral expansion: $\Delta X=+14.78\%$, $\Delta H_x=-4.17\%$; Proportional: $\Delta X=+18.85\%$, $\Delta H_x=-0.00\%$; Incumbent amplification: $\Delta X=+26.34\%$, $\Delta H_x=+4.77\%$ & Ten synthetic seeds; fixed active set \\
Chain dispersion versus resilience & Identical 60/40 chain shares with alternative route/component structures & Chain HHI $0.52$ throughout; worst route loss $40\%,20\%,10\%,10\%$; shared-component loss reaches $40\%$ in the final case & Static synthetic stress construction, not observed bridge failure \\
\bottomrule
\end{tabularx}
\end{table}

%% file: tables/tab_theory_evidence_bridge.tex
\begin{table}[t]
\centering
\caption{Theory-to-evidence bridge.}
\label{tab:theory-evidence-bridge}
\AVTableSetup
\begin{tabularx}{\textwidth}{P{0.20\textwidth}P{0.25\textwidth}P{0.29\textwidth}Y}
\toprule
Theoretical result & Simulation output & Observable empirical counterpart & Current empirical status \\
\midrule
Network amplification & Direct and indirect response, distance profile, and convergence & Active position-holder addresses, total actions, and third-party participation & Partial counterpart; direct exposure and economic-interaction links are not identified \\
Growth--concentration divergence & Total activity and activity HHI under equal-mass allocations & Holder HHI and action-side HHI reported with participation & Directly assessable at address level; mechanism remains noncausal \\
Dispersion--resilience divergence & Chain HHI, worst route loss, and shared-component loss & Chain shares and verified infrastructure-route dependencies & Venue dispersion is observable; route-removal outcome is not currently measurable \\
\bottomrule
\end{tabularx}
\end{table}

%% file: sections/04_data_open_science.tex
\section{Measurement: from protocol events to networks}\label{sec:data}

\subsection{Observed events and analysis windows}

The unit of observation is a decoded Aave V3 Pool event. We analyze Supply, Withdraw, Borrow, Repay, and LiquidationCall events across all observed reserve assets. This choice measures how frequently addresses participate in protocol actions; it does not equate heterogeneous token quantities with a common monetary value.

Ethereum and Arbitrum are observed for event weeks $-16$ through $+16$ around verified GHO reserve activations: 15 July 2023 on Ethereum and 2 July 2024 on Arbitrum. The resulting panels contain 118,806 and 1,837,410 retained Pool events, respectively. The common-calendar analysis aligns the Arbitrum and Gnosis series to the verified Arbitrum execution time, 2 July 2024 at 15:40:32 UTC, and covers 33 complete seven-day windows from 12 March through 29 October 2024. It contains the same 1,837,410 Arbitrum events and 81,814 Gnosis events. All 33 chain-weeks are present; no weekly outcome is interpolated. Exact Pool contracts, block bounds, UTC timestamps, event interfaces, provider checks, and extraction limitations are documented in the appendices and supplementary materials.

\input{tables/tab02_data_network_definition}

Figure~\ref{fig:research-design} records the measurement sequence and the boundary between quantities observed in the present data and quantities that require additional evidence.

\input{figures/fig01_research_design_v2}

\subsection{Why protocol roles matter}

A blockchain transaction can call a router that internally reaches the Aave Pool. Treating the top-level sender and receiver as the economic relation would therefore confuse execution routing with protocol participation. We instead decode the Pool receipt log and retain its native roles. Each event identifies the address carrying out the action (the \emph{action-side address}) and the address whose lending position changes (the \emph{position-holder address}); the Pool records the event but is not automatically treated as a participant. Figure~\ref{fig:tx-log-network} illustrates the distinction.

\input{figures/fig04_tx_call_log_network}

Weekly networks preserve action type and reserve identity. For chain $c$ and week $t$, the two primary common-calendar outcomes are
\begin{align}
Y^{P}_{ct} &= \log\!\left(1+\mathrm{ActivePositionHolderAddresses}_{ct}\right),
\label{eq:empirical-participation}\\
Y^{A}_{ct} &= \mathrm{PositionHolderEventHHI}_{ct}
=\sum_i s_{ict}^{2},
\qquad
s_{ict}=\frac{n_{ict}}{\sum_j n_{jct}},
\label{eq:empirical-hhi}
\end{align}
where $n_{ict}$ is address $i$'s retained Pool-event count. The first outcome measures address-level participation breadth; the second measures the concentration of observed activity among position-holder addresses. Higher HHI indicates that fewer addresses account for more observed actions. The legacy machine-field name \texttt{beneficiary\_address} is rendered throughout the paper as \emph{position-holder address}. An address is an observable blockchain identifier rather than a verified person, institution, or independently controlled economic actor, so participation breadth does not equal a unique-user count.

Adjacent-week entrants, retained addresses, and exits reveal whether a participation stock reflects persistent use or rapid turnover. Self-directed Pool events remain in breadth and activity-HHI calculations because they are observed protocol actions. Structural measures instead use only non-self delegated or third-party role relations, and self-loops are excluded from topology because a self-directed action does not reveal delegation. Weekly aggregation uses the half-open UTC intervals documented with the data; missing calendar cells would be retained and reported rather than silently interpolated, although none occur in the 33-week common-calendar panel.

The analysis also separates role from scale. Action-side and position-holder concentration answer different economic questions. Likewise, raw PageRank HHI falls mechanically as the number of nodes increases, so cross-graph interpretation uses scale-adjusted quantities alongside coreness and persistence. Appendix~\ref{app:indicator-registry} provides the complete 39-indicator registry, formulas, units, and measurement conventions.

\subsection{Addresses, actors, and infrastructure}

An address can represent a person, exchange system, treasury, router, bridge, or contract. Address-level concentration is therefore observable, while economic-actor concentration requires verified evidence about common control. We report conservative actor-concentration bounds rather than infer ownership from behavior, timing, or bytecode similarity.

Infrastructure dependence requires another missing relation: verified links from economic activity to bridges, facilitators, messaging paths, or routes. Contract incidence alone cannot establish this mapping. Route concentration and removal-loss measures are therefore defined by the theory but reserved for a future route-event layer. This distinction is central to the framework: it identifies what would make a stronger claim possible rather than replacing missing relations with assumptions.

\subsection{Evidentiary scope}

Each reported result is tied to a stated source, observation window, network definition, and measure. Checks of event uniqueness, temporal boundaries, role semantics, and graph construction establish the domain in which a statistic is meaningful. The paper distinguishes measured outcomes from simulations, bounds, diagnostic comparisons, and quantities for which the required data are unavailable. This discipline is substantive: it determines whether an observed pattern supports an address-level description, an actor-level interpretation, a claim about infrastructure, or a causal conclusion.

%% file: tables/tab02_data_network_definition.tex
\begin{table*}[t]
\centering
\caption{Protocol-event and network construction rules.}
\label{tab:data-network-definition}
\AVTableSetup
\begin{tabularx}{0.98\textwidth}{>{\centering\arraybackslash}p{1.05cm}P{0.18\textwidth}YP{0.27\textwidth}}
\toprule
\AVTableHeader
\multicolumn{2}{l}{\textbf{Object}} & \textbf{Operational definition} & \textbf{Interpretation boundary} \\
\midrule
\AVTableIcon[AVBlue]{document} & \textbf{Observation} & Canonical Aave V3 Pool \texttt{Supply}, \texttt{Withdraw}, \texttt{Borrow}, \texttt{Repay}, or \texttt{LiquidationCall} log & Not an arbitrary top-level transaction \\
\AVTableIcon[AVViolet]{users} & \textbf{Action side} & Protocol-native caller or actor field for the action family & May be an EOA, router, adapter, or contract \\
\AVTableIcon[AVTeal]{users} & \textbf{Position holder} & \texttt{onBehalfOf} for Supply/Borrow; \texttt{user} for Withdraw/Repay/LiquidationCall & An observed address, not a verified person or institution \\
\AVTableIcon[AVTeal]{network} & \textbf{Edge} & Directed action-side $\rightarrow$ position-holder pair, weighted by event count & Pool is the verified log emitter, not automatically a node \\
\AVTableIcon[AVBlue]{clock} & \textbf{Time} & UTC event week relative to the verified activation date & Week 0 is separated from clean pre and post windows \\
\AVTableIcon[AVViolet]{chart} & \textbf{Aggregate panel} & All five actions and all reserves; action and reserve labels retained & Heterogeneous raw token amounts are never summed as value \\
\AVTableIcon[AVTeal]{network} & \textbf{Network-structure graph} & Non-self delegated/third-party address-role projection & Self-directed events stay in breadth/HHI but not topology \\
\AVTableIcon[AVOrange]{filter} & \textbf{Actor layer} & Primary-source must-links plus assumption-indexed bounds & Bytecode or behavioral similarity does not prove ownership \\
\AVTableIcon[AVOrange]{bridge} & \textbf{Infrastructure layer} & Verified route-event and dependency records & Not yet measurable; contract mediation is only a proxy \\
\bottomrule
\end{tabularx}
\end{table*}

%% file: figures/fig01_research_design_v2.tex
\begin{figure}[!htbp]
\centering
\suspendrefereelines
\begin{tikzpicture}[x=1cm,y=1cm,font=\AVFigureFont,line cap=round,line join=round]
\path[use as bounding box] (0,0) rectangle (12.85,7.62);
\tikzset{
  stage/.style={rounded corners=2.4mm,draw=AVRule,line width=.72pt,fill=white},
  stage title/.style={font=\AVFigureSmallFont\bfseries,text=white,align=center},
  body title/.style={font=\AVFigureSmallFont\bfseries,text=AVInk,align=center},
  body copy/.style={font=\AVFigureTinyFont,text=AVInk,align=center,anchor=north},
  support/.style={rounded corners=2.1mm,draw=AVRule,line width=.62pt,fill=AVLight},
  stage flow/.style={-{Latex[length=1.20mm,width=.95mm]},line width=.70pt,
    shorten >=.6pt,shorten <=.6pt},
  rail flow/.style={-{Latex[length=1.20mm,width=.95mm]},line width=.70pt,
    shorten >=.4pt,shorten <=.4pt}
}

\node[anchor=west,font=\AVFigureTitleFont,text=AVInk,text width=124mm,align=left]
  at (.18,7.40) {From protocol events to supported decentralization claims};
\node[anchor=west,font=\AVFigureTinyFont,text=AVInk,text width=124mm,align=left]
  at (.18,6.88) {Events $\rightarrow$ roles $\rightarrow$ weekly networks $\rightarrow$ measures $\rightarrow$ supported conclusions};

\draw[stage] (.10,1.58) rectangle (3.00,6.55);
\fill[AVNavy,rounded corners=2.2mm] (.10,5.92) rectangle (3.00,6.55);
\node[stage title] at (1.55,6.24) {1 · Events};
\node[inner sep=0pt] at (1.55,4.83)
  {\AVClipArt[15.5mm]{asset-protocol-event-extraction}};
\node[body title,text width=27mm] at (1.55,3.73) {Verified Pool events};
\node[body copy,text width=26mm] at (1.55,3.43)
  {emitter + signature\\block + log index};
\node[av/tag,fill=AVNavy,font=\AVFigureTinyFont] at (1.55,2.20)
  {separate event dates};

\draw[stage] (3.34,1.58) rectangle (6.24,6.55);
\fill[AVBlue,rounded corners=2.2mm] (3.34,5.92) rectangle (6.24,6.55);
\node[stage title] at (4.79,6.24) {2 · Roles};
\node[inner sep=0pt] at (4.79,4.82)
  {\AVClipArt[15.5mm]{asset-address-role-entity-decoding}};
\node[body title,text width=27mm] at (4.79,3.73) {Protocol-native roles};
\node[body copy,text width=26mm] at (4.79,3.43)
  {action $\rightarrow$ holder\\reserve + week};
\node[av/tag,fill=AVOrange,font=\AVFigureTinyFont] at (4.79,2.20)
  {address $\neq$ actor};

\draw[stage] (6.58,1.58) rectangle (9.48,6.55);
\fill[AVTeal,rounded corners=2.2mm] (6.58,5.92) rectangle (9.48,6.55);
\node[stage title] at (8.03,6.24) {3 · Measures};
\node[inner sep=0pt] at (8.03,4.78)
  {\AVClipArt[21mm]{asset-hero-cross-chain-observatory}};
\node[body title,text width=27mm] at (8.03,3.73) {Weekly role graphs};
\node[body copy,text width=25mm] at (8.03,3.43)
  {ETH 118,806\\ARB 1,837,410};
\node[av/tag,fill=AVTeal,font=\AVFigureTinyFont] at (8.03,2.20)
  {HHI + topology};

\draw[stage] (9.82,1.58) rectangle (12.75,6.55);
\fill[AVViolet,rounded corners=2.2mm] (9.82,5.92) rectangle (12.75,6.55);
\node[stage title] at (11.285,6.24) {4 · Scope};
\node[inner sep=0pt] at (11.285,4.90)
  {\AVClipArt[15.5mm]{asset-evidence-boundary}};
\node[body title] at (11.285,3.79) {Supported claims};
\node[body copy,text=AVTeal,text width=25mm,font=\AVFigureTinyFont\bfseries]
  at (11.285,3.49) {Measured now\\HHI + topology};
\draw[AVRule,line width=.55pt] (10.08,2.70)--(12.49,2.70);
\node[av/tag,fill=AVOrange,text width=22.5mm,align=center,
  font=\AVFigureTinyFont\bfseries,inner xsep=1.5mm,inner ysep=1.15mm]
  at (11.285,2.12) {needs routes + actors\\+ causality};

\draw[stage flow,draw=AVBlue] (3.04,4.18)--(3.30,4.18);
\draw[stage flow,draw=AVBlue] (6.28,4.18)--(6.54,4.18);
\draw[stage flow,draw=AVViolet] (9.52,4.18)--(9.78,4.18);

\draw[support,dashed,draw=AVBlue] (.10,.10) rectangle (4.45,1.25);
\node[inner sep=0pt] at (.80,.675)
  {\AVClipArt[13.5mm]{asset-strategic-network-mechanism}};
\node[anchor=west,font=\AVFigureSmallFont\bfseries,text=AVInk] at (1.52,.84)
  {Mechanism lens};
\node[anchor=west,font=\AVFigureTinyFont,text=AVInk,text width=27mm,align=left]
  at (1.52,.47) {model illustration;\\not estimated};
\draw[rail flow,draw=AVBlue,dashed] (4.28,1.22)
  .. controls (4.42,1.36) and (4.63,1.45) .. (4.78,1.55);

\draw[support] (4.72,.10) rectangle (12.75,1.25);
\node[inner sep=0pt] at (5.43,.675)
  {\AVClipArt[14mm]{asset-documented-analysis}};
\node[av/tag,fill=AVNavy,font=\AVFigureTinyFont\bfseries] at (6.68,.86)
  {data};
\draw[rail flow,draw=AVBlue] (7.12,.85)--(7.58,.85);
\node[av/tag,fill=AVBlue,font=\AVFigureTinyFont\bfseries] at (8.18,.86)
  {validation};
\draw[rail flow,draw=AVViolet] (8.92,.85)--(9.34,.85);
\node[av/tag,fill=AVViolet,font=\AVFigureTinyFont\bfseries] at (9.86,.86)
  {limits};
\node[anchor=west,font=\AVFigureTinyFont,text=AVInk,text width=61mm,align=left]
  at (6.22,.37) {documented inputs, validation checks, and stated evidence limits};
\draw[rail flow,draw=AVTeal] (11.285,1.56)--(11.285,1.28);
\end{tikzpicture}
\resumerefereelines
\caption{Protocol-event measurement sequence and evidence boundary. Verified Pool events are decoded into typed weekly address--role networks and four decentralization dimensions. The mechanism panel is a model illustration rather than an estimated formation process. Ethereum and Arbitrum counts are observed event totals; participation, event-frequency concentration, and network-structure diagnostics are descriptive. Route dependence, verified economic-actor linkage, and causal effects require additional evidence.}
\label{fig:research-design}
\end{figure}

%% file: figures/fig04_tx_call_log_network.tex
\begin{figure}[!htbp]
\centering
\suspendrefereelines
\begin{tikzpicture}[font=\AVFigureFont,line cap=round,line join=round]
\tikzset{
  mapcard/.style={rounded corners=2mm,draw=AVRule,line width=.7pt,fill=white,
    minimum width=29mm,minimum height=40mm,align=center,inner sep=1.5mm},
  mapflow/.style={-{Latex[length=1.55mm,width=1.05mm]},draw=AVNavy,line width=.82pt,
    shorten >=.5pt,shorten <=.5pt},
  roleflow/.style={-{Latex[length=1.55mm,width=1.05mm]},draw=AVTeal,line width=.88pt,
    shorten >=.5pt,shorten <=.5pt}
}
\path[use as bounding box] (-.12,-5.02) rectangle (12.76,2.15);
\node[mapcard] (tx) at (1.48,0) {};
\node[mapcard] (call) at (4.70,0) {};
\node[mapcard,draw=AVBlue] (log) at (7.92,0) {};
\node[mapcard,draw=AVTeal] (edge) at (11.14,0) {};
\pic[av/icon slate,scale=.54] at ($(tx.north)+(0,-.62)$) {av/document};
\pic[av/icon,scale=.54] at ($(call.north)+(0,-.62)$) {av/code};
\node[inner sep=0pt] at ($(log.north)+(0,-.72)$)
  {\AVClipArt[13.5mm]{asset-protocol-event-extraction}};
\node[inner sep=0pt] at ($(edge.north)+(0,-.72)$)
  {\AVClipArt[13.5mm]{asset-address-role-entity-decoding}};
\node[anchor=north,align=center,text width=26mm] at ($(tx.north)+(0,-1.30)$)
  {\textbf{Transaction}\\sender, target\\calldata\\\textcolor{AVSlate}{routing context}};
\node[anchor=north,align=center,text width=26mm] at ($(call.north)+(0,-1.30)$)
  {\textbf{Internal call}\\router or adapter\\calls Pool};
\node[anchor=north,align=center,text width=26mm] at ($(log.north)+(0,-1.46)$)
  {\textbf{Pool event log}\\emitter + signature\\block, tx, log index};
\node[anchor=north,align=center,text width=26mm] at ($(edge.north)+(0,-1.46)$)
  {\textbf{Typed role edge}\\action side $\rightarrow$ holder\\action, reserve, week};
\draw[mapflow] (tx.east)--(call.west);
\draw[mapflow] (call.east)--(log.west);
\draw[roleflow] (log.east)--(edge.west);

\node[av/risk,text width=54mm,minimum height=13mm,align=center,inner sep=1.7mm] (emitter) at (3.02,-3.15) {
\textbf{Emitter rule}\quad The Pool proves the event source; it is not automatically a network node.};
\node[text width=54mm,minimum height=16mm,align=center,inner sep=2.4mm] (roles) at (9.62,-3.15) {
\textbf{Indexed-role rule}\\[1mm]Supply/Borrow: \texttt{topics[2]} $=$ \texttt{onBehalfOf}; other actions use canonical \texttt{user}.};
\draw[AVOrange,densely dashed,line width=.8pt] (log.south west)--(emitter.north east);
\draw[AVTeal,densely dashed,line width=.8pt]
  (log.south east)--++(0,-.18)--(6.50,-2.18);

\draw[AVRule,line width=.6pt] (.18,-4.16)--(12.46,-4.16);
\node[anchor=west,text=AVSlate] at (.30,-4.52) {caller graph};
\draw[AVSlate,line width=.7pt,dashed] (2.35,-4.52)--(3.25,-4.52);
\node[anchor=center,text=AVSlate] at (4.10,-4.52) {$\neq$};
\draw[AVTeal,line width=1pt,-{Latex[length=1.55mm,width=1.05mm]}] (4.72,-4.52)--(5.72,-4.52);
\node[anchor=west,text=AVNavy] at (5.92,-4.52) {protocol-native address-role network};
\end{tikzpicture}
\resumerefereelines
\caption{Transaction--call--log--role mapping. A top-level transaction and its internal call path provide execution context, while the canonical Pool receipt log supplies the economic action and indexed roles. The Pool is the verified emitter, not automatically a graph node. Each resulting edge is identified by chain, block, transaction hash, and log index and remains typed by action, reserve, and week.}
\label{fig:tx-log-network}
\end{figure}

%% file: sections/05_causal_inference.tex
\section{Empirical results: mechanism-consistent patterns and comparative boundaries}\label{sec:causal}

\subsection{Prespecified mechanism-to-evidence classification}\label{subsec:interpretation-framework}

The analytical and simulated mechanisms permit every joint direction in Table~\ref{tab:activity-concentration-classification}. This classification is stated before reporting the empirical estimates so that observed signs do not redefine a favored mechanism. The simulation-results table and the classification table are related but not duplicates: the former verifies that the synthetic model can generate the permitted regimes, whereas the latter fixes the empirical sign-to-interpretation rule. The labels are descriptive regime names: \emph{consistent with} denotes compatibility with a joint pattern, not proof of the behavioral process, an estimate of $\beta$, or a chain-specific prediction generated by the simulation.

\begin{table}[t]
\centering
\caption{Prespecified classification of joint activity and concentration changes.}
\label{tab:activity-concentration-classification}
\AVTableSetup
\begin{tabularx}{\textwidth}{P{0.21\textwidth}P{0.25\textwidth}Y}
\toprule
Activity response & Concentration response & Theory-consistent descriptive interpretation \\
\midrule
\rowcolor{AVTeal!14}
Rises & HHI falls & \textbf{Peripheral expansion} \AVTableMeta{(Ethereum sign cell)} \\
Rises & HHI approximately unchanged & Proportional growth \\
\rowcolor{AVOrange!14}
Rises & HHI rises & \textbf{Incumbent amplification} \AVTableMeta{(Arbitrum sign cell)} \\
Falls & HHI falls & Contraction with relative equalization \\
Falls & HHI approximately unchanged & Proportional contraction \\
Falls & HHI rises & Concentrated contraction or peripheral exit \\
\bottomrule
\end{tabularx}
\begin{minipage}{0.97\textwidth}
\AVTableNoteFont\emph{Notes:} Tint marks the two cells matched by the observed chain-relative signs; all six cells remain prespecified possibilities. The Arbitrum label is applied in the text only as ``closer to incumbent amplification'' because its participation increase is small.
\end{minipage}
\end{table}

The empirical rule uses active-address breadth as the observable counterpart to the simulation's activity response, but the two are not the same estimand: the simulation measures total synthetic activity, whereas the longitudinal panels count active position-holder addresses and separately measure their event-share HHI. The Ethereum issuance and Arbitrum expansion also occur on different chains, dates, and institutional baselines; they are parallel chain-relative descriptions, not a treated-versus-control comparison. The remaining measures are structural qualifiers, not additional conditions that must be forced to confirm the regime. Participation may broaden while structural position concentrates, activity HHI may fall while infrastructure dependence rises, and address counts may increase without greater economic-entity diversity. We therefore report each evidence layer separately rather than convert them into a scalar decentralization score.

\subsection{Chain-specific mechanism-consistent empirical patterns}\label{subsec:chain-relative-evidence}

The Ethereum and Arbitrum panels contain 1,956,216 Aave V3 Pool events. Self-directed events represent 76.79\% of Ethereum events and 76.99\% of Arbitrum events; the remaining 27,579 and 422,834 events form the delegated address-role topology. Table~\ref{tab:aggregate-sample} reports the complete fields used in the main analysis. Fields unavailable for one or more chains are preserved in Appendix Table~\ref{tab:extended-sample-coverage-audit} rather than displayed as if they were comparable outcomes.

\input{tables/tab03_aggregate_sample}

Figure~\ref{fig:real-v5-cross-chain} provides the paper's central empirical comparison. Excluding activation week zero, mean weekly active position-holder addresses increase from 762.7 to 1,456.4 on Ethereum (91.0\%) and from 7,990.3 to 8,123.4 on Arbitrum (1.7\%). Mean weekly position-holder-event HHI falls from 0.00956 to 0.00655 on Ethereum ($-31.5\%$) but rises from 0.00932 to 0.01474 on Arbitrum ($+58.1\%$) over the same chain-relative windows. Pooling events within the pre and post periods instead yields Ethereum position-holder HHI values of 0.005647 and 0.003555, a 37.0\% decline. Because HHI is nonlinear, this pooled-period statistic answers a different question from the mean weekly comparison; it is reported because the pooled address distribution is the observed input to the actor-bound analysis in Appendix~\ref{app:real-v4-partial-identification}.

\input{figures/fig05_real_v5_cross_chain}

Read through the prespecified typology, Ethereum combines substantial address growth with declining mean HHI, a joint pattern consistent with \emph{peripheral expansion at the address level}. Arbitrum combines nearly unchanged breadth with rising concentration, a mixed pattern closer to \emph{incumbent amplification} than to broad peripheral expansion. The qualification matters: a 1.7\% increase is not evidence of substantial expansion, and neither label identifies the process that generated the pattern. Supporting distributional measures point in the same descriptive direction---Ethereum requires more addresses to account for 51\% of events after the focal week, whereas Arbitrum requires fewer---but these remain all-action, all-reserve patterns around different chain-specific dates. They do not establish a GHO effect.

The next evidence layers explain and qualify those two primary regimes. Membership flows are the first qualifier. Figure~\ref{fig:address-state-flows} shows substantial entrant and exit shares on both chains, alongside modest increases in prior-week retention. A high active-address count therefore need not imply a stable participant base. These are membership-state transitions, not observations of edge rewiring, so they inform acquisition and persistence but do not estimate the network-formation response $dG/dp$ discussed in Section~\ref{subsec:dynamic-network-interpretation}.

\input{figures/fig05d_address_state_flows}

Role-aware topology is the second qualifier. The full delegated graphs contain 8,891 Ethereum nodes and 50,693 Arbitrum nodes. Action-side strength is substantially more concentrated than position-holder strength on Ethereum, whereas some measures reverse across chains. Figure~\ref{fig:action-layer-profile} shows that supplying, borrowing, repaying, and liquidating generate distinct topologies. Liquidation, despite its small event share, produces the largest maximum-$k$-core share on Ethereum. An aggregate network should therefore preserve its typed layers rather than treating every edge as economically interchangeable.

\input{tables/tab05_real_v5_topology}
\input{figures/fig05b_action_layer_profile}

The full-graph core diagnostics make the scale distinction concrete. Ethereum has 8,891 nodes and 10,896 edges; its maximum $k$-core has $k_{\max}=6$ and 14 nodes. Arbitrum has 50,693 nodes and 69,582 edges; its maximum $k$-core has $k_{\max}=8$ and 106 nodes. The fitted Borgatti--Everett (BE) core contains two nodes on each chain, the algorithmic minimum. Its overlap with the maximum-$k$ core is 0.143 on Ethereum and 0.019 on Arbitrum, while the full-node Spearman correlations between Rombach coreness and $k$-core number are 0.521 and 0.578. Figure~\ref{fig:real-v5-core-periphery} displays a deterministic 160-node backbone for visual inspection only; all counts, fits, overlaps, and correlations come from the complete graphs.

\input{figures/fig06_real_v5_core_periphery}

Global dispersion and local operational cores can coexist. After adjusting PageRank HHI for node count, the effective PageRank-node fractions are 99.41\% on Ethereum and 98.01\% on Arbitrum. Figure~\ref{fig:centrality-roles} therefore focuses on the two diagnostics that distinguish the substantive structural claim: the shares of full address-role graphs assigned to discrete cores and the role composition of full versus high-coreness nodes. Panel A directly labels Ethereum and Arbitrum with redundant fill and hatch encodings; Panel B uses a separate blue--orange--teal role palette. The former black/white ambiguity is removed, and the figure no longer asks nearly overlapping PageRank and continuous-coreness curves to carry the main storyline.

\input{figures/fig05c_centrality_roles}

Public labels associate available core addresses with protocol-adjacent helper, gateway, or migration functions. This supports only a narrow infrastructure-oriented interpretation of the observed address-role core. It does not identify ownership, natural persons, economic-actor control, route resilience, or the effect of removing a node. Structural diagnostics consequently qualify the Ethereum and Arbitrum joint regimes; they do not replace the primary participation--HHI classification or produce a blockchain ranking.

\subsection{Common-calendar comparative evidence and identification boundary}\label{subsec:arbitrum-gnosis-comparison}

The chain-relative panels answer how each outcome moved around its own GHO event. A separate common-calendar design asks whether Arbitrum's path around the 2 July 2024 expansion differed from contemporaneous movement on Aave's Gnosis deployment. Gnosis did not receive the focal 2024 Arbitrum expansion and therefore removes some common calendar variation, but this role does not make its path an assumed counterfactual. The analysis estimates three related objects and reports them as comparative longitudinal evidence.

First, on the 24 included weeks---clean pre weeks $-16$ to $-9$ and post weeks $+1$ to $+16$---the conventional two-chain interaction is
\begin{equation}
Y_{ct}=\alpha+\gamma\mathrm{ARB}_{c}+\lambda\mathrm{Post}_{t}
+\beta\bigl(\mathrm{ARB}_{c}\times\mathrm{Post}_{t}\bigr)+\varepsilon_{ct}.
\label{eq:did-style}
\end{equation}
The coefficient $\beta$ is algebraically the change in the Arbitrum-minus-Gnosis mean difference; its existence does not establish that the causal estimand is identified. We estimate the equivalent weekly-gap regression so serial dependence is modeled in one aggregate time series.

Second, let $G_t=Y_{\mathrm{ARB},t}-Y_{\mathrm{GNO},t}$ and $D_t=\ind\{t\geq1\}$. The controlled interrupted-time-series specification is
\begin{equation}
G_t=\theta_0+\theta_1t+\theta_2D_t+\theta_3(t-T_0)D_t+u_t,
\qquad T_0=0,
\label{eq:controlled-its}
\end{equation}
where $\theta_1$ is the clean-pre differential slope, $\theta_2$ is the fitted comparative level shift at the excluded event-week reference, and $\theta_3$ is the post-expansion differential slope change. Segmented and controlled interrupted-time-series methods motivate this separation of level and slope changes \cite{wagner2002segmented,bernal2017interrupted,bernal2018controls}. Because weeks $-8$ through 0 are omitted, $\theta_2$ is an extrapolated level difference across that interval rather than an observed week-0 discontinuity.

Third, the weekly dynamic comparative contrast is
\begin{equation}
C_k=\left(Y_{\mathrm{ARB},k}-Y_{\mathrm{GNO},k}\right)
-\overline{\left(Y_{\mathrm{ARB}}-Y_{\mathrm{GNO}}\right)}_{k=-16:-9}.
\label{eq:dynamic-comparative-contrast}
\end{equation}
These centered arithmetic gaps display timing and heterogeneity without being relabeled as event-study treatment effects.

Figure~\ref{fig:arbitrum-gnosis-comparative-dynamics} places the observed paths, controlled-ITS fits, and centered gaps on one page. For log participation, Arbitrum's clean-pre-to-post change is +0.0314 while Gnosis changes by $-1.9519$, yielding a comparative coefficient of +1.9833. Most of that contrast is contemporaneous Gnosis movement. For HHI, the corresponding changes are +0.00501 and +0.02343, yielding $-0.01842$: concentration rises on both chains, but less on Arbitrum. The negative common-calendar HHI contrast therefore does not contradict the chain-relative incumbent-amplification pattern; the estimands use different clocks, and neither says that Arbitrum HHI fell. The controlled-ITS estimates are likewise conditional on the chosen trend specification: the week-0-referenced level and slope shifts are +4.0396 and +0.1845 per week for log participation, and $-0.02849$ and $-0.003515$ per week for HHI.

\input{figures/fig07_arbitrum_gnosis_comparative_dynamics}
\input{tables/tab06_arbitrum_gnosis_main_results}
\FloatBarrier

The comparison is economically informative but specification-sensitive. Excluding no anticipation weeks reduces the participation comparative coefficient to +1.3414 and changes the HHI coefficient to +0.00595; four- and eight-week exclusions move both coefficients, and the controlled-ITS terms change sign under some alternatives. By contrast, the $+8$, $+12$, and $+16$ post horizons preserve the primary coefficient directions. Appendix~\ref{app:benchmark-did} reports these results, differential pre-slopes, and placebo calculations.

Difference-in-differences supports causal interpretation only when the comparison path supplies the relevant counterfactual trend and anticipation, spillovers, and concurrent shocks are adequately addressed \cite{roth2023trending}. Here the Gnosis participation path differs during the clean pre-period, its HHI series contains large weekly movements, and most of the arithmetic participation contrast is the contemporaneous Gnosis decline. Failure to reject a pre-trend test would not establish parallel trends because such tests can have low power and conditioning analysis on a pretest can distort inference \cite{roth2022pretest}. We therefore report raw paths, differential slopes, sensitivity ranges, and model-based HAC uncertainty rather than present significance stars as identification evidence.

Serial correlation can materially understate uncertainty in difference-in-differences applications \cite{bertrand2004trust}. The reported Bartlett/Newey--West standard errors model temporal dependence in the weekly Arbitrum-minus-Gnosis series \cite{newey1987simple}; they use fixed lags---three for the mean-change model and two for controlled ITS---without a finite-sample correction. With one focal chain, one comparison chain, and 24 aggregate gaps, they do not reproduce conventional many-cluster treatment-assignment inference. Methods for a small number of treated units still require an informative collection of untreated units and do not resolve a one-comparison-chain design \cite{conley2011inference}. The evidence supports a widening participation gap and a smaller relative HHI increase on Arbitrum, not an identified GHO effect.

\medskip
\noindent\textbf{Theory-led synthesis of observed multidimensional change.}\par\smallskip

Table~\ref{tab:multidimensional-synthesis} returns the empirical sequence to the theoretical question. The first row classifies the primary joint patterns. The following rows show why participant turnover, action-layer structure, local cores, and common-calendar movement are complementary evidence rather than alternative labels for the same outcome. Compatibility is not confirmation: the present data establish address-level and comparative patterns, not the behavioral mechanism or a causal effect of GHO.

\input{tables/tab07_multidimensional_synthesis}

The current evidence thus supports two sharply stated but bounded conclusions. Ethereum's chain-relative pattern is consistent with address-level peripheral expansion, whereas Arbitrum's pattern is closer to incumbent amplification than broad peripheral expansion. Whether network formation produced, delayed, or counteracted either pattern remains an open mechanism question taken up in Section~\ref{sec:conclusion}.

%% file: tables/tab03_aggregate_sample.tex
\begin{table*}[t]
\centering
\caption{Audited longitudinal and benchmark samples.}
\label{tab:aggregate-sample}
\AVTableSetup
\begin{tabularx}{0.98\textwidth}{P{0.16\textwidth}Y Y Y}
\toprule
\AVTableHeader
\textbf{Audit item} & \textbf{Ethereum} & \textbf{Arbitrum} & \textbf{Gnosis comparison} \\
\midrule
Event clock & 15 Jul 2023 & 2 Jul 2024 & 2 Jul 2024 (Arbitrum-aligned) \\
Observation window & Event weeks $-16{:}+16$ & Event weeks $-16{:}+16$ & Event weeks $-16{:}+16$ \\
Pool-event count & 118,806 & 1,837,410 & 81,814 \\
Common outcome coverage & 33/33 weeks: active position-holder addresses and position-holder-event HHI & 33/33 weeks: active position-holder addresses and position-holder-event HHI & 33/33 weeks: active position-holder addresses and position-holder-event HHI \\
Permitted interpretation & Chain-relative descriptive pattern; no causal effect & Chain-relative description and common-calendar comparison; no causal effect & Contemporaneous comparison diagnostic; not an assumed counterfactual \\
\bottomrule
\end{tabularx}
\begin{minipage}{0.96\textwidth}
\AVTableNoteFont\emph{Notes:} Data documentation, not an empirical result. Gnosis is aligned to Arbitrum's verified activation date; its own GHO activation occurred in 2025. The series removes some common calendar variation but is not assumed to supply the counterfactual path needed for a definitive causal interpretation. All units are addresses or address roles, not verified economic actors. Appendix Table~\ref{tab:extended-sample-coverage-audit} preserves the extended coverage audit, including unavailable fields; no missing value is interpreted as zero.
\end{minipage}
\end{table*}

%% file: figures/fig05_real_v5_cross_chain.tex
\begin{figure}[!htbp]
\centering
\suspendrefereelines
\includegraphics[width=0.98\textwidth]{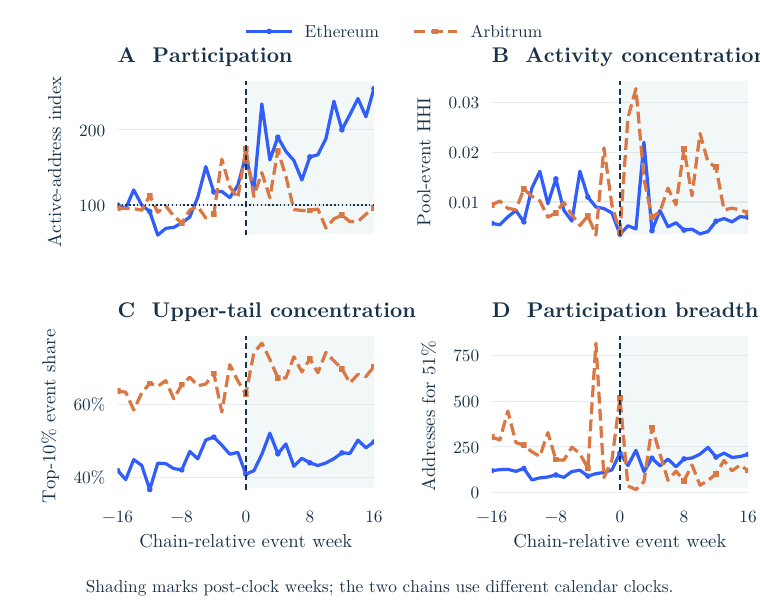}
\resumerefereelines
\caption{Four complementary address-level trajectories around chain-specific institutional clocks. Panel A indexes each chain's weekly active position-holder-address count to its own mean over weeks $-16$ to $-1$. Panel B reports the HHI of equally weighted Pool-event frequency, where larger values indicate greater concentration. Panel C reports the share of Pool events assigned to the top 10\% of active position-holder addresses, and Panel D reports the minimum number of active position-holder addresses needed to reach 51\% of events. The vertical dashed line marks week 0 and light shading marks post-clock weeks. Ethereum and Arbitrum use different verified clocks and different calendar years; the lines are descriptive trajectories, not a treated--control comparison. All actions and reserves are aggregated, addresses are not verified economic actors, and no GHO-specific or causal effect is shown.}
\label{fig:real-v5-cross-chain}
\end{figure}

%% file: figures/fig05d_address_state_flows.tex
\begin{figure}[t]
\centering
\suspendrefereelines
\includegraphics[width=0.98\textwidth]{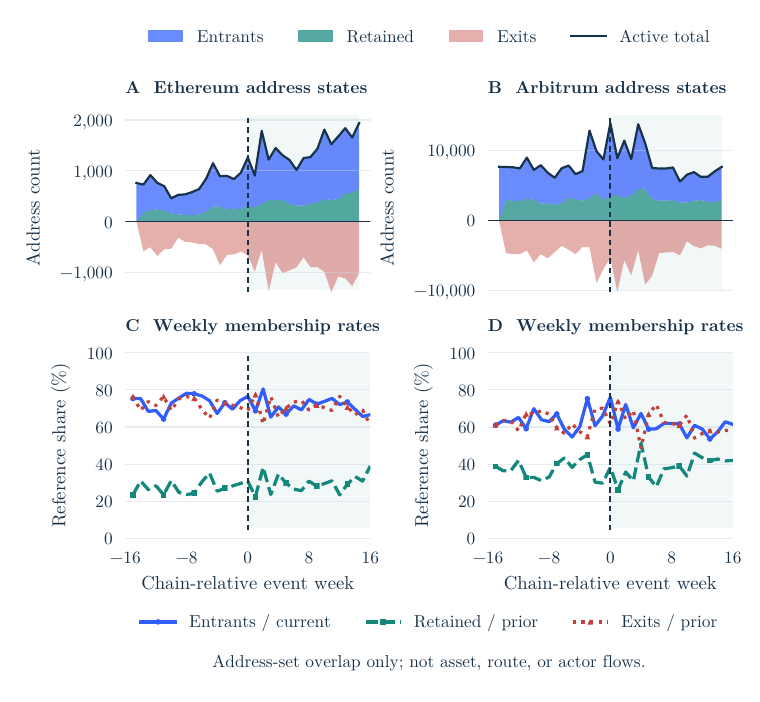}
\resumerefereelines
\caption{Observed adjacent-week position-holder-address state evolution. Panels A and B decompose each current weekly active set into addresses retained from the previous week and entrants; exits from the previous active set are plotted below zero, and the dark line is the current active total. Thus, for week $t$, entrants plus retained addresses equal the current active count, while exits plus retained addresses equal the week-$t-1$ active count. Panels C and D report entrants as a share of the current active set and retained and exiting addresses as shares of the prior active set. Week $-16$ has no prior-week rate and is omitted from the rate panels. The dashed line marks week 0 and shading marks post-clock weeks. Ethereum and Arbitrum use different verified clocks and calendar years; all actions and reserves are aggregated. These are address-set membership transitions, not people, value transfers, bridge-route flows, cross-chain movements, GHO-specific effects, or causal estimates.}
\label{fig:address-state-flows}
\end{figure}

%% file: tables/tab05_real_v5_topology.tex
\begin{table}[htbp]
\centering
\caption{Audited full-window address-role topology with scale diagnostics.}
\label{tab:real-v5-topology}
\AVTableSetup
\begin{tabularx}{0.94\linewidth}{Yrr}
\toprule
\AVTableHeader
\textbf{Measure} & \textbf{Ethereum} & \textbf{Arbitrum} \\
\midrule
\AVTableSubheader
\multicolumn{3}{l}{\emph{Scale and role assignment}} \\
Pool events & 118,806 & 1,837,410 \\
Delegated or third-party events & 27,579 & 422,834 \\
Role-network nodes & 8,891 & 50,693 \\
Unique directed edges & 10,896 & 69,584 \\
\addlinespace[2pt]
\AVTableSubheader
\multicolumn{3}{l}{\emph{Role-weight and scale-aware diagnostics}} \\
Raw action-side strength HHI & 0.4237 & 0.1668 \\
Raw position-holder-side strength HHI & 0.0008 & 0.0067 \\
Raw PageRank HHI & 0.000113 & 0.000020 \\
PageRank HHI $\times$ node count & 1.0059 & 1.0203 \\
Effective PageRank-node fraction & 99.41\% & 98.01\% \\
Maximum $k$-core share & 0.157\% & 0.209\% \\
Persistent maximum-core share & 2.294\% & 0.152\% \\
\addlinespace[2pt]
\AVTableSubheader
\multicolumn{3}{l}{\emph{Core-definition diagnostics}} \\
BE binary-core nodes (minimum allowed) & 2 & 2 \\
BE nodes contained in maximum-$k$ core & 100\% & 100\% \\
Rombach--$k$-core Spearman $\rho$ & 0.521 & 0.578 \\
\bottomrule
\end{tabularx}
\begin{tablenotes}[flushleft]
\AVTableNoteFont
\item \emph{Notes:} A directed edge runs from action-side to position-holder address and is weighted by Pool-event count. Self-directed events are excluded from topology. PageRank HHI has lower bound $1/n$; multiplying by $n$ gives 1 under a uniform distribution, and its reciprocal is the effective-node fraction. The two-node Borgatti--Everett (BE) solutions hit the algorithm's imposed minimum and are fully contained in the maximum-$k$ cores, so unequal-set Jaccard values are not interpreted as membership disagreement. The BE nodes are interpreted only as an address-role infrastructure-core diagnostic; Appendix~\ref{app:be-core-infrastructure} lists their address evidence and labeling limits. Raw graph size and density differ; no scalar cross-chain structural ranking is claimed.
\end{tablenotes}
\end{table}

%% file: figures/fig05b_action_layer_profile.tex
\begin{figure}[!htbp]
\centering
\suspendrefereelines
\includegraphics[width=0.98\textwidth]{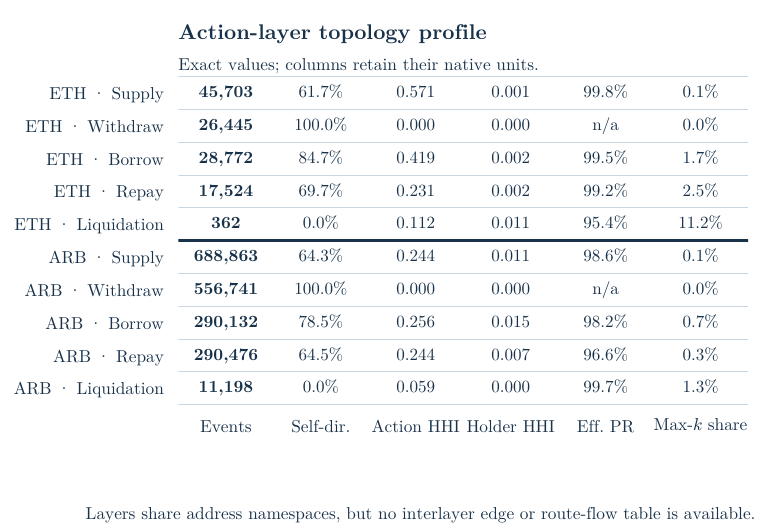}
\resumerefereelines
\caption{Action-layer topology profile for the audited Pool-event system. Rows separate Supply, Withdraw, Borrow, Repay, and Liquidation on Ethereum and Arbitrum; columns report event scale, self-directed-event share, action-side and position-holder-side strength HHI, the scale-adjusted effective PageRank-node fraction, and maximum-$k$-core node share. Printed values are the estimands; cell color is normalized separately within each column and supports comparison of visual rank only. Withdraw is fully self-directed under the decoded role map, so removing self-loops leaves no delegated topology and the PageRank fraction is not applicable. Layers reuse the same address namespace, but the available compact output contains neither verified interlayer edges nor bridge-route flows. The figure therefore compares typed intralayer summaries and does not claim a multiplex community, economic-actor, or route-dependence result.}
\label{fig:action-layer-profile}
\end{figure}

%% file: figures/fig05c_centrality_roles.tex
\begin{figure}[!htbp]
\centering
\suspendrefereelines
\includegraphics[width=0.98\textwidth]{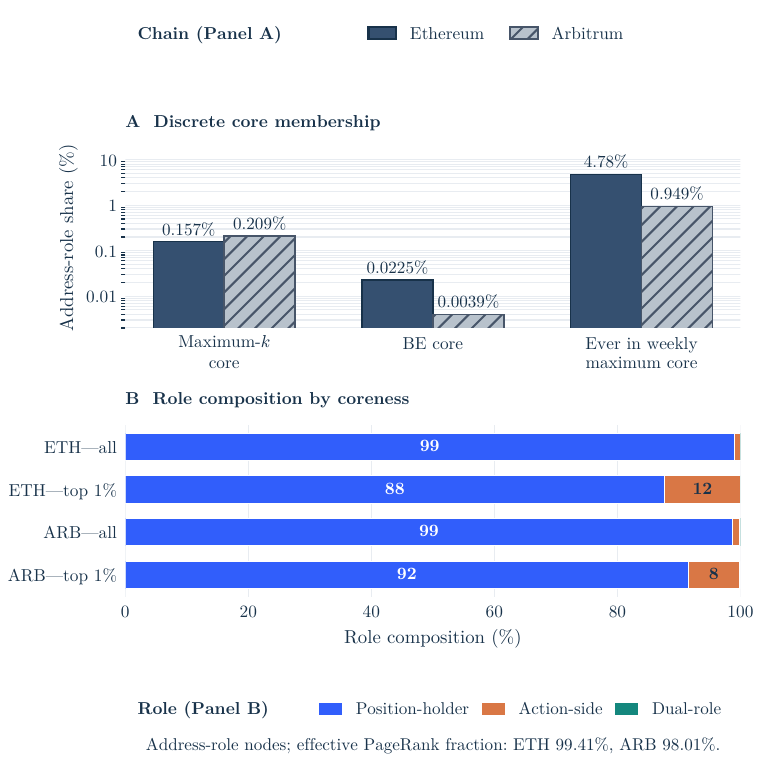}
\resumerefereelines
\caption{Global centrality dispersion coexists with small operational cores and role-specific high-coreness tails. Panel A reports the full-graph address shares in the maximum-$k$ core, the fitted Borgatti--Everett (BE) core, and any weekly maximum core on a logarithmic scale. Ethereum is shown with solid navy bars and Arbitrum with hatched slate bars; each value is printed directly above its bar. The two-node BE solutions equal the algorithmic lower bound. Panel B compares the full-graph role composition with roles at or above each chain's 99th percentile of Rombach coreness; blue, orange, and teal denote position-holder, action-side, and dual-role semantics. The effective PageRank-node fractions remain 99.41\% on Ethereum and 98.01\% on Arbitrum, so the figure focuses on the local-core and role evidence that qualifies that diffuse global mass. Statistics use all 8,891 Ethereum and 50,693 Arbitrum address-role nodes, not the 160-node display backbone. Address roles do not establish ownership or independent economic actors, and neither panel is a causal, dynamic-network, or removal-dependence estimate.}
\label{fig:centrality-roles}
\end{figure}

%% file: figures/fig07_arbitrum_gnosis_comparative_dynamics.tex
\begin{figure*}[t]
\centering
\includegraphics[width=\textwidth]{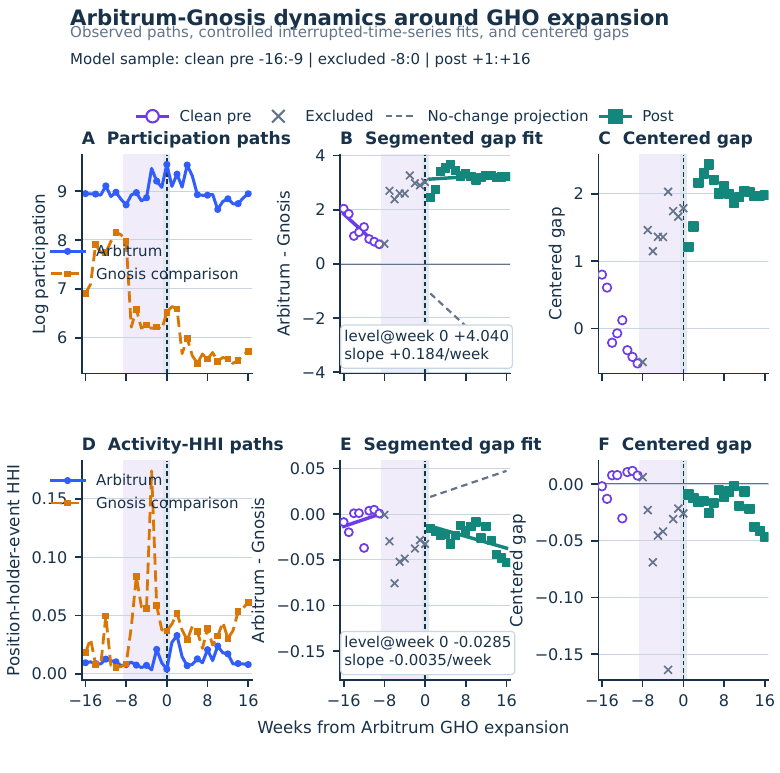}
\caption{Arbitrum--Gnosis comparative dynamics around the 2 July 2024 GHO expansion. Panels A/D show observed weekly log participation and position-holder-event HHI; B/E show controlled interrupted-time-series fits to the Arbitrum-minus-Gnosis gap; and C/F show gaps centered on their clean-pre means. Purple shading marks the excluded anticipation window ($-8$ to $-1$) and grey marks excluded week 0. Fitted segments stop at the included clean-pre ($-16$ to $-9$) and post ($+1$ to $+16$) boundaries; the dashed line is the clean-pre no-change projection. Week-0 level shifts are extrapolated across the omitted interval. These are comparative longitudinal estimates with descriptive HAC uncertainty, not identified treatment effects.}
\label{fig:arbitrum-gnosis-comparative-dynamics}
\end{figure*}

%% file: tables/tab06_arbitrum_gnosis_main_results.tex
\begin{table*}[t]
\centering
\caption{Arbitrum--Gnosis comparative longitudinal estimates and interpretation.}
\label{tab:arbitrum-gnosis-main-results}
\AVTableSetup
\begin{tabularx}{0.99\textwidth}{P{0.20\textwidth}*{5}{>{\centering\arraybackslash}X}}
\toprule
\AVTableHeader
\textbf{Outcome} & \textbf{Arbitrum}\newline\textbf{change} & \textbf{Gnosis}\newline\textbf{change} & \textbf{Relative}\newline\textbf{change} & \textbf{cITS level}\newline\textbf{shift} & \textbf{cITS slope}\newline\textbf{shift} \\
\midrule
Log participation & +0.0314 & $-1.9519$ & \textbf{+1.9833}\newline (0.1916) & \textbf{+4.0396}\newline (0.4190) & \textbf{+0.1845}\newline (0.0302) \\
\multicolumn{6}{P{0.94\textwidth}}{\AVTableNoteFont\emph{Reading:} The Gnosis decline supplies most of the arithmetic contrast; this is not an Arbitrum treatment effect.}\\[2pt]
Position-holder-event HHI & +0.00501 & +0.02343 & \textbf{$-0.01842$}\newline (0.00540) & \textbf{$-0.02849$}\newline (0.01267) & \textbf{$-0.003515$}\newline (0.001078) \\
\multicolumn{6}{P{0.94\textwidth}}{\AVTableNoteFont\emph{Reading:} HHI rises on both chains and rises less on Arbitrum; the negative contrast is not an absolute decline.}\\
\bottomrule
\end{tabularx}
\begin{minipage}{0.97\textwidth}
\AVTableNoteFont\emph{Notes:} Direct empirical output. Log participation is $\log(1+\text{active position-holder addresses})$. Changes compare the clean pre-period (weeks $-16$ to $-9$) with the post-period (weeks $+1$ to $+16$); weeks $-8$ to $-1$ and week 0 are excluded. Parentheses contain Bartlett/Newey--West HAC standard errors without a finite-sample correction: lag 3 for the comparative mean-change coefficient and lag 2 for the controlled interrupted-time-series (cITS) coefficients. Each regression uses 24 weekly Arbitrum-minus-Gnosis gaps. The cITS level coefficient is referenced to excluded week 0. No $p$-values or significance stars are reported: with one focal chain, one comparison chain, 24 aggregate weekly gaps, and no many-cluster assignment design, such quantities would be model-conditional and would not validate the counterfactual. The HAC standard errors describe modeled serial dependence in the realized gap series, not treatment-assignment uncertainty.
\end{minipage}
\end{table*}

%% file: tables/tab07_multidimensional_synthesis.tex
\begin{table*}[!t]
\centering
\caption{Theory-led synthesis of the current empirical evidence.}
\label{tab:multidimensional-synthesis}
\AVTableSetup
\begin{tabularx}{0.99\textwidth}{P{0.18\textwidth}P{0.30\textwidth}P{0.25\textwidth}Y}
\toprule
\AVTableHeader
Evidence layer & Observed result & Role in the storyline & Claim boundary \\
\midrule
Primary joint pattern & Ethereum: mean weekly breadth +91.0\%; position-holder-event HHI $-31.5\%$. Arbitrum: breadth +1.7\%; HHI +58.1\%. & Ethereum is consistent with address-level peripheral expansion; Arbitrum is closer to incumbent amplification than to broad peripheral expansion. & The labels classify joint signs. They do not identify a behavioral mechanism, a GHO effect, or economic actors. \\
Membership turnover & Both chains show substantial entry and exit alongside only modestly higher prior-week retention. & Explains why a larger participation stock need not represent a persistent participant base. & Entry, retention, and exit are address-state transitions, not edge rewiring or an estimate of $dG/dp$. \\
Role-aware structure & Action layers differ; global PageRank mass is diffuse while small maximum-$k$ and BE cores and role-specific coreness tails remain. & Qualifies the primary regimes: breadth and activity concentration do not determine operational position. & Address roles and algorithmic cores do not establish common control, ownership, or removal dependence. \\
Common-calendar comparison & Log-participation contrast +1.9833; HHI contrast $-0.01842$. Gnosis supplies most participation movement; HHI rises on both chains and less on Arbitrum. & Adds a contemporaneous comparative diagnostic without overturning the chain-relative regime classification. & Differential paths, anticipation sensitivity, and one comparison chain preclude a treatment-effect interpretation. \\
Infrastructure boundary & Contract-role labels provide narrow protocol-adjacent diagnostics; no verified route-event panel or failure/substitution experiment is available. & Keeps user-side expansion distinct from shared-route or shared-component dependence. & Route resilience and systemic dependence remain unestimated. \\
\bottomrule
\end{tabularx}
\AVTableNoteFont\emph{Notes:} ``Consistent with'' and ``closer to'' denote compatibility between the prespecified simulation typology and observed signs; neither phrase confirms a mechanism. Position holders are protocol-observed addresses, not verified people or economic entities. Weekly and pooled HHI remain separate because HHI is nonlinear in aggregation.
\end{table*}

%% file: sections/06_conclusion.tex
\section{Discussion and Future Research}\label{sec:conclusion}

\subsection{What the fixed-network simulation explains---and what it omits}\label{subsec:dynamic-network-interpretation}

The evidence answers the title question conditionally rather than categorically. Decentralized finance can broaden address-level access while activity, structural position, or infrastructure remains concentrated. In the primary chain-relative comparison, Ethereum's 91.0\% increase in mean weekly participation and 31.5\% decline in mean weekly position-holder-event HHI are consistent with address-level peripheral expansion. Arbitrum's 1.7\% increase in breadth and 58.1\% increase in HHI are closer to incumbent amplification than to broad peripheral expansion. These statements map observed signs to the prespecified simulation typology; they are neither chain-specific predictions from the simulation nor identified effects of GHO.

The Arbitrum--Gnosis comparison complements this interpretation rather than reversing it. On the common 2024 calendar, the log-participation gap widens by +1.9833, largely because Gnosis participation declines, while the HHI contrast is $-0.01842$ because HHI rises on both chains and rises less on Arbitrum. Different clocks and estimands can therefore produce a negative comparative HHI coefficient alongside a positive chain-relative Arbitrum change. The design's differential paths, anticipation sensitivity, and single comparison chain make network adjustment one possible explanation, not a resolved causal mechanism.

The theoretical baseline deliberately holds the interaction matrix fixed to isolate activity adjustment on an existing network. A dynamic extension would instead model
\[
G_{t+1}=\mathcal F(G_t,x_t,p_t,c_t),
\]
where $p_t$ denotes protocol conditions and $c_t$ chain and infrastructure conditions. When differentiable, the total response can be decomposed as
\[
\frac{dx}{dp}=
\underbrace{(I-\beta G)^{-1}\frac{db}{dp}}_{\text{activity response on the existing network}}
+\underbrace{(I-\beta G)^{-1}\beta\frac{dG}{dp}x}_{\text{network-formation response}}.
\]
The first term is the implemented benchmark; the second captures entry, exit, and rewiring. Figure~\ref{fig:address-state-flows} supplies evidence about entry, retention, and exit, but not edge formation or $dG/dp$. It should therefore be read as motivation for a dynamic model rather than as validation of one.

Network-formation research makes several conditional mechanisms plausible \cite{jackson1996strategic,jackson2008social}. Preferential attachment, liquidity complementarities, switching costs, and incumbent reputation may strengthen an existing core. Lower access costs, new user niches, or multichain entry may exert a countervailing force by adding peripheral participants. Common bridges, sequencers, facilitators, or service providers may simultaneously create a cross-layer infrastructure core. The literature does not imply inevitable centralization, and the present results do not establish that peripheral expansion delays it. A sharper future hypothesis is therefore falsifiable: conditional on comparable protocol shocks, peripheral entry slows the growth of activity or structural concentration relative to settings dominated by incumbent amplification. Testing that proposition requires observed network formation, repeated events, and credible comparisons.

\subsection{Cross-disciplinary implications}\label{subsec:cross-disciplinary-implications}

For economics and finance, the organizing contribution is to connect the network multiplier to distributional and dependency outcomes rather than treating aggregate activity as the endpoint. Growth can change participation breadth, activity concentration, structural position, and shared-component dependence in different directions. Future welfare analysis should consequently model heterogeneous entry costs, liquidity complementarities, switching costs, and governance power alongside adoption.

For computer science and blockchain systems, reproducible transaction records are necessary but not sufficient. Economic claims depend on call semantics, contract roles, authorization paths, bridge and facilitator routes, oracle and sequencer dependencies, and evidence linking addresses to decision makers. A route-aware system model would make it possible to distinguish chain diversification from genuine execution-path redundancy and to evaluate failure, substitution, and recovery.

For network science, the results motivate temporal and multilayer analysis in which lending, stablecoin, governance, bridge, and infrastructure relations remain distinct. Memory-aware communities can identify persistent cores and role transitions \cite{clemente2024memory}; higher-order representations can preserve events that jointly involve users, contracts, assets, and infrastructure \cite{iacopini2024temporal,zhang2023higherorder}; and temporal aggregation can be selected according to its effect on modeled dynamics \cite{allen2024compressing}. These methods become evidentially useful only after their interlayer relations are observed rather than inferred from layout or protocol familiarity.

For protocol governance and policy, the implication is operational: adoption dashboards should report participation, activity distribution, structural position, and infrastructure dependence separately, with an explicit unit of observation and evidence status. More users need not mean more decentralization, and a less concentrated user-side outcome need not imply resilience to the loss of a common service. The present framework is best understood as a reproducible baseline and organizing research agenda, not as a claim to have resolved every dimension.

\subsection{Future evidence and identification roadmap}\label{subsec:future-systemic}

Figure~\ref{fig:future-research-roadmap} turns the present claim boundary into a cross-disciplinary program of falsifiable extensions. New data, technology, and application scenarios enable three discipline-specific paths from required evidence to advanced methods and testable outcomes; the detailed six-field roadmap is preserved in Appendix~\ref{app:future-systemic}.

\input{figures/fig08_future_research_roadmap}

The first identification priority is replication across additional same-semantic markets and repeated expansion events. Credible comparison support should be assessed before estimation; anticipation windows and outcomes should be prespecified; and longer pre-periods should be retained. Group-time, interaction-weighted, and imputation event-study methods become relevant only with multiple adoption cohorts and appropriate not-yet-treated comparisons \cite{callaway2021difference,sun2021estimating,goodmanbacon2021difference,borusyak2024revisiting}. Bounded-trend sensitivity can state how conclusions change under departures from parallel trends \cite{rambachan2023credible}. Synthetic comparisons require a larger verified donor pool, stable pre-event fit, predictor balance, and placebo evidence. If a venue later requires model-conditional $p$-values, they should appear only beside these design limitations; they cannot substitute for comparison validity.

The second priority is a dynamic, heterogeneous multilayer network. Lending, GHO issuance and transfer, bridge, governance, and infrastructure relations should form separate technological or institutional layers linked by evidence-supported entities. Counterfactual removal and cascading-failure simulations can then ask whether liquidity and participation remain resilient when a bridge, facilitator, oracle, sequencer, or major liquidity provider fails. Multilayer models with higher-order couplings provide a foundation for studying dependencies that operate at different timescales \cite{nicoletti2024information}. Route events, substitution behavior, and failure observations are prerequisites; the current contract-role and core diagnostics cannot be relabeled as resilience estimates.

The third priority is actor and delegation provenance. Autonomous agents create both a methodological opportunity and an economic-actor problem. Language-model agents may supplement---not replace---the mathematical model and rule-based simulation by representing heterogeneous suppliers, borrowers, liquidators, arbitrageurs, and governance participants with memory and adaptive strategies. EconAgent illustrates this possibility \cite{li2024econagent}, but validation against observed behavior remains essential. More fundamentally, a single principal may operate many agent-controlled wallets, while many users may delegate to the same model, strategy, wallet infrastructure, or service provider. Address-level participation could then expand while effective decision making becomes more concentrated. Qualitative research maps possible DeFi agent roles \cite{ante2026autonomous}, but it does not establish their prevalence in the Aave records studied here.

The measured decentralization vector remains
\[
\boldsymbol{\Delta}_t=(\mathcal P_t,\mathcal A_t,\mathcal S_t,\mathcal I_t).
\]
Once independently evidenced, a future framework could add
\[
\boldsymbol{\Delta}_t^{+}=(\mathcal P_t,\mathcal A_t,\mathcal S_t,\mathcal I_t,\mathcal D_t),
\]
where $\mathcal D_t$ denotes concentration in effective decision making among principals, autonomous agents, models, strategies, and service providers. Candidate indicators include provider or strategy HHI, principal concentration across agent-operated wallets, shared-dependency concentration, and the share of autonomous versus human-confirmed execution. These quantities require verified labels, delegation and authorization records, smart-wallet permissions, intent attestations, auditable execution traces, and independently validated off-chain evidence. Appendix~\ref{app:future-systemic} provides the staged technical requirements.

\subsection{Conclusion}\label{subsec:conclusion-summary}

This study provides an evidence-gated account of decentralization that links economic mechanism to observable protocol networks. The theory explains why aggregate growth can coexist with lower, unchanged, or higher concentration and why cross-chain dispersion need not reduce common-component dependence. The empirical evidence then identifies two bounded address-level regimes: Ethereum is consistent with peripheral expansion, whereas Arbitrum is closer to incumbent amplification. Membership turnover, typed action layers, local cores, and the Arbitrum--Gnosis comparison explain why neither regime is a complete decentralization ranking.

The durable contribution is therefore an organizing framework and reproducible baseline for research at the intersection of economics and finance, computer science, and network science. It makes previously conflated questions separately measurable, states where current evidence ends, and defines the actor, route, dynamic-network, and comparative evidence needed to test the next generation of claims.

%% file: figures/fig08_future_research_roadmap.tex
\begin{figure*}[!t]
\centering
\begin{tikzpicture}[x=1mm,y=1mm,
  header/.style={av/card,minimum width=39mm,minimum height=11mm,font=\AVFigureSmallFont\bfseries,text width=32mm,align=center},
  enabler/.style={av/planned,minimum width=38mm,minimum height=11mm,font=\AVFigureSmallFont,text width=32mm},
  item/.style={av/planned,minimum width=38mm,minimum height=12.5mm,font=\AVFigureSmallFont,text width=32mm,inner sep=1.5mm},
  roadmap flow/.style={av/pending-flow,draw=AVViolet!76},
  synthesis/.style={av/card,draw=AVTeal!75,fill=AVTeal!5,minimum width=126mm,minimum height=14mm,font=\AVFigureSmallFont,text width=117mm}
]
\node[enabler] (data) at (21,82) {\textbf{New data}\\ verified entities, routes, cohorts, and delegation traces};
\node[enabler] (tech) at (64,82) {\textbf{New technology}\\ auditable agents, provenance, and dependency instrumentation};
\node[enabler] (scenarios) at (107,82) {\textbf{New application scenarios}\\ repeated rollouts, outages, and governance interventions};

\node[header,draw=AVBlue!75,fill=AVMist] (eh) at (21,64) {Economics \&\\ finance};
\node[header,draw=AVViolet!70,fill=AVLilac] (ch) at (64,64) {Computer /\\ computational science};
\node[header,draw=AVTeal!75,fill=AVTeal!6] (nh) at (107,64) {Network\\ science};

\node[item,draw=AVBlue!70] (ee) at (21,48) {\textbf{Evidence target}\\ repeated rollouts plus entity and welfare panels};
\node[item,draw=AVBlue!70] (em) at (21,31) {\textbf{Advanced method}\\ cohort designs plus formation and welfare models};
\node[item,draw=AVBlue!70] (eo) at (21,14) {\textbf{Falsifiable outcome}\\ entry slows, leaves unchanged, or reverses concentration growth};

\node[item,draw=AVViolet!72] (ce) at (64,48) {\textbf{Evidence target}\\ route, authorization, execution, and recovery traces};
\node[item,draw=AVViolet!72] (cm) at (64,31) {\textbf{Advanced method}\\ provenance-aware stress tests and agent ablations};
\node[item,draw=AVViolet!72] (co) at (64,14) {\textbf{Falsifiable outcome}\\ independent paths recover, or shared providers propagate loss};

\node[item,draw=AVTeal!76] (ne) at (107,48) {\textbf{Evidence target}\\ typed temporal, multilayer, and higher-order events};
\node[item,draw=AVTeal!76] (nm) at (107,31) {\textbf{Advanced method}\\ memory-aware roles, dynamic communities, and cascades};
\node[item,draw=AVTeal!76] (no) at (107,14) {\textbf{Falsifiable outcome}\\ persistent cores disperse, remain stable, or relocate across layers};

\foreach \a/\b in {data/eh,tech/ch,scenarios/nh,eh/ee,ee/em,em/eo,ch/ce,ce/cm,cm/co,nh/ne,ne/nm,nm/no}{\draw[roadmap flow] (\a.south)--(\b.north);}

\node[synthesis] (s) at (64,-4) {\textbf{Interdisciplinary synthesis:} bounded causal conclusions \enspace$\bullet$\enspace systemic resilience \enspace$\bullet$\enspace economic-actor decentralization \enspace$\bullet$\enspace agency/delegation concentration \enspace$\bullet$\enspace protocol and policy design};
\draw[roadmap flow] (eo.south)--(s.north west);
\draw[roadmap flow] (co.south)--(s.north);
\draw[roadmap flow] (no.south)--(s.north east);
\end{tikzpicture}
\caption{\textbf{From current evidence to next-generation DeFi decentralization research.} The diagram is a prospective research architecture, not a summary of completed causal, actor-level, route-level, or autonomous-agent analyses. Dashed boxes and arrows mark data, technology, methods, and tests that require new validation. Each disciplinary path ends in an explicitly falsifiable outcome rather than an assumed direction of change.}
\label{fig:future-research-roadmap}
\end{figure*}

%% file: appendix/appendix_guide.tex
The appendix follows the logic of the evidence rather than the order in which files were produced. Source and sample records come first; measurement and unit-of-observation choices follow; model and simulation checks are then kept separate from empirical robustness; comparative identification is audited before the final evidence-status and future-research module. This order lets an interdisciplinary reader move from what is documented, to what is measured, to what is inferred, and finally to what remains prospective.

\begingroup
\AVTableSetup
\setlength{\tabcolsep}{3pt}
\setlength{\LTleft}{0pt}
\setlength{\LTright}{0pt}
\renewcommand{\arraystretch}{1.18}
\begin{longtable}{P{0.10\linewidth}P{0.24\linewidth}P{0.35\linewidth}P{0.21\linewidth}}
\caption{Navigation map for the appendix.}
\label{tab:appendix-navigation}\\
\toprule
\AVTableHeader
\textbf{Module} & \textbf{Reader question} & \textbf{Main-text connection} & \textbf{Primary status} \\
\midrule
\endfirsthead
\multicolumn{4}{c}{\tablename\ \thetable\ (continued)}\\
\toprule
\AVTableHeader
\textbf{Module} & \textbf{Reader question} & \textbf{Main-text connection} & \textbf{Primary status} \\
\midrule
\endhead
\midrule
\multicolumn{4}{r}{Continued on next page}\\
\endfoot
\bottomrule
\endlastfoot
A & How is the appendix organized and how should evidence labels be read? & Paper-wide claim and literature architecture & Mixed / guide \\
B & Which sources, clocks, and samples produce the observed panels? & Institutional setting and data construction & \AVVerifiedTag \\
C & What is measured, displayed, analyzed, or bounded, and at which unit? & Four-dimensional measurement framework & \AVObservedTag\ / \AVBoundedTag \\
D & Do the numerical procedures reproduce the formal mechanism? & Model and simulation & \AVSimulatedTag \\
E & How robust are longitudinal membership and structural patterns? & Chain-relative empirical results & \AVObservedTag\ / \AVDiagnosticTag \\
F & What does the comparative design support under its diagnostics? & Arbitrum--Gnosis comparison & \AVDiagnosticTag \\
G & Which claims are current, reproducible, pending, or future? & Discussion and future-research roadmap & Mixed / \AVPendingTag \\
\end{longtable}
\noindent\AVTableNoteFont\emph{Reading key:} \AVVerifiedTag\ denotes a documented source or procedure; \AVObservedTag\ a reported empirical quantity at its stated unit; \AVDiagnosticTag\ a specification or structural check that does not itself identify causality; \AVSimulatedTag\ a mechanism result; \AVBoundedTag\ an assumption-indexed set; \AVPendingTag\ a prespecified but unimplemented extension; and \AVUnavailableTag\ a claim for which required verified inputs do not exist. Figure or table coverage is a presentation attribute, not an evidence category.
\endgroup

%% file: appendix/literature_coding_audit.tex
\subsection{Fixed-corpus literature coding audit}\label{app:literature-coding}

The main-text literature synthesis is claim-led and includes both foundational and recent boundary-setting work through 2026. The earlier fixed 18-work method--domain--evidence coding remains committed in \texttt{generated/data/literature\_method\_domain\_evidence.csv}, \texttt{literature\_sankey\_nodes.csv}, and \texttt{literature\_sankey\_links.csv} for reproducibility. Because that historical corpus is neither exhaustive nor a measure of intellectual weight, the former Sankey is not rendered, numbered, or used to imply citation, influence, importance, evidence quality, or causal strength. Main-text Figure~\ref{fig:literature-landscape} is the current selective chronological synthesis.

%% file: appendix/source_audit_tables.tex
\suspendrefereelines
\begingroup
\setstretch{1.02}
\AVTableSetup
\setlength{\tabcolsep}{3pt}
\begin{longtable}{P{0.145\linewidth}P{0.255\linewidth}P{0.06\linewidth}P{0.115\linewidth}P{0.34\linewidth}}
\caption{Complete event-source record and role in the analysis.}\label{tab:event-source-audit}\\
\toprule
\textbf{Time / chain} & \textbf{Event / source} & \textbf{G} & \textbf{Record} & \textbf{Causal role} \\
\midrule
\endfirsthead
\multicolumn{5}{l}{\tablename\ \thetable\ continued}\\
\toprule
\textbf{Time / chain} & \textbf{Event / source} & \textbf{G} & \textbf{Record} & \textbf{Causal role} \\
\midrule
\endhead
\bottomrule
\endfoot
2022-03-16\par\AVTableMeta{Multichain} & \textbf{Aave v3 launch}\par\AVTableMeta{Official protocol changelog} & B+ & \href{https://aave.com/docs/resources/changelog}{public source} & Protocol-regime boundary \\
2022-07-07\par\AVTableMeta{Ethereum} & \textbf{Introducing GHO forum post}\par\AVTableMeta{Official governance forum / concept announcement} & B & \href{https://governance.aave.com/t/introducing-gho/8730}{public source} & Anticipation-window start candidate \\
2023-01-27\par\AVTableMeta{Ethereum} & \textbf{Aave v3 Ethereum Core market deployed}\par\AVTableMeta{Official protocol changelog} & B+ & \href{https://aave.com/docs/resources/changelog}{public source} & Concurrent market-regime covariate \\
2023-06-06\par\AVTableMeta{Ethereum} & \textbf{GHO mainnet ARFC posted}\par\AVTableMeta{Official governance forum / ARFC} & B & \href{https://governance.aave.com/t/arfc-gho-mainnet-launch/13574}{public source} & Anticipation-window start for mainnet launch \\
2023-07-15\newline 14:02:59Z\par\AVTableMeta{Ethereum} & \textbf{GHO launch payload executed}\par\AVTableMeta{On-chain governance execution / market activation} & A+ & \href{https://etherscan.io/tx/0xae8e542d4fdb5a6a33eeb129bb80f9bf23a1ceb3ef5f6caed1fd634ae3730c0b}{public source} & Primary policy treatment \\
2023-07-15\newline 14:03:47Z\par\AVTableMeta{Ethereum} & \textbf{First verified GHO borrow/mint}\par\AVTableMeta{On-chain user transaction / Aave Borrow and Mint logs} & A- & \href{https://etherscan.io/tx/0x97858017903ffbe793a4c5789293f170719f5a9d918db52e31e7f74e1129dd75}{public source} & Post-treatment implementation check and outcome onset \\
2023-07-16\par\AVTableMeta{Ethereum} & \textbf{Aave changelog records GHO launch}\par\AVTableMeta{Official protocol changelog / public announcement} & B+ & \href{https://aave.com/docs/resources/changelog}{public source} & Post-treatment publicity or attention shock \\
2023-08-16\par\AVTableMeta{Base} & \textbf{Aave v3 Base market deployed}\par\AVTableMeta{Official protocol changelog} & B+ & \href{https://aave.com/docs/resources/changelog}{public source} & Destination-market readiness shock \\
2023-11-06\par\AVTableMeta{Gnosis} & \textbf{Aave v3 Gnosis market deployed}\par\AVTableMeta{Official protocol changelog} & B+ & \href{https://aave.com/docs/resources/changelog}{public source} & Destination-market readiness shock \\
2024-01-29\par\AVTableMeta{Ethereum} & \textbf{GHO Stability Module introduced}\par\AVTableMeta{Official protocol changelog} & B+ & \href{https://aave.com/docs/resources/changelog}{public source} & Mechanism or confounder event \\
2024-05-08\par\AVTableMeta{Arbitrum} & \textbf{Cross-chain GHO ARFC names Arbitrum as first destination}\par\AVTableMeta{Official governance forum / ARFC} & B & \href{https://governance.aave.com/t/arfc-gho-cross-chain-launch/17616}{public source} & Anticipation-window start for Arbitrum \\
2024-06-23\newline 00:25:05Z\par\AVTableMeta{Arbitrum} & \textbf{GHO token and CCIP pool payload executed}\par\AVTableMeta{On-chain governance / infrastructure transaction} & A & \href{https://arbiscan.io/tx/0x3c5b32821cd57712e686488f3ea1963b58a12b7dab715328e292d239df001c3a}{public source} & Technical-readiness clock \\
2024-07-02\newline 15:40:32Z\par\AVTableMeta{Arbitrum} & \textbf{GHO reserve activated in Aave Arbitrum}\par\AVTableMeta{On-chain governance execution / market activation} & A+ & \href{https://arbiscan.io/tx/0x8cad096ec3dfdc94be9376f32d1a77909e6afe074575b2a3beaca88eb46ec8a6}{public source} & Primary policy treatment \\
2024-07-02\newline 15:45:24Z\par\AVTableMeta{Arbitrum} & \textbf{First verified successful CCIP GHO delivery}\par\AVTableMeta{On-chain cross-chain delivery transaction} & A- & \href{https://arbiscan.io/tx/0xe7bdb7e7ccd5ca9750e3d7aca7c2c57531e8fa99bc39b80b60f80529e3d39357}{public source} & Technical-availability robustness clock \\
2024-07-02\newline 16:01:29Z\par\AVTableMeta{Arbitrum} & \textbf{First verified user GHO borrow}\par\AVTableMeta{On-chain user transaction / Aave Borrow logs} & A- & \href{https://arbiscan.io/tx/0xf29752a7d517637c6576ec034a01252848f75c470c14a4b5a099eddaaabf90e6}{public source} & Post-treatment implementation check and outcome onset \\
2024-10-07\par\AVTableMeta{Base} & \textbf{GHO-on-Base ARFC posted}\par\AVTableMeta{Official governance forum / ARFC} & B & \href{https://governance.aave.com/t/arfc-launch-gho-on-base-set-aci-as-emissions-manager-for-rewards/19338}{public source} & Anticipation-window start for Base \\
2024-10-07\par\AVTableMeta{Avalanche} & \textbf{GHO-on-Avalanche ARFC posted}\par\AVTableMeta{Official governance forum / ARFC} & B & \href{https://governance.aave.com/t/arfc-launch-gho-on-avalanche-set-aci-as-emissions-manager-for-rewards/19339}{public source} & Anticipation-window start for Avalanche \\
2025-01-07\par\AVTableMeta{Mantle} & \textbf{Aave v3 Mantle ARFC posted}\par\AVTableMeta{Official governance forum / market ARFC} & B & \href{https://governance.aave.com/t/arfc-deploy-aave-v3-on-mantle/20542}{public source} & Anticipation-window start for Mantle market \\
2025-02-10\newline 15:28:47Z\par\AVTableMeta{Base} & \textbf{GHO and CCIP configuration payload executed}\par\AVTableMeta{On-chain governance / infrastructure transaction} & A & \href{https://basescan.org/tx/0x9de14a0ba0726c5856825c578a5e2b93fd40100d518a06ef65cebc139ee4b13e}{public source} & Technical-readiness clock \\
2025-02-10\newline 16:19:31Z\par\AVTableMeta{Base} & \textbf{GHO reserve listed and borrowing enabled in Aave Base}\par\AVTableMeta{On-chain governance execution / market activation} & A+ & \href{https://basescan.org/tx/0x0e26c649e7421d9e2bf10f61c511d396b4c503963f2bf8682b9a606471c95e1b}{public source} & Primary policy treatment \\
2025-02-10\newline 16:27:25Z\par\AVTableMeta{Base} & \textbf{First verified user GHO borrow}\par\AVTableMeta{On-chain user transaction / Aave Borrow logs} & A- & \href{https://basescan.org/tx/0x0706d479ee55af0b7b40ac1190646f65fdba881724a6a681a4fc4a6d59e177e0}{public source} & Post-treatment implementation check and outcome onset \\
2025-03-13\par\AVTableMeta{Gnosis} & \textbf{GHO-on-Gnosis ARFC posted}\par\AVTableMeta{Official governance forum / ARFC} & B & \href{https://governance.aave.com/t/arfc-launch-gho-on-gnosis-chain/21379}{public source} & Anticipation-window start for Gnosis \\
2025-06-25\newline 20:21:16Z\par\AVTableMeta{Avalanche} & \textbf{GHO and CCIP configuration payload executed}\par\AVTableMeta{On-chain governance / infrastructure transaction} & A & \href{https://snowscan.xyz/tx/0x02dccf97560254656ce0563558a0f84e740e7fd6bc40da5353607c2e2511f0ba}{public source} & Technical-readiness clock \\
2025-06-25\newline 21:06:11Z\par\AVTableMeta{Avalanche} & \textbf{GHO reserve listed and borrowing enabled in Aave Avalanche}\par\AVTableMeta{On-chain governance execution / market activation} & A+ & \href{https://snowscan.xyz/tx/0xfc950ab0642b944a0d81d312af8743bf0a9bce6fa7b16304a98c5fc6604c204e}{public source} & Primary policy treatment \\
2025-06-25\newline 21:40:46Z\par\AVTableMeta{Avalanche} & \textbf{First verified user GHO borrow}\par\AVTableMeta{On-chain user transaction / Aave Borrow logs} & A- & \href{https://snowscan.xyz/tx/0xfcc2ec076368b1a149d9323555fdfb1666078a34121f0637c36a0ff43dcd2f48}{public source} & Post-treatment implementation check and outcome onset \\
2025-08-05\newline 14:48:10Z\par\AVTableMeta{Gnosis} & \textbf{GHO and CCIP configuration payload executed}\par\AVTableMeta{On-chain governance / infrastructure transaction} & A & \href{https://gnosisscan.io/tx/0x19f54b6b17728d7aff96e9d169e2205cc1a6c3b959dafcaa34de13f054f8ef8b}{public source} & Technical-readiness clock \\
2025-08-05\newline 17:18:10Z\par\AVTableMeta{Gnosis} & \textbf{GHO reserve listed and borrowing enabled in Aave Gnosis}\par\AVTableMeta{On-chain governance execution / market activation} & A+ & \href{https://gnosisscan.io/tx/0xa83f67826954a6aac0ffc1d9b09885bade5c15d61cc49c06c73d9b4042142163}{public source} & Primary policy treatment \\
2025-08-05\newline 19:38:40Z\par\AVTableMeta{Gnosis} & \textbf{First verified user GHO borrow}\par\AVTableMeta{On-chain user transaction / Aave Borrow logs} & A- & \href{https://gnosisscan.io/tx/0x94756d22fa5f5b394a089d57b5f8df16a824fbf2c141ed16a1e66497bf9c0f3c}{public source} & Post-treatment implementation check and outcome onset \\
2026-01-29\newline 08:22:32Z\par\AVTableMeta{Mantle} & \textbf{GHO and CCIP configuration payload executed}\par\AVTableMeta{On-chain governance / infrastructure transaction} & A & \href{https://mantlescan.xyz/tx/0x682add8cb17a8d3ffd05e05d4b229022608ab682c109f027538d4e7cae4bc0eb}{public source} & Technical-readiness clock \\
2026-02-11\newline 10:10:50Z\par\AVTableMeta{Mantle} & \textbf{Aave v3 Mantle market activated with GHO reserve}\par\AVTableMeta{On-chain governance execution / market activation} & A+ & \href{https://mantlescan.xyz/tx/0x2ce73734a150fabab836ab33b2af537343a1f149d5bf44b5f6886ba30fb57497}{public source} & Primary policy treatment \\
2026-02-12\newline 16:16:50Z\par\AVTableMeta{Mantle} & \textbf{First verified user GHO borrow}\par\AVTableMeta{On-chain user transaction / Aave Borrow logs} & A- & \href{https://mantlescan.xyz/tx/0xe2f4e8c7ca5a6b312f3643d1f1f4bf23132c9eb371bafcc4156089c0d73c5874}{public source} & Post-treatment implementation check and outcome onset \\
\end{longtable}
\endgroup

\begingroup
\setstretch{1.02}
\AVTableSetup
\setlength{\tabcolsep}{3pt}
\begin{longtable}{P{0.205\linewidth}P{0.09\linewidth}P{0.52\linewidth}P{0.12\linewidth}}
\caption{Complete acquisition-source catalog and delivery-risk audit.}\label{tab:source-catalog}\\
\toprule
\textbf{Source / scope} & \textbf{E/D} & \textbf{Trust and limitation} & \textbf{Record} \\
\midrule
\endfirsthead
\multicolumn{4}{l}{\tablename\ \thetable\ continued}\\
\toprule
\textbf{Source / scope} & \textbf{E/D} & \textbf{Trust and limitation} & \textbf{Record} \\
\midrule
\endhead
\bottomrule
\endfoot
\path{aave_gho_docs}\par\AVTableMeta{Scope: GHO mechanism and deployed contracts} & B+/A & High for Aave's maintained description of GHO, facilitator mechanics, CCIP design, and published contract links. Limit: Mutable documentation cannot replace historical block-level activation or event records. & \href{https://aave.com/docs/ecosystem/gho}{public source} \\
\path{aave_changelog}\par\AVTableMeta{Scope: Protocol launch chronology} & B+/A & High for official day-level launch chronology and public-attention dates. Limit: Mutable and usually day-level; not proof of exact on-chain execution. & \href{https://aave.com/docs/resources/changelog}{public source} \\
\path{aave_governance_forum}\par\AVTableMeta{Scope: Governance concepts and ARFCs} & B/A & High for proving when an intention or proposal became public and could affect expectations. Limit: Posts may be edited and do not prove approval, payload execution, or market availability. & \href{https://governance.aave.com/}{public source} \\
\path{aavekit_v3_graphql}\par\AVTableMeta{Scope: Current Aave V3 market and reserve catalog} & B+/A & High for current market, pool, reserve, and token metadata returned by Aave's official interface. Limit: Mutable current-state API with caching; not a historical event archive or treatment evidence. & \href{https://api.v3.aave.com/graphql}{public source} \\
\path{aave_address_book}\par\AVTableMeta{Scope: Maintained protocol contract registry} & B+/A & High when the dated source record is identified and addresses are checked against deployed bytecode. Limit: Current web references can change; human-readable constants still require on-chain validation. & \href{https://github.com/aave-dao/aave-address-book}{public source} \\
\path{aave_v3_core_interface}\par\AVTableMeta{Scope: Canonical Aave V3 Pool event ABI} & B+/A & High for event signatures, indexed fields, and ABI types when the official source file and retrieval date are documented. Limit: Source code defines decoding semantics but does not prove log completeness or historical contract bytecode; both require RPC checks. & \href{https://github.com/aave/aave-v3-core/blob/782f51917056a53a2c228701058a6c3fb233684a/contracts/interfaces/IPool.sol}{public source} \\
\path{aave_protocol_subgraphs}\par\AVTableMeta{Scope: Aave event-indexing code and deployment records} & B+/A & High for schema, mappings, and advertised deployment identifiers when the dated source files are documented. Limit: Source documentation does not provide query results; indexed data may lag or omit events. & \href{https://github.com/aave/protocol-subgraphs}{public source} \\
\path{the_graph_gateway}\par\AVTableMeta{Scope: Aave subgraph query delivery} & A-/B & High for indexed records when deployment ID, indexed block, query, and retained response are documented and reconciled to raw logs. Limit: Requires access credentials or payment; indexer lag, schema changes, and mapping errors remain possible. & \href{https://thegraph.com/docs/en/subgraphs/querying/introduction/}{public source} \\
\path{ethereum_bigquery_public}\par\AVTableMeta{Scope: Ethereum token-transfer warehouse} & A-/B & High for reproducible large-scale indexed transfers when the query, table schema, partition, and archived output are documented. Limit: Derived index rather than consensus RPC; access needs a billing project and schemas can change. & \href{https://cloud.google.com/blockchain-analytics/docs/supported-datasets}{public source} \\
\path{ethereum_rpc_drpc}\par\AVTableMeta{Scope: Ethereum consensus records} & A+/B & Very high for immutable blocks, receipts, and logs after chain-ID and cross-provider checks. Limit: Third-party delivery may throttle, return transient nulls, or change free-plan limits; retries and a second provider are mandatory. & \href{https://eth.drpc.org}{public source} \\
\path{ethereum_rpc_mevblocker}\par\AVTableMeta{Scope: Ethereum historical Pool logs and headers} & A+/B & Very high for immutable Ethereum headers and logs after chain-ID checks and exact independent-provider samples. Limit: Public delivery can throttle or change archival policy; every queried range is documented and independently sampled. & \href{https://rpc.mevblocker.io}{public source} \\
\path{ethereum_rpc_blast}\par\AVTableMeta{Scope: Ethereum consensus validation samples} & A+/B- & Very high for immutable records when they exactly match the primary provider. Limit: Historical eth\_getLogs requests were limited to ten blocks at the audit date; suitable only for fixed validation samples. & \href{https://eth-mainnet.public.blastapi.io}{public source} \\
\path{ethereum_rpc_nodies}\par\AVTableMeta{Scope: Ethereum historical state validation} & A+/B & Very high for immutable historical bytecode when chain ID, block, address, and a second provider agree exactly. Limit: Community delivery can throttle or change archival availability; it is used as an independent validation provider rather than the sole log source. & \href{https://eth-pokt.nodies.app}{public source} \\
\path{arbitrum_official_rpc}\par\AVTableMeta{Scope: Arbitrum consensus records} & A+/B+ & Very high for immutable Arbitrum blocks, receipts, and logs after chain-ID validation. Limit: Availability and historical-range limits can change; raw output and retries must be recorded. & \href{https://arb1.arbitrum.io/rpc}{public source} \\
\path{base_official_rpc}\par\AVTableMeta{Scope: Base consensus records} & A+/B+ & Very high for immutable Base blocks, receipts, and logs after chain-ID validation. Limit: Public endpoint is rate limited and not guaranteed for large archival queries. & \href{https://mainnet.base.org}{public source} \\
\path{avalanche_official_rpc}\par\AVTableMeta{Scope: Avalanche C-Chain consensus records} & A+/B+ & Very high for immutable C-Chain blocks, receipts, and logs after chain-ID validation. Limit: Historical-range and rate limits can change; a complete retrieval record is required. & \href{https://api.avax.network/ext/bc/C/rpc}{public source} \\
\path{gnosis_public_rpc}\par\AVTableMeta{Scope: Gnosis Chain consensus records} & A+/B & Very high for immutable Gnosis blocks, receipts, and logs after chain-ID validation. Limit: Community endpoint availability and backend composition can change; independent checks are required. & \href{https://rpc.gnosischain.com}{public source} \\
\path{mantle_official_rpc}\par\AVTableMeta{Scope: Mantle consensus records} & A+/B+ & Very high for immutable Mantle blocks, receipts, and logs after chain-ID validation. Limit: Rate limits and archive support can change; raw responses and failure logs must be retained. & \href{https://rpc.mantle.xyz}{public source} \\
\path{evm_explorer_pages}\par\AVTableMeta{Scope: Human-readable transaction and contract mirrors} & A/B+ & High for public inspection of a specific transaction, decoded logs, and contract page when identified by transaction hash. Limit: Explorer labels and decoders are mutable third-party annotations; RPC is authoritative for analysis. & \href{https://etherscan.io/}{public source} \\
\path{chainlink_ccip_docs}\par\AVTableMeta{Scope: Cross-chain GHO infrastructure mechanism} & B+/A & High for CCIP protocol definitions, chain selectors, and official integration documentation. Limit: Documentation cannot prove that a specific GHO route was active or used at a given block. & \href{https://docs.chain.link/ccip}{public source} \\
\path{bergermann_tudisco_core}\par\AVTableMeta{Scope: Nonlinear multilayer core-periphery method} & A/A & High for the published method when equations, parameters, convergence diagnostics, and implementation details are documented. Limit: Method validity does not validate entity labels, edge construction, or causal interpretation in this application. & \href{https://link.aps.org/doi/10.1103/yr23-71yc}{public source} \\
\end{longtable}
\endgroup
\resumerefereelines
\noindent\AVTableNoteFont Every displayed source label is a live hyperlink to the complete public URL. Literal URLs, access requirements, trust statements, and limitations are preserved in the canonical CSV files and the generated Markdown audit.

%% file: appendix/real_v2_ethereum_audit.tex
\subsection{Audited Ethereum real-data extraction}\label{app:real-v2-ethereum}
The symmetric Ethereum extraction covers event weeks $-16$ through $+16$ around the governance-controlled GHO activation block. It is an address-level descriptive input, not a causal estimate or an entity-level result.

\begin{table}[htbp]
\centering
\caption{Audited Ethereum Aave V3 Pool extraction.}
\label{tab:real-v2-ethereum-audit}
\AVTableSetup
\begin{tabular}{lr}
\toprule
\textbf{Audit item} & \textbf{Value} \\
\midrule
First block & 16,904,946 \\
Last block & 18,549,157 \\
Pool events & 118,806 \\
Transactions & 103,171 \\
Distinct addresses & 15,762 \\
Reserve addresses & 25 \\
Raw partitions & 165 \\
Boundary checks matching & 4/4 \\
Log samples matching & 4/4 \\
Causal estimate produced & No \\
\bottomrule
\end{tabular}
\end{table}

\begin{table}[htbp]
\centering
\caption{Action composition in the 33-week study panel.}
\label{tab:real-v2-ethereum-actions}
\AVTableSetup
\begin{tabular}{lrrr}
\toprule
\textbf{Action} & \textbf{Events} & \textbf{Weekly minimum} & \textbf{Weekly maximum} \\
\midrule
Borrow & 28,772 & 330 & 1,883 \\
Liquidation & 362 & 0 & 74 \\
Repay & 17,524 & 182 & 1,207 \\
Supply & 45,703 & 512 & 2,422 \\
Withdraw & 26,445 & 317 & 1,728 \\
\bottomrule
\end{tabular}
\begin{minipage}{0.90\linewidth}
\AVTableNoteFont\emph{Notes:} Counts are descriptive. Addresses are not persons or entities; raw reserve amounts are not combined across assets or labelled USD. No treatment effect is estimated in this extraction stage.
\end{minipage}
\end{table}
All four sampled boundary records, all four sampled log windows, and the Pool
bytecode at block 17,699,249 matched across independent providers. Complete
source grades, delivery limitations, public URLs, and access requirements appear
in Appendix~\ref{tab:source-catalog}.

%% file: appendix/measurement_glossary_and_claim_map.tex
\subsection{Measurement glossary and claim map}\label{app:indicator-registry}

The measurement system is classified on three separate axes: substantive dimension, research function, and evidence status. Presentation coverage---prose, table, or figure---is recorded separately because ``visual'' is not a competing scientific category. Table~\ref{tab:measurement-use-index} supplies the compact crosswalk used in the main text; the complete 39-indicator registry then remains grouped by participation, activity distribution, network structure, and infrastructure dependence. ``Observed'' means computed from documented address or address-role data; it does not imply economic-actor identification or causality. ``Bounded'' means the analysis reports an assumption-indexed set rather than a point estimate. ``Simulated'' denotes the mechanism simulation, while ``unavailable'' means that the required verified input does not exist.

\input{tables/tabA_measurement_use_index}

\begingroup
\AVTableSetup
\setlength{\tabcolsep}{3pt}
\setlength{\LTleft}{0pt}
\setlength{\LTright}{0pt}
\renewcommand{\arraystretch}{1.18}
\begin{longtable}{P{0.24\linewidth}P{0.43\linewidth}P{0.23\linewidth}}
\caption{Claim map for this study.}
\label{tab:claim-map}\\
\toprule
\AVTableHeader
\textbf{Claim class} & \textbf{Permitted interpretation} & \textbf{Status} \\
\midrule
\endfirsthead
\multicolumn{3}{c}{\tablename\ \thetable\ (continued)}\\
\toprule
\AVTableHeader
\textbf{Claim class} & \textbf{Permitted interpretation} & \textbf{Status} \\
\midrule
\endhead
\midrule
\multicolumn{3}{r}{Continued on next page}\\
\endfoot
\bottomrule
\endlastfoot
Longitudinal address outcomes & Chain-relative breadth and event-share concentration & \AVObservedTag \\
Address-role topology & Structural position in the delegated projection & \AVObservedTag \\
Arbitrum--Gnosis comparison & Common-calendar DiD-style change, controlled ITS, and dynamic contrasts & \AVDiagnosticTag \\
Actor concentration & Assumption-indexed identified set only; entity decentralization is defined but not point-measured & \AVBoundedTag \\
Infrastructure-core diagnostic & BE core nodes with public address-label audit; no route-removal loss & \AVDiagnosticTag \\
Mechanism simulation & Comparative statics, not fitted evidence & \AVSimulatedTag \\
Subgraph reconciliation & Independent future validation protocol & \AVPendingTag \\
Route dependence / causal ATT & No reported result & \AVUnavailableTag \\
\end{longtable}
\endgroup
\input{tables/tabA_network_measure_glossary}

%% file: tables/tabA_measurement_use_index.tex
\begingroup
\AVTableSetup
\setlength{\tabcolsep}{3pt}
\setlength{\LTleft}{0pt}
\setlength{\LTright}{0pt}
\renewcommand{\arraystretch}{1.18}
\begin{longtable}{P{0.22\linewidth}P{0.42\linewidth}P{0.28\linewidth}}
\caption{Measure-use index: research function, evidence status, and presentation coverage.}
\label{tab:measurement-use-index}\\
\toprule
\AVTableHeader
\textbf{Measure family} & \textbf{Scientific classification} & \textbf{Presentation coverage} \\
\midrule
\endfirsthead
\multicolumn{3}{c}{\tablename\ \thetable\ (continued)}\\
\toprule
\AVTableHeader
\textbf{Measure family} & \textbf{Scientific classification} & \textbf{Presentation coverage} \\
\midrule
\endhead
\midrule
\multicolumn{3}{r}{Continued on next page}\\
\endfoot
\bottomrule
\endlastfoot
Active-address breadth and transitions & \textbf{Dimension:} Participation. \textbf{Function:} Data construction and descriptive analysis. \textbf{Status:} \AVObservedTag. & Main text, figures, and tables \\
Event-frequency HHI and distribution summaries & \textbf{Dimension:} Activity distribution. \textbf{Function:} Descriptive analysis. \textbf{Status:} \AVObservedTag. & Main text, figures, and tables \\
Typed topology, centrality, coreness, and persistence & \textbf{Dimension:} Network structure. \textbf{Function:} Network construction and structural analysis. \textbf{Status:} \AVObservedTag\ / \AVDiagnosticTag. & Main figures; detailed appendix tables \\
Economic-actor concentration & \textbf{Dimension:} Activity distribution and unit boundary. \textbf{Function:} Partial identification. \textbf{Status:} \AVBoundedTag. & Appendix figure and tables \\
Arbitrum--Gnosis contrasts and sensitivity ranges & \textbf{Dimensions:} Participation and activity distribution. \textbf{Function:} Comparative diagnostic. \textbf{Status:} \AVDiagnosticTag. & Main figure and table; appendix diagnostics \\
Network multiplier and growth--concentration regimes & \textbf{Dimensions:} Mechanism across dimensions. \textbf{Function:} Formal analysis and simulation. \textbf{Status:} \AVSimulatedTag. & Main model figures and tables \\
Route loss, shared components, and substitution & \textbf{Dimension:} Infrastructure dependence. \textbf{Function:} Simulation and conceptual extension. \textbf{Status:} \AVSimulatedTag\ / \AVUnavailableTag. & Theory figure; future appendix design \\
Entity, delegation, and decision-provider concentration & \textbf{Dimensions:} Economic actors and agency. \textbf{Function:} Conceptual extension. \textbf{Status:} \AVPendingTag\ / \AVUnavailableTag. & Future-system figure and appendix roadmap \\
\end{longtable}
\noindent\AVTableNoteFont\emph{Notes:} Research functions are non-exclusive workflow roles: a measure family may span data construction, descriptive or structural analysis, comparative diagnostics, formal analysis or simulation, partial identification, and conceptual extension. A measure can appear in prose, a table, and a figure simultaneously; ``visual'' therefore describes coverage, not scientific status. The complete registry below remains grouped by the paper's four substantive dimensions.
\endgroup

%% file: tables/tabA_network_measure_glossary.tex
\suspendrefereelines
\begingroup
\setstretch{1.02}
\AVTableSetup
\setlength{\tabcolsep}{3pt}
\setlength{\LTleft}{0pt}
\setlength{\LTright}{0pt}
\renewcommand{\arraystretch}{1.16}
\begin{longtable}{>{\raggedright\arraybackslash}p{0.235\linewidth}>{\raggedright\arraybackslash}p{0.27\linewidth}>{\raggedright\arraybackslash}p{0.425\linewidth}}
\caption{Network-measure glossary: formulas, interpretation, evidence, and visualization coverage.}
\label{tab:network-measure-glossary}\\
\toprule
\AVTableHeader
\textbf{Measure} & \textbf{Notation} & \textbf{Interpretation, evidence, and source} \\
\midrule
\endfirsthead
\multicolumn{3}{c}{\tablename\ \thetable\ (continued)}\\
\toprule
\AVTableHeader
\textbf{Measure} & \textbf{Notation} & \textbf{Interpretation, evidence, and source} \\
\midrule
\endhead
\midrule
\multicolumn{3}{r}{Continued on next page}\\
\endfoot
\bottomrule
\endlastfoot
\AVTableSubheader
\multicolumn{3}{l}{\textbf{Participation decentralization}} \\
\textbf{Active observed participants}\par\vspace{1pt}\AVTableMeta{Source field: num\_\allowbreak{}nodes (when the node set is a period's active addresses)} & \(\displaystyle\begin{gathered}N_t=|V_t|\end{gathered}\) & Number of distinct observed position-holder addresses active in period t; not a count of people or independent economic actors. \emph{Direction:} Higher means broader observed participation, conditional on the address caveat. \par\vspace{1pt}\AVTableMeta{Unit: position-holder address} \par\vspace{1pt}\AVObservedTag; empirical output; Fig.~\ref{fig:real-v5-cross-chain}; \cite{newman2018networks} \\\\
\textbf{Entrants}\par\vspace{1pt}\AVTableMeta{Source field: new nodes} & \(\displaystyle\begin{gathered}E_t=|V_t\setminus V_{t-1}|\end{gathered}\) & Addresses active in period t but not in t-1. \emph{Direction:} Higher means more observed address entry; it need not mean new economic actors. \par\vspace{1pt}\AVTableMeta{Unit: position-holder address} \par\vspace{1pt}\AVObservedTag; empirical output; Fig.~\ref{fig:address-state-flows}; \cite{holme2012temporal} \\\\
\textbf{Exits}\par\vspace{1pt}\AVTableMeta{Source field: departing nodes} & \(\displaystyle\begin{gathered}X_t^{-}=|V_{t-1}\setminus V_t|\end{gathered}\) & Addresses active in t-1 but absent in t. \emph{Direction:} Higher means more observed address exit. \par\vspace{1pt}\AVTableMeta{Unit: position-holder address} \par\vspace{1pt}\AVObservedTag; empirical output; Fig.~\ref{fig:address-state-flows}; \cite{holme2012temporal} \\\\
\textbf{Retained participants and retention rate}\par\vspace{1pt}\AVTableMeta{Source field: retained nodes} & \(\displaystyle\begin{gathered}R_t=|V_t\cap V_{t-1}|\\[-1pt] r_t=R_t/|V_{t-1}|\end{gathered}\) & Addresses appearing in two adjacent periods and their share of the prior-period set. \emph{Direction:} Higher r\_t means greater address persistence, not necessarily actor persistence. \par\vspace{1pt}\AVTableMeta{Unit: position-holder address / share} \par\vspace{1pt}\AVObservedTag; empirical output; Fig.~\ref{fig:address-state-flows}; \cite{holme2012temporal} \\\\
\textbf{Effective number of participants}\par\vspace{1pt}\AVTableMeta{Source field: inverse HHI} & \(\displaystyle\begin{gathered}N^{\ast}=H^{-1}\\[-1pt] H=\sum_i s_i^2\end{gathered}\) & Number of equal-share participants that would generate the observed concentration. \emph{Direction:} Higher means a more even activity distribution. \par\vspace{1pt}\AVTableMeta{Unit: effective address count} \par\vspace{1pt}\AVObservedTag; empirical output; derived; no standalone visual; \cite{hirschman1964paternity} \\\\
\textbf{Cross-chain address-set overlap}\par\vspace{1pt}\AVTableMeta{Source field: Jaccard overlap} & \(\displaystyle\begin{gathered}J_t=\frac{|V_{E,t}\cap V_{A,t}|}{|V_{E,t}\cup V_{A,t}|}\end{gathered}\) & Overlap of Ethereum and Arbitrum position-holder-address sets at the same chain-relative event week. \emph{Direction:} Higher means more repeated addresses across panels; because calendar dates differ it is not contemporaneous migration. \par\vspace{1pt}\AVTableMeta{Unit: share} \par\vspace{1pt}\AVObservedTag; empirical output; no standalone visual; \cite{jaccard1901distribution} \\\\
\textbf{Multilayer participation coefficient}\par\vspace{1pt}\AVTableMeta{Source field: cross-layer participation} & \(\displaystyle\begin{gathered}P_i=1-\sum_{\ell}(k_{i\ell}/k_i)^2\end{gathered}\) & Dispersion of node i's connections across network layers. \emph{Direction:} Higher means connections are spread across more layers. \par\vspace{1pt}\AVTableMeta{Unit: node score in [0,1]} \par\vspace{1pt}\AVUnavailableTag; no reported output; no standalone visual; \cite{kivela2014multilayer} \\\\
\addlinespace[2pt]
\AVTableSubheader
\multicolumn{3}{l}{\textbf{Activity distribution}} \\
\textbf{Total observed activity}\par\vspace{1pt}\AVTableMeta{Source field: total\_\allowbreak{}value /\allowbreak{} event\_\allowbreak{}count} & \(\displaystyle\begin{gathered}X_t=\sum_i x_{it}\end{gathered}\) & Total action count in the empirical panels or equilibrium activity in the simulation; heterogeneous token amounts are not summed across reserves. \emph{Direction:} Scale measure, not a decentralization score by itself. \par\vspace{1pt}\AVTableMeta{Unit: events or model activity} \par\vspace{1pt}\AVObservedTag; empirical output; Fig.~\ref{fig:simulation-outcomes}; \cite{newman2018networks} \\\\
\textbf{Pool-event-frequency Herfindahl--Hirschman index}\par\vspace{1pt}\AVTableMeta{Source field: activity\_\allowbreak{}hhi} & \(\displaystyle\begin{gathered}H_t=\sum_i s_{it}^2\\[-1pt] s_{it}=x_{it}/\sum_jx_{jt}\end{gathered}\) & Squared-share concentration of equally weighted Pool events assigned to the stated observed unit. \emph{Direction:} Higher means greater event-frequency concentration; address HHI is not actor HHI or value concentration. \par\vspace{1pt}\AVTableMeta{Unit: index in [1/n,1]} \par\vspace{1pt}\AVObservedTag; empirical output; Fig.~\ref{fig:real-v5-cross-chain}, Fig.~\ref{fig:real-v4-partial-identification}; \cite{hirschman1964paternity} \\\\
\textbf{Economic-actor HHI period bounds and change envelope}\par\vspace{1pt}\AVTableMeta{Source field: entity-adjusted HHI} & \(\displaystyle\begin{gathered}H_t^a\leq H_t^g\leq1\\[-1pt] \mathcal B_\Delta\subseteq[H_1^a-1,1-H_0^a]\end{gathered}\) & Period-specific actor-HHI bounds are sharp; subscripts 0 and 1 denote pre and post periods. Subtracting them gives a conservative outer envelope unless one common controller partition is optimized jointly. \emph{Direction:} An outer envelope crossing zero does not prove that both signs are jointly feasible. \par\vspace{1pt}\AVTableMeta{Unit: period bounds / outer envelope} \par\vspace{1pt}\AVBoundedTag; partial-identification output; Fig.~\ref{fig:real-v4-partial-identification}; \cite{molinari2020partial,binette2022entity} \\\\
\textbf{Normalized Shannon entropy}\par\vspace{1pt}\AVTableMeta{Source field: entropy} & \(\displaystyle\begin{gathered}S_t=-\frac{\sum_i s_{it}\log s_{it}}{\log n_t}\end{gathered}\) & Evenness of the activity-share distribution. \emph{Direction:} Higher means more even activity. \par\vspace{1pt}\AVTableMeta{Unit: index in [0,1]} \par\vspace{1pt}\AVPendingTag; derivable; not reported; no standalone visual; \cite{shannon1948theory} \\\\
\textbf{Gini coefficient}\par\vspace{1pt}\AVTableMeta{Source field: activity\_\allowbreak{}gini} & \(\displaystyle\begin{gathered}D_t=\sum_{ij}|x_{it}-x_{jt}|\\[-1pt] G_t=D_t/(2n_tX_t)\end{gathered}\) & Mean pairwise inequality in observed activity; the numerator aggregates pairwise absolute differences and the activity total provides the normalization. \emph{Direction:} Higher means greater inequality. \par\vspace{1pt}\AVTableMeta{Unit: index in [0,1]} \par\vspace{1pt}\AVObservedTag; empirical output; no standalone visual; \cite{gini1912variabilita} \\\\
\textbf{Top-p activity share}\par\vspace{1pt}\AVTableMeta{Source field: top1pct\_\allowbreak{}share /\allowbreak{} top10pct\_\allowbreak{}share} & \(\displaystyle\begin{gathered}T_{p,t}=\sum_{j=1}^{\lceil pn_t\rceil}s_{(j)t}\end{gathered}\) & Share held by the largest p fraction of observed units; p is 1\% or 10\% in the analysis panel. \emph{Direction:} Higher means greater top-tail concentration. \par\vspace{1pt}\AVTableMeta{Unit: share} \par\vspace{1pt}\AVObservedTag; empirical output; no standalone visual; \cite{newman2018networks} \\\\
\textbf{Minimum 51\% coalition}\par\vspace{1pt}\AVTableMeta{Source field: nakamoto\_\allowbreak{}51} & \(\displaystyle\begin{gathered}S_{k,t}=\sum_{j=1}^{k}s_{(j)t}\\[-1pt] \mathcal K_t=\{k:S_{k,t}\geq.51\}\\[-1pt] K_{.51,t}=\min\mathcal K_t\end{gathered}\) & Smallest number of observed units whose cumulative ordered share reaches 51\%; the calligraphic K indexed by t is the admissible threshold set. \emph{Direction:} Higher means control of the measured activity is spread across more units. \par\vspace{1pt}\AVTableMeta{Unit: address count} \par\vspace{1pt}\AVObservedTag; empirical output; no standalone visual; \cite{zhang2022sokdecentralization} \\\\
\textbf{Chain concentration}\par\vspace{1pt}\AVTableMeta{Source field: chain\_\allowbreak{}hhi} & \(\displaystyle\begin{gathered}H_{\ell}=\sum_{\ell}s_{\ell}^{2}\end{gathered}\) & Concentration of activity across chain layers. \emph{Direction:} Higher means activity is concentrated on fewer chains. \par\vspace{1pt}\AVTableMeta{Unit: index} \par\vspace{1pt}\AVSimulatedTag; simulated data; Fig.~\ref{fig:simulation-outcomes}; \cite{kivela2014multilayer} \\\\
\addlinespace[2pt]
\AVTableSubheader
\multicolumn{3}{l}{\textbf{Network structure}} \\
\textbf{Graph nodes and edges}\par\vspace{1pt}\AVTableMeta{Source field: num\_\allowbreak{}nodes /\allowbreak{} num\_\allowbreak{}edges} & \(\displaystyle\begin{gathered}n=|V|\\[-1pt] m=|E|\end{gathered}\) & Size of the address-role graph and number of unique observed ties. \emph{Direction:} Scale measures; neither alone ranks decentralization. \par\vspace{1pt}\AVTableMeta{Unit: nodes / edges} \par\vspace{1pt}\AVObservedTag; empirical output; Fig.~\ref{fig:real-v5-core-periphery}; \cite{newman2018networks} \\\\
\textbf{Graph density}\par\vspace{1pt}\AVTableMeta{Source field: Relative degree} & \(\displaystyle\begin{gathered}\delta_G=\chi m/[n(n-1)]\\[-1pt] \chi\in\{1,2\}\end{gathered}\) & Fraction of possible ties that are observed; chi equals 1 for a directed graph and 2 for an undirected graph. \emph{Direction:} Direction is context-dependent; density normally falls mechanically as sparse graphs grow. \par\vspace{1pt}\AVTableMeta{Unit: share} \par\vspace{1pt}\AVPendingTag; derivable; not reported; no standalone visual; \cite{newman2018networks} \\\\
\textbf{Weak components and giant-component share}\par\vspace{1pt}\AVTableMeta{Source field: Components\_\allowbreak{}cnt /\allowbreak{} giant\_\allowbreak{}com\_\allowbreak{}ratio} & \(\displaystyle\begin{gathered}C=|\mathcal C(G)|\\[-1pt] g=|V^{\star}|/n\end{gathered}\) & Number of disconnected weak components and the fraction of nodes in the largest one; the component family is C(G) and the largest node set is V-star. \emph{Direction:} Larger g means greater reachability; it is not necessarily greater decentralization. \par\vspace{1pt}\AVTableMeta{Unit: count / share} \par\vspace{1pt}\AVObservedTag; giant share reported; no standalone visual; \cite{newman2018networks} \\\\
\textbf{Directed reciprocity}\par\vspace{1pt}\AVTableMeta{Source field: directed\_\allowbreak{}reciprocity} & \(\displaystyle\begin{gathered}r=m_{\leftrightarrow}/m\end{gathered}\) & Fraction of directed edges that participate in a reciprocated dyad. \emph{Direction:} Higher means more bidirectional address-role interaction; economic meaning depends on edge semantics. \par\vspace{1pt}\AVTableMeta{Unit: share} \par\vspace{1pt}\AVObservedTag; empirical output; no standalone visual; \cite{garlaschelli2004reciprocity} \\\\
\textbf{Degree moments and top-hub ratio}\par\vspace{1pt}\AVTableMeta{Source field: Degree mean /\allowbreak{} Degree std /\allowbreak{} Top10Degree mean /\allowbreak{} Top10Degree std /\allowbreak{} Top10 Degree mean ratio} & \(\displaystyle\begin{gathered}\bar d=n^{-1}\sum_i d_i\\[-1pt] \sigma_d^2=n^{-1}\sum_i(d_i-\bar d)^2\\[-1pt] Q_{10}=\bar d_{10}/\bar d\end{gathered}\) & Average and dispersion of observed neighbors and the top-ten hub multiple; d-bar-10 is the mean degree of the ten highest-degree nodes. \emph{Direction:} Higher dispersion or Q10 indicates a more unequal degree distribution. \par\vspace{1pt}\AVTableMeta{Unit: degree / ratio} \par\vspace{1pt}\AVPendingTag; derivable; not reported; no standalone visual; \cite{freeman1979centrality} \\\\
\textbf{Degree-centrality mean and dispersion}\par\vspace{1pt}\AVTableMeta{Source field: DCmean /\allowbreak{} DCstd} & \(\displaystyle\begin{gathered}c_i^d=d_i/(n-1)\\[-1pt] \sigma_c^2=n^{-1}\sum_i(c_i^d-\bar c)^2\end{gathered}\) & Normalized degree and cross-node dispersion used as a structural benchmark. \emph{Direction:} Higher dispersion indicates greater hub inequality. \par\vspace{1pt}\AVTableMeta{Unit: centrality / dispersion} \par\vspace{1pt}\AVPendingTag; derivable; not reported; no standalone visual; \cite{freeman1979centrality} \\\\
\textbf{Weighted out- and in-strength HHI}\par\vspace{1pt}\AVTableMeta{Source field: weighted\_\allowbreak{}out\_\allowbreak{}degree\_\allowbreak{}hhi /\allowbreak{} weighted\_\allowbreak{}in\_\allowbreak{}degree\_\allowbreak{}hhi} & \(\displaystyle\begin{gathered}p_i^{\rightarrow}=w_i^{\rightarrow}/\sum_jw_j^{\rightarrow}\\[-1pt] H^{\rightarrow}=\sum_i(p_i^{\rightarrow})^2\\[-1pt] p_i^{\leftarrow}=w_i^{\leftarrow}/\sum_jw_j^{\leftarrow}\\[-1pt] H^{\leftarrow}=\sum_i(p_i^{\leftarrow})^2\end{gathered}\) & Concentration of action-side and position-holder-side Pool-event weights; the arrowed quantities are the corresponding normalized strength shares. \emph{Direction:} Higher means more event weight is concentrated on fewer address roles. \par\vspace{1pt}\AVTableMeta{Unit: index} \par\vspace{1pt}\AVObservedTag; empirical output; no standalone visual; \cite{hirschman1964paternity} \\\\
\textbf{Local clustering moments and transitivity}\par\vspace{1pt}\AVTableMeta{Source field: Cluster\_\allowbreak{}mean /\allowbreak{} Cluster\_\allowbreak{}std /\allowbreak{} Transitivity} & \(\displaystyle\begin{gathered}C_i=2e_i/[d_i(d_i-1)]\\[-1pt] T=3\Delta_G/\Lambda_G\end{gathered}\) & Local and global closure of the undirected projection; Delta-G counts triangles and Lambda-G counts connected triples. \emph{Direction:} Higher means greater triadic closure; it does not have a universal decentralization sign. \par\vspace{1pt}\AVTableMeta{Unit: coefficient} \par\vspace{1pt}\AVPendingTag; derivable; not reported; no standalone visual; \cite{watts1998collective} \\\\
\textbf{Eigenvector and closeness moments}\par\vspace{1pt}\AVTableMeta{Source field: eig\_\allowbreak{}mean /\allowbreak{} eig\_\allowbreak{}std /\allowbreak{} closeness\_\allowbreak{}mean /\allowbreak{} closeness\_\allowbreak{}std} & \(\displaystyle\begin{gathered}A\mathbf c=\lambda_1\mathbf c\\[-1pt] c_i^{\ell}=(n-1)/\sum_{j\neq i}d_{ij}\end{gathered}\) & Recursive-neighbor importance and inverse geodesic distance, summarized by their mean and dispersion. \emph{Direction:} Higher dispersion indicates concentration of the stated centrality; disconnected-graph conventions must be reported. \par\vspace{1pt}\AVTableMeta{Unit: centrality} \par\vspace{1pt}\AVPendingTag; derivable; not reported; no standalone visual; \cite{freeman1979centrality,newman2018networks} \\\\
\textbf{Weighted PageRank and concentration}\par\vspace{1pt}\AVTableMeta{Source field: maximum\_\allowbreak{}pagerank\_\allowbreak{}share /\allowbreak{} pagerank\_\allowbreak{}hhi} & \(\displaystyle\begin{gathered}\mathbf q=(1-\eta)\mathbf 1/n\\[-1pt] \boldsymbol\pi=\mathbf q+\eta P^\top\boldsymbol\pi\\[-1pt] H_{\pi}=\sum_i\pi_i^2\end{gathered}\) & Stationary importance under weighted directed transitions and its concentration; eta is the damping parameter and q is the teleport vector. \emph{Direction:} Higher PageRank HHI means PageRank mass is concentrated on fewer nodes. \par\vspace{1pt}\AVTableMeta{Unit: node score / index} \par\vspace{1pt}\AVObservedTag; empirical output; Fig.~\ref{fig:real-v5-core-periphery}; \cite{brin1998anatomy} \\\\
\textbf{Community modularity}\par\vspace{1pt}\AVTableMeta{Source field: Modularity} & \(\displaystyle\begin{gathered}B_{ij}=A_{ij}-k_ik_j/(2m)\\[-1pt] \delta_{ij}=\mathbf 1\{g_i=g_j\}\\[-1pt] Q=(2m)^{-1}\sum_{ij}B_{ij}\delta_{ij}\end{gathered}\) & Excess within-community connectivity relative to the configuration benchmark; the residual edge term is weighted by a binary same-community indicator. \emph{Direction:} Higher means a stronger partition, not automatically more decentralization. \par\vspace{1pt}\AVTableMeta{Unit: score} \par\vspace{1pt}\AVObservedTag; display-backbone sensitivity only; Fig.~\ref{fig:display-community-audit}; full-graph output unavailable; \cite{newman2006modularity,traag2019leiden} \\\\
\textbf{Core number, maximum k-core, and maximum-core share}\par\vspace{1pt}\AVTableMeta{Source field: maximum\_\allowbreak{}k\_\allowbreak{}core /\allowbreak{} maximum\_\allowbreak{}core\_\allowbreak{}node\_\allowbreak{}share} & \(\displaystyle\begin{gathered}c_i=\max\{k:i\in G_k\}\\[-1pt] I_i^{\star}=\mathbf 1\{c_i=k_{\star}\}\\[-1pt] S_{\star}=n^{-1}\sum_i I_i^{\star}\end{gathered}\) & Nested degree-based cohesion in the unweighted undirected projection; the maximum nonempty shell is paired with a binary membership indicator. \emph{Direction:} Higher coreness means deeper cohesion; the maximum shell and its share must be read together. \par\vspace{1pt}\AVTableMeta{Unit: node shell / count / share} \par\vspace{1pt}\AVObservedTag; empirical output; Fig.~\ref{fig:real-v5-core-periphery}; \cite{batagelj2003cores} \\\\
\textbf{Borgatti--Everett binary core and fit}\par\vspace{1pt}\AVTableMeta{Source field: cp\_\allowbreak{}test\_\allowbreak{}pvalue /\allowbreak{} cp\_\allowbreak{}significance /\allowbreak{} core\_\allowbreak{}cnt /\allowbreak{} avg\_\allowbreak{}core\_\allowbreak{}neighbor} & \(\displaystyle\begin{gathered}B_{ij}=\mathbf 1\{b_i+b_j\geq1\}\\[-1pt] \rho_B=\rho(A,B)\end{gathered}\) & Fits a dense core with core--periphery ties and sparse periphery--periphery ties; the fit is the correlation between observed and ideal adjacency, and the audit requires at least two core nodes. \emph{Direction:} Larger fit means closer agreement with the binary ideal; core count is definition-sensitive. \par\vspace{1pt}\AVTableMeta{Unit: binary node class / fit} \par\vspace{1pt}\AVObservedTag; empirical output; Fig.~\ref{fig:real-v5-core-periphery}; \cite{borgatti2000models,kojaku2018core} \\\\
\textbf{Rombach continuous coreness}\par\vspace{1pt}\AVTableMeta{Source field: continuous coreness robustness} & \(\displaystyle\begin{gathered}R_{\alpha\beta}=\sum_{ij}A_{ij}C_iC_j\\[-1pt] \chi_i\propto\sum_{\alpha,\beta}C_i(\alpha,\beta)R_{\alpha\beta}\end{gathered}\) & Quality-weighted continuous coreness aggregated over the documented alpha-beta grid. \emph{Direction:} Higher means a stronger continuous core position under this definition. \par\vspace{1pt}\AVTableMeta{Unit: node score in [0,1]} \par\vspace{1pt}\AVObservedTag; empirical output; Fig.~\ref{fig:real-v5-core-periphery}; \cite{rombach2014core} \\\\
\textbf{Temporal maximum-core persistence}\par\vspace{1pt}\AVTableMeta{Source field: persistent coreness} & \(\displaystyle\begin{gathered}P_i^{k}=\frac{\sum_t\mathbf 1\{c_{it}=k_{\star t}\}}{\sum_t\mathbf 1\{i\in V_t\}}\end{gathered}\) & Fraction of a node's active topology weeks spent in that week's maximum shell. \emph{Direction:} Higher means a more persistent address-role core position. \par\vspace{1pt}\AVTableMeta{Unit: share} \par\vspace{1pt}\AVObservedTag; empirical output; no standalone visual; \cite{holme2012temporal} \\\\
\textbf{Betweenness brokerage}\par\vspace{1pt}\AVTableMeta{Source field: brokerage} & \(\displaystyle\begin{gathered}b_i=\sum_{s\neq i\neq t}\sigma_{st}(i)/\sigma_{st}\end{gathered}\) & Share of shortest paths between other nodes that pass through i. \emph{Direction:} Higher means greater geodesic brokerage under the chosen path weights. \par\vspace{1pt}\AVTableMeta{Unit: node score} \par\vspace{1pt}\AVPendingTag; derivable; not reported; no standalone visual; \cite{freeman1979centrality} \\\\
\textbf{Temporal community persistence}\par\vspace{1pt}\AVTableMeta{Source field: community persistence} & \(\displaystyle\begin{gathered}J_{gh,t}=\frac{|C_{g,t}\cap C_{h,t+1}|}{|C_{g,t}\cup C_{h,t+1}|}\\[-1pt] P_{g,t}^{\mathcal C}=\max_h J_{gh,t}\end{gathered}\) & Best adjacent-period Jaccard match for a detected community. \emph{Direction:} Higher means more stable community membership, conditional on the algorithm. \par\vspace{1pt}\AVTableMeta{Unit: share} \par\vspace{1pt}\AVPendingTag; no reported output; no standalone visual; \cite{holme2012temporal,traag2019leiden} \\\\
\textbf{Network multiplier centrality and structural HHI}\par\vspace{1pt}\AVTableMeta{Source field: structural\_\allowbreak{}hhi} & \(\displaystyle\begin{gathered}z=(I-\alpha G)^{-1}\mathbf 1\\[-1pt] H_z=\sum_i(z_i/\sum_jz_j)^2\end{gathered}\) & Model-implied amplification of activity through network complementarities and its concentration. \emph{Direction:} Higher Hz means multiplier influence is more concentrated. \par\vspace{1pt}\AVTableMeta{Unit: model score / index} \par\vspace{1pt}\AVSimulatedTag; simulated data; Fig.~\ref{fig:simulation-outcomes}; \cite{ballester2006keyplayer} \\\\
\addlinespace[2pt]
\AVTableSubheader
\multicolumn{3}{l}{\textbf{Infrastructure dependence}} \\
\textbf{Route HHI and effective routes}\par\vspace{1pt}\AVTableMeta{Source field: route\_\allowbreak{}hhi} & \(\displaystyle\begin{gathered}H_R=\sum_k q_k^2\\[-1pt] N_R^{\ast}=H_R^{-1}\end{gathered}\) & Concentration of verified interchain activity across independent routes. \emph{Direction:} Higher route HHI means greater route concentration; a higher effective route count means more dispersion. \par\vspace{1pt}\AVTableMeta{Unit: index / effective route count} \par\vspace{1pt}\AVSimulatedTag\newline\AVUnavailableTag; simulated data only; Fig.~\ref{fig:simulation-outcomes} (simulation); \cite{belchior2021survey} \\\\
\textbf{Bridge or facilitator share}\par\vspace{1pt}\AVTableMeta{Source field: bridge/facilitator shares} & \(\displaystyle\begin{gathered}q_k=x_k/\sum_jx_j\end{gathered}\) & Share of verified interchain activity assigned to infrastructure k. \emph{Direction:} Higher maximum share means greater single-infrastructure dependence. \par\vspace{1pt}\AVTableMeta{Unit: share} \par\vspace{1pt}\AVUnavailableTag; no verified route table; no standalone visual; \cite{belchior2021survey} \\\\
\textbf{Targeted route-removal loss}\par\vspace{1pt}\AVTableMeta{Source field: max\_\allowbreak{}route\_\allowbreak{}removal\_\allowbreak{}loss} & \(\displaystyle\begin{gathered}L_k=(X-X^{(-k)})/X\end{gathered}\) & Fraction of feasible activity lost when infrastructure k and its dependent links are removed. \emph{Direction:} Higher means greater functional dependence on k. \par\vspace{1pt}\AVTableMeta{Unit: share} \par\vspace{1pt}\AVSimulatedTag\newline\AVUnavailableTag; simulated data only; Fig.~\ref{fig:simulation-outcomes} (simulation); \cite{albert2000error} \\\\
\textbf{Edge-disjoint route redundancy}\par\vspace{1pt}\AVTableMeta{Source field: disjoint paths} & \(\displaystyle\begin{gathered}\lambda_{uv}=|\mathcal P_{uv}^{\star}|\end{gathered}\) & Cardinality of a maximum family of pairwise edge-disjoint routes between the relevant endpoints. \emph{Direction:} Higher means more topological redundancy, conditional on verified route edges. \par\vspace{1pt}\AVTableMeta{Unit: path count} \par\vspace{1pt}\AVUnavailableTag; no verified route table; no standalone visual; \cite{ford1956maximal} \\\\
\textbf{Minimum cut}\par\vspace{1pt}\AVTableMeta{Source field: minimum cut} & \(\displaystyle\begin{gathered}\mathcal S_{uv}=\{S:u\in S,v\notin S\}\\[-1pt] \kappa_{uv}=\min_{S\in\mathcal S_{uv}}\sum_{e\in\delta(S)}w_e\end{gathered}\) & Smallest verified route capacity whose removal separates the two endpoints; the calligraphic set enumerates all separating node partitions. \emph{Direction:} Higher means greater cut-based redundancy. \par\vspace{1pt}\AVTableMeta{Unit: capacity or edge count} \par\vspace{1pt}\AVUnavailableTag; no verified route table; no standalone visual; \cite{ford1956maximal} \\\\
\textbf{Expected infrastructure\allowbreak-failure loss}\par\vspace{1pt}\AVTableMeta{Source field: expected route loss} & \(\displaystyle\begin{gathered}\mathcal L=\sum_k\pi_kL_k\end{gathered}\) & Failure-probability-weighted activity loss across infrastructure components. \emph{Direction:} Higher means greater expected infrastructure exposure. \par\vspace{1pt}\AVTableMeta{Unit: expected share} \par\vspace{1pt}\AVDiagnosticTag; no reported output; no standalone visual; \cite{acemoglu2015systemic} \\\\
\addlinespace[2pt]
\end{longtable}
\noindent\begin{minipage}{0.98\linewidth}
\AVTableNoteFont\emph{Notes:} ``Observed'' means a reported empirical result at the stated address or address-role unit; it does not establish an economic-actor or causal claim. ``Simulated'' denotes the mechanism simulation only. ``Pending'' includes additional benchmarks that are derivable from the documented event rows but are not reported in this study. ``Unavailable'' means the necessary verified entity, multilayer, or route-event input does not exist. A missing standalone figure does not imply missing data; the evidence status and data source remain explicit in the glossary.
\end{minipage}
\endgroup
\resumerefereelines

%% file: appendix/real_v3_entity_audit.tex
\subsection{Contract-role and entity audit}\label{app:real-v3-entities}
This audit checks historical runtime code at every participant address's first
and last observed blocks and applies only documented primary-source labels. Runtime-code
families measure shared technical templates, not common economic ownership.

\begin{table}[!ht]
\centering
\AVTableSetup
\caption{Ethereum contract-role and entity-label audit.}
\label{tab:real-v3-entity-audit}
\begin{tabular}{lr}
\toprule
\textbf{Audit item} & \textbf{Value} \\
\midrule
Participant addresses & 15,762 \\
Addresses with observed runtime code & 1,128 (7.2\%) \\
Deduplicated event--address incidences & 148,437 \\
Contract-address incidence share & 38.1\% \\
Primary-source curated labels & 10 \\
Curated-label incidence coverage & 21.9\% \\
Runtime-code template families & 213 \\
Cross-provider code checks & 32/32 \\
Economic-actor incidence coverage & 0.0\% \\
Economic-actor coverage sufficient & No \\
Causal estimate produced & No \\
\bottomrule
\end{tabular}
\begin{minipage}{0.92\linewidth}
\AVTableNoteFont\emph{Notes:} Empty runtime code is not interpreted as a natural person. Curated mappings identify protocol or asset infrastructure; they do not identify the terminal beneficiary behind an adapter-mediated action.
\end{minipage}
\end{table}

\begin{table}[!ht]
\centering
\AVTableSetup
\caption{Unadjusted event-time measurement means.}
\label{tab:real-v3-period-means}
\begin{tabular}{@{}lrrrr@{}}
\toprule
\textbf{Period} & \textbf{Eff. addresses} & \textbf{Curated entity} & \textbf{Contract share} & \textbf{Aave share} \\
\midrule
Pre ($-16$ to $-1$) & 27.04 & 18.31 & 38.0\% & 22.7\% \\
Week 0 & 44.47 & 25.55 & 32.7\% & 19.3\% \\
Post ($+1$ to $+16$) & 38.14 & 20.24 & 38.1\% & 21.7\% \\
\bottomrule
\end{tabular}
\begin{minipage}{0.92\linewidth}
\AVTableNoteFont\emph{Notes:} These period means are descriptive and unadjusted. They are not event-study or difference-in-differences estimates.
\end{minipage}
\end{table}
\input{figures/fig03_real_v3_measurement}

Contract addresses represent only about 7.2\% of observed addresses but carry
38.1\% of event--address incidences. Ten official protocol or asset labels cover
21.9\% of incidences. This gap shows why raw address counts and terminal-user
counts are not interchangeable. Because economic-actor incidence coverage is
0\%, all entity-adjusted values are sensitivities; the primary entity and
causal interpretations remain unsupported.

%% file: figures/fig03_real_v3_measurement.tex
\begin{figure}[htbp]
\centering
\suspendrefereelines
\definecolor{AddressBlue}{HTML}{315EFB}
\definecolor{EntityCoral}{HTML}{7C3AED}
\definecolor{ContractTeal}{HTML}{00A6A6}
\definecolor{ProtocolPurple}{HTML}{FF6B35}
\begin{tikzpicture}[x=.935cm,y=1cm,font=\AVFigureFont]
\path[use as bounding box] (-0.55,-5.55) rectangle (12.85,4.35);
\draw[AddressBlue,very thick] (6.85,4.05)--(7.45,4.05);
\node[av/legend label] at (7.63,4.05) {addresses};
\draw[EntityCoral,very thick,dashed] (9.15,4.05)--(9.75,4.05);
\node[av/legend label] at (9.93,4.05) {curated sensitivity};
\node[av/panel title] at (0,3.62) {A. Effective activity breadth};
\draw[gray!45] (0,0) rectangle (12.48,3.2);
\draw[ProtocolPurple,dashed] (6.24,0) -- (6.24,3.2);
\draw[AddressBlue,very thick] plot[smooth] coordinates {(0.0000,1.3050) (0.3900,1.0880) (0.7800,1.2741) (1.1700,1.2545) (1.5600,1.2686) (1.9500,1.0361) (2.3400,1.2561) (2.7300,1.1993) (3.1200,1.2376) (3.5100,1.6179) (3.9000,1.6372) (4.2900,1.6711) (4.6800,1.7512) (5.0700,2.0409) (5.4600,1.5670) (5.8500,1.8697) (6.2400,2.3719) (6.6300,1.8701) (7.0200,2.7809) (7.4100,1.8740) (7.8000,2.2897) (8.1900,1.9319) (8.5800,2.0761) (8.9700,2.0100) (9.3600,1.9120) (9.7500,1.7410) (10.1400,1.6863) (10.5300,1.5116) (10.9200,1.7064) (11.3100,1.3695) (11.7000,2.6967) (12.0900,2.4956) (12.4800,2.5936)};
\draw[EntityCoral,very thick,dashed] plot[smooth] coordinates {(0.0000,0.8097) (0.3900,0.8138) (0.7800,0.7617) (1.1700,0.7298) (1.5600,0.8139) (1.9500,0.8148) (2.3400,0.9141) (2.7300,0.7966) (3.1200,0.9737) (3.5100,1.0136) (3.9000,1.0006) (4.2900,1.1598) (4.6800,1.2567) (5.0700,1.3604) (5.4600,1.1728) (5.8500,1.2290) (6.2400,1.3626) (6.6300,1.1612) (7.0200,1.5005) (7.4100,1.4102) (7.8000,1.2268) (8.1900,1.1083) (8.5800,1.1568) (8.9700,1.1890) (9.3600,1.0585) (9.7500,1.1424) (10.1400,1.0331) (10.5300,0.8990) (10.9200,0.7016) (11.3100,0.6763) (11.7000,0.9592) (12.0900,1.0398) (12.4800,1.0077)};
\node[anchor=east] at (-0.12,0) {0}; \node[anchor=east] at (-0.12,1.6) {30}; \node[anchor=east] at (-0.12,3.2) {60};
\begin{scope}[yshift=-4.75cm]
\draw[ContractTeal,very thick] (6.30,4.05)--(6.90,4.05);
\node[av/legend label] at (7.08,4.05) {contracts};
\draw[ProtocolPurple,very thick,dash dot] (9.45,4.05)--(10.05,4.05);
\node[av/legend label] at (10.23,4.05) {curated labels};
\node[av/panel title] at (0,3.62) {B. Contract and curated-label incidence};
\draw[gray!45] (0,0) rectangle (12.48,3.2);
\draw[ProtocolPurple,dashed] (6.24,0) -- (6.24,3.2);
\draw[ContractTeal,very thick] plot[smooth] coordinates {(0.0000,1.7386) (0.3900,1.8357) (0.7800,1.8994) (1.1700,1.7199) (1.5600,1.8591) (1.9500,2.1042) (2.3400,2.1990) (2.7300,1.9800) (3.1200,1.9844) (3.5100,1.9747) (3.9000,1.9095) (4.2900,2.3128) (4.6800,2.3401) (5.0700,2.2053) (5.4600,2.1998) (5.8500,2.2017) (6.2400,1.7430) (6.6300,1.8739) (7.0200,1.9889) (7.4100,2.2014) (7.8000,1.8805) (8.1900,1.9988) (8.5800,1.9151) (8.9700,1.9927) (9.3600,2.1476) (9.7500,2.0104) (10.1400,2.0000) (10.5300,2.0850) (10.9200,2.1223) (11.3100,2.1649) (11.7000,2.0839) (12.0900,2.0400) (12.4800,1.9909)};
\draw[ProtocolPurple,very thick,dash dot] plot[smooth] coordinates {(0.0000,1.3460) (0.3900,1.3451) (0.7800,1.3887) (1.1700,1.4210) (1.5600,1.3483) (1.9500,1.3038) (2.3400,1.1816) (2.7300,1.3364) (3.1200,1.1499) (3.5100,1.1680) (3.9000,1.1981) (4.2900,1.0120) (4.6800,1.0158) (5.0700,0.9855) (5.4600,1.0816) (5.8500,1.0569) (6.2400,1.0297) (6.6300,1.1180) (7.0200,0.9429) (7.4100,0.7879) (7.8000,1.0867) (8.1900,1.1384) (8.5800,1.1088) (8.9700,1.0879) (9.3600,1.1623) (9.7500,1.1179) (10.1400,1.1797) (10.5300,1.2752) (10.9200,1.4511) (11.3100,1.4791) (11.7000,1.2237) (12.0900,1.1618) (12.4800,1.1879)};
\node[anchor=east] at (-0.12,0) {0\%}; \node[anchor=east] at (-0.12,1.6) {30\%}; \node[anchor=east] at (-0.12,3.2) {60\%};
\foreach \x/\label in {0/-16,3.12/-8,6.24/0,9.36/+8,12.48/+16}{\draw (\x,0)--(\x,-0.08) node[below=0.8mm]{\label};}
\node at (6.24,-0.65) {Event week relative to GHO activation};
\end{scope}
\end{tikzpicture}
\resumerefereelines
\caption{Address breadth and contract-incidence sensitivity across event time.}
\label{fig:real-v3-measurement}
\begin{minipage}{0.92\linewidth}
\AVFigureSmallFont\emph{Notes:} Curated sensitivity collapses only primary-source protocol/asset labels and leaves all unresolved addresses separate. Contract and Aave shares are event--address incidence shares. Lines are descriptive.
\end{minipage}
\end{figure}

%% file: appendix/real_v4_partial_identification.tex
\subsection{Partial identification under unresolved economic ownership}\label{app:real-v4-partial-identification}
Entity resolution is an inferential problem rather than a clerical replacement of addresses with people \cite{binette2022entity}. We therefore report assumption-indexed identified sets in the sense of partial-identification analysis \cite{molinari2020partial}.

Let $j=1,\ldots,N$ index Aave Pool events and let $a(j)$ denote the event's position-holder address. For each address $a$, let $n_a=\sum_j\mathbf{1}\{a(j)=a\}$ and $q_a=n_a/N$. The economic actors induce an unobserved partition, and $H=\sum_g s_g^2$ is the actor-level event-frequency HHI.

\begin{proposition}[Sharp period-specific logical and stable-address bounds]
If each event may belong to a different underlying actor, $H\in[1/N,1]$. If each address has one controller within the reported group but a controller may operate multiple addresses, $H\in[\sum_a q_a^2,1]$. If verified must-link relations are imposed, replace addresses in the lower endpoint with their connected components.
\end{proposition}
\begin{proof}
Merging two groups with shares $x$ and $y$ changes the HHI by $(x+y)^2-x^2-y^2=2xy\geq0$. The finest admissible partition therefore attains the lower endpoint and a single group attains the upper endpoint. Both endpoints are feasible under the stated assumptions.
\end{proof}

The stable-address model is not assumption-free: a custodial address may represent many underlying users. Conversely, one actor may control many addresses. Shared runtime code, co-timing, transaction similarity, and common counterparties are not accepted as ownership evidence. These bounds are sharp within each period. Their arithmetic difference is only a conservative outer envelope unless one common cross-period controller partition is imposed and optimized jointly.

The treatment clock is the successful governance execution at block 17,699,249 on 15 July 2023 at 14:02:59 UTC. Aave's 16 July changelog entry is retained as a distinct publicity record, not substituted for the on-chain activation.

\begin{table}[!ht]
\centering
\AVTableSetup
\caption{Position-holder address measurements and actor-level HHI bounds.}
\label{tab:real-v4-bounds}
\begin{tabular}{@{}lrrr@{}}
\toprule
\textbf{Quantity} & \textbf{Pre ($-16{:}-1$)} & \textbf{Post ($+1{:}+16$)} & \textbf{Full} \\
\midrule
Events & 38,663 & 76,232 & 118,806 \\
Position-holder addresses & 6,275 & 11,291 & 15,351 \\
Address-proxy HHI & 0.005647 & 0.003555 & 0.003581 \\
Effective addresses & 177.09 & 281.26 & 279.22 \\
Contract position-holder share & 23.70\% & 23.49\% & 23.39\% \\
Stable-address HHI set & $[0.005647,1]$ & $[0.003555,1]$ & $[0.003581,1]$ \\
\bottomrule
\end{tabular}
\begin{minipage}{0.94\linewidth}
\AVTableNoteFont\emph{Notes:} Percentages are position-holder-event shares. The position holder is \texttt{onBehalfOf} for Supply and Borrow and \texttt{user} for Withdraw, Repay, and LiquidationCall. The machine field \texttt{beneficiary\_address} identifies the protocol-observed position-holder address; it is not a verified person or institution.
\end{minipage}
\end{table}
\input{figures/fig04_real_v4_partial_identification}

%% file: figures/fig04_real_v4_partial_identification.tex
\begin{figure}[!ht]
\centering
\suspendrefereelines
\begin{tikzpicture}[
  font=\AVFigureFont,
  panel/.style={rounded corners=2.4mm,draw=AVRule,line width=.7pt,fill=white,
    minimum width=37mm,minimum height=56mm,inner sep=2mm},
  copy/.style={anchor=north,align=center,text width=32mm,inner sep=0pt},
  arrow/.style={-{Latex[length=1.65mm,width=1.12mm]},line width=0.82pt,
    shorten >=.6pt,shorten <=.6pt}
]
\node[panel,draw=AVTeal,fill=AVTeal!5] (address) at (0,0) {};
\node[panel,draw=AVViolet,fill=AVLilac!46] (bounds) at (4.25,0) {};
\node[panel,draw=AVOrange,dashed,fill=AVWarm!55] (boundary) at (8.50,0) {};

\node[inner sep=0pt] at ($(address.north)+(0,-.82)$)
  {\AVClipArt[15mm]{asset-address-role-entity-decoding}};
\node[copy] at ($(address.north)+(0,-1.67)$)
  {\textbf{Observed address proxy}\\118,806 position-holder events\\Pre HHI: 0.005647\\Post HHI: 0.003555\\\textcolor{AVTeal}{\bfseries -37.0\% address HHI}};

\draw[AVViolet,line width=1.0pt] ($(bounds.north)+(-.62,-.78)$)--($(bounds.north)+(.62,-.78)$);
\draw[AVViolet,line width=1.0pt] ($(bounds.north)+(-.62,-.64)$)--($(bounds.north)+(-.62,-.92)$);
\draw[AVViolet,line width=1.0pt] ($(bounds.north)+(.62,-.64)$)--($(bounds.north)+(.62,-.92)$);
\draw[AVOrange,line width=.9pt] ($(bounds.north)+(0,-.55)$)--($(bounds.north)+(0,-1.01)$);
\node[font=\AVFigureTinyFont\bfseries,text=AVOrange] at ($(bounds.north)+(0,-1.18)$) {0};
\node[copy] at ($(bounds.north)+(0,-1.67)$)
  {\textbf{Conservative change envelope}\\Period bounds are sharp\\Post--pre outer envelope:\\$[-0.996445,0.994353]$\\\textcolor{AVViolet}{\bfseries Not a joint sharp set}};

\node[inner sep=0pt] at ($(boundary.north)+(0,-.82)$)
  {\AVClipArt[15mm]{asset-evidence-boundary}};
\node[copy] at ($(boundary.north)+(0,-1.67)$)
  {\textbf{Evidence limit}\\0 accepted actor must-links\\23.39\% contract incidence\\7.43\% curated-label incidence\\\textcolor{AVOrange}{\bfseries No signed actor claim}};

\draw[arrow,draw=AVTeal] (address.east)--(bounds.west);
\draw[arrow,draw=AVViolet] (bounds.east)--(boundary.west);
\node[av/tag,fill=AVTeal] at (2.13,3.28) {Assumption bounds};
\node[av/tag,fill=AVViolet] at (6.38,3.28) {Fail-closed rule};

\node[av/card,draw=AVRule,fill=AVLight,text width=114mm,align=center,
  below=8mm of bounds] {\textbf{Interpretation:} the address proxy becomes less concentrated. Subtracting separate period bounds permits incompatible controller partitions, so the displayed envelope is not a sharp joint change set or a treatment effect.};
\end{tikzpicture}
\resumerefereelines
\caption{Address-level pattern and economic-actor evidence boundary. The protocol-observed position-holder field records the Aave position-holder address, not a verified actor. Period bounds are sharp under their stated assumptions; the displayed change envelope is conservative and no signed actor-change claim is produced.}
\label{fig:real-v4-partial-identification}
\end{figure}

%% file: appendix/model_translation_tables.tex
\subsection{Model-to-protocol and model-to-evidence translations}\label{app:model-translation}

The theoretical model deliberately abstracts from protocol-specific accounting. Tables~\ref{tab:model-protocol-translation} and~\ref{tab:model-network-translation} document the permitted interpretation and prevent the synthetic economic-interaction network from being silently equated with an observed transaction graph.

\begin{table}[!htbp]
\centering
\caption{Translation from protocol conditions to direct-benefit changes.}
\label{tab:model-protocol-translation}
\AVTableSetup
\begin{tabularx}{\textwidth}{P{0.18\textwidth}P{0.27\textwidth}P{0.31\textwidth}Y}
\toprule
Protocol context & Model representation & Illustrative components of $\Delta b_i$ & Non-claim \\
\midrule
GHO issuance on Ethereum & $b^1=b^0+\Delta b$ on the issuance venue & Minting access, borrowing-cost differences, liquidity, repayment, liquidation, arbitrage, or integration opportunities & Does not estimate actor exposure or the historical effect of issuance \\
GHO expansion to Arbitrum & Heterogeneous $\Delta b_i$ plus a destination venue & Transaction costs, destination liquidity and users, incentives, integrations, arbitrage, bridge access, and infrastructure exposure & Does not represent Aave becoming multichain; Aave was already multichain \\
Route stress & Removal of a synthetic route or shared component & Immediate loss of reachability before substitution & Does not measure a deployed CCIP, bridge, router, oracle, or sequencer failure \\
\bottomrule
\end{tabularx}
\end{table}

\begin{table}[!htbp]
\centering
\caption{Distinction between the theoretical and empirical networks.}
\label{tab:model-network-translation}
\AVTableSetup
\begin{tabularx}{\textwidth}{P{0.20\textwidth}P{0.34\textwidth}Y}
\toprule
Object & Construction and meaning & Authorized use \\
\midrule
Theoretical $G$ & Directed weighted economic influence; $g_{ij}>0$ means that $j$'s activity enters $i$'s incentive and the influence edge is $j\rightarrow i$ & Comparative statics and theorem-consistent synthetic simulations \\
Empirical weekly network & Directed action-side-address to position-holder relation derived from typed Pool events & Address-level participation, activity, roles, and structural summaries \\
Entity-control graph & Would link addresses under common control using independently verified evidence & Not identified by the present Pool-event data \\
Infrastructure-dependency graph & Would link routes to shared providers, contracts, governance, or technical components & Required for empirical route-removal and resilience analysis; not presently available \\
\bottomrule
\end{tabularx}
\end{table}

The analogy is mechanism-oriented: a protocol event changes direct incentives, the modeled economic network shapes propagation, the recipients of the response determine concentration, and the venue of activity does not reveal route independence. No equality between model actors and observed addresses is assumed.

%% file: appendix/formal_model_details.tex
\subsection{Formal model details and comparative statics}\label{app:formal-model}

Let $N=\{1,\ldots,n\}$ denote potential economic actors, $V$ chain--market venues, and $G=(g_{ij})$ a possibly directed nonnegative feasible-interaction network with zero diagonal. The convention is explicit: $g_{ij}>0$ means that actor $j$'s activity enters actor $i$'s incentive, so the corresponding influence edge is $j\rightarrow i$. Actor $i$ chooses an outside option or venue $v_i\in V$ and activity $x_i\geq0$. Policy and infrastructure conditions $p$ enter net private benefit through
\begin{equation}
b_i(v_i;p)=a_i+\lambda_{v_i}+\tau_i s_{v_i}-c_{iv_i}-\rho_i r_{v_i}-\delta_i d_{k(v_i)},
\label{eq:net-benefit-app}
\end{equation}
where the terms represent baseline use value, venue value, heterogeneous stablecoin access gains, execution/search costs, risk, and private infrastructure exposure. Conditional on $G$ and venue choices,
\begin{equation}
u_i=b_i(v_i;p)x_i-\frac12x_i^2+\beta x_i\sum_{j\ne i}g_{ij}x_j-\sum_{j\ne i}\kappa_{ij}g_{ij}.
\label{eq:utility-app}
\end{equation}
\subsubsection{Proofs of the preparatory lemmas}

\begin{proof}[Proof of Lemma~\ref{lem:optimal-response}]
Holding $x_{-i}$, $G$, and $p$ fixed, differentiation of Equation~\eqref{eq:payoff-main} gives
\[
\frac{\partial u_i}{\partial x_i}
=b_i(p)-x_i+\beta\sum_jg_{ij}x_j,
\qquad
\frac{\partial^2u_i}{\partial x_i^2}=-1<0.
\]
Strict concavity makes the stationary point unique. Imposing $x_i\geq0$ yields Equation~\eqref{eq:best-response}; when the solution is interior, the truncation does not bind and stacking the responses gives $x=b(p)+\beta Gx$.
\end{proof}

\begin{proof}[Proof of Lemma~\ref{lem:walk-expansion}]
Because $\beta\rho(G)<1$, the matrix series $\sum_{k=0}^{\infty}(\beta G)^k$ converges and equals $(I-\beta G)^{-1}$. Nonnegativity of $\beta$ and $G$ makes every term elementwise nonnegative. Standard matrix multiplication gives $(G^k)_{ij}$ as the sum of the products of edge weights over all length-$k$ walks from $j$ to $i$.
\end{proof}

\begin{proof}[Proof of Lemma~\ref{lem:share-change}]
Since $q_i=x_i/X$, the quotient rule gives
\[
\frac{\partial q_i}{\partial t}
=\frac{m_iX-x_iM}{X^2}
=\frac{m_i-q_iM}{X}.
\]
The sign statement follows because $X>0$.
\end{proof}

\begin{proof}[Proof of Lemma~\ref{lem:location-route}]
The route allocation does not change the fixed chain-share vector. Since the nonnegative route exposures sum to $s$, their maximum is at least their mean $s/R$ and at most their sum $s$. Equal allocation attains the lower bound, whereas $(s,0,\ldots,0)$ attains the upper bound.
\end{proof}

\subsubsection{Proof of Theorem~\ref{thm:equilibrium-multiplier}}

If $\beta\rho(G)<1$ and the active-set net-benefit vector is positive, the activity subgame has the unique interior equilibrium reported in Equation~\eqref{eq:equilibrium}. For a scalar policy component $p_h$ that changes net benefits while $G$ and the active set are fixed,
\begin{equation}
\frac{\partial x^{*}}{\partial p_h}=(I-\beta G)^{-1}\frac{\partial b}{\partial p_h}.
\end{equation}
The inverse admits the nonnegative walk expansion $\sum_{m=0}^{\infty}\beta^mG^m$, which gives the network-multiplier interpretation.

\begin{proof}
The first-order condition from Equation~\eqref{eq:utility-app} is $x=b+\beta Gx$, hence $(I-\beta G)x=b$. Because $\beta\rho(G)<1$, the matrix $I-\beta G$ is nonsingular and its inverse is the convergent Neumann series $\sum_{m=0}^{\infty}\beta^mG^m$. This proves existence and uniqueness of $x^{*}=(I-\beta G)^{-1}b$. Every term of the series is elementwise nonnegative, so $b>0$ implies $x^{*}>0$ and validates the interior solution. Differentiating the equilibrium system while holding $G$ and the active set fixed gives Equation~\eqref{eq:network-multiplier-main}. If $\partial b/\partial p_h\geq0$, elementwise nonnegativity of the inverse gives $\partial x^{*}/\partial p_h\geq0$. An entry is strictly positive whenever a directly affected actor is connected to it by a positive-weight walk.
\end{proof}

\subsubsection{Proof of Theorem~\ref{thm:growth-concentration}}

For $H_x=\sum_i(x_i/X)^2$ and $m_i=\partial x_i^{*}/\partial p_h$,
\begin{equation}
\frac{\partial H_x}{\partial p_h}=\frac{2}{X}\sum_i(q_i-H_x)m_i.
\label{eq:hhi-derivative-app}
\end{equation}
Thus positive aggregate marginal activity $\sum_i m_i$ does not determine the sign of concentration change. Concentration rises when gains are sufficiently tilted toward actors whose initial activity shares exceed the concentration-weighted benchmark.

\begin{proof}
Write $H_x=\sum_i x_i^2/X^2$, where $X=\sum_i x_i$. Differentiation with respect to $p_h$ gives
\[
\frac{\partial H_x}{\partial p_h}
=\frac{2\sum_i x_i m_i}{X^2}
-\frac{2\sum_i x_i^2}{X^3}\sum_i m_i.
\]
Using $q_i=x_i/X$ and $H_x=\sum_iq_i^2$ reduces this expression to Equation~\eqref{eq:hhi-derivative-app}. The derivative is positive, zero, or negative exactly as $\sum_i(q_i-H_x)m_i$ is positive, zero, or negative. By contrast, $\partial X/\partial p_h=\sum_i m_i$, so its sign imposes no restriction on the covariance-like weighted sum governing concentration.
\end{proof}

\subsubsection{Proof of Theorem~\ref{thm:dispersion-resilience}}

The chain-dispersion comparison in the main text assumes a new destination share $s\in(0,1/2]$. Chain HHI falls from one to $(1-s)^2+s^2$. Direct route exposure equals $s$ only when all destination activity is immediately unreachable after route failure and cannot continue locally or substitute. Equal, non-overlapping exposure across $R$ independent routes gives maximum direct exposure $s/R$. These conditions do not identify endogenous equilibrium loss $L_k$.

Route loss and shared-component loss are distinct. If route $k$ has exposure $r_k$, its immediate loss is $r_k$. If component $h$ supports route set $K_h$, its immediate component exposure is $\sum_{k\in K_h}r_k$ when those route exposures do not overlap at the activity level. Consequently, four nominal routes can each have exposure $s/4$ while one component common to all four retains exposure $s$. Neither quantity incorporates behavioral substitution or recovery.

\begin{proof}
The initial chain HHI is one. After the shift it is $(1-s)^2+s^2=1-2s(1-s)$, so the decline is $2s(1-s)>0$. Under the stated no-substitution stress assumption, failure of the sole route disconnects the entire destination share, giving direct exposure $s$. With equal allocation across $R$ independent and non-overlapping routes, each route carries $s/R$, so loss of any single route directly exposes at most $s/R$. No conclusion about equilibrium reallocation after failure follows without behavioral and route-substitution primitives.
\end{proof}

\subsubsection{Boundary cases and robustness scope}

The three theorems are conditional comparative statics, not empirical treatment effects. If $\beta\rho(G)\geq1$, the Neumann-series argument and stable interior equilibrium can fail. If a shock changes the active set, Theorem~\ref{thm:growth-concentration} applies piecewise within regions with a fixed active set; entry or exit creates a boundary at which a one-sided comparison is required. Negative shocks or strategic substitutes can reverse the monotonicity conclusion in Theorem~\ref{thm:equilibrium-multiplier}. Route overlap, correlated failures, local continuation, and endogenous substitution can make $s/R$ an invalid exposure calculation. Appendix~\ref{app:simulation-sensitivity} therefore varies the network-complementarity parameter and records rejected unstable configurations, while the main text treats route resilience as theoretically illustrated but empirically unmeasured.

A reduced-form social objective can additionally account for common infrastructure exposure:
\begin{equation}
W(G,p)=\sum_i u_i(x^{*};G,p)-\omega\sum_k\pi_k(X-X^{-k}),
\end{equation}
where $\pi_k$ is a failure probability or stress weight and $\omega$ converts lost feasible activity into social cost. Neither this welfare object nor empirical route-removal loss is estimated in the present study.

%% file: appendix/simulation_protocol.tex
\subsection{Agent-based simulation protocol}\label{app:simulation-protocol}

The simulation follows a purpose--entities--state variables--process--validation structure consistent with reproducible agent-based model description \cite{grimm2020odd}. Code, machine-readable outputs, chart-source CSV files, generated LaTeX values, and a SHA-256 manifest accompany the pinned companion release. The Overleaf bundle retains the manuscript-facing assets and provenance record without duplicating the chart-source CSV files.

\subsubsection{Purpose, entities, and state variables}

The purpose is theorem validation: to expose the adjustment path implied by the fixed-network best response and to construct controlled counterexamples for growth--concentration and dispersion--resilience divergence. The model contains 40 rule-based Mesa agents. Each agent stores a nonnegative activity level $x_i$ and reads its direct benefit $b_i$, feedback strength $\beta$, and row $i$ of the normalized influence matrix $G$. Agents are not identified wallets, persons, institutions, or language-model systems.

Direct benefits are positive log-normal draws with log mean $-0.35$ and log standard deviation $0.45$. Every topology contains a directed ring to guarantee support. Additional edge probabilities are disclosed in the executable source: the primary core--periphery generator uses probabilities 0.62 within the core, 0.22 across the core boundary, and 0.055 within the periphery; the sparse generator uses 0.075; and the modular generator uses 0.34 within and 0.025 between four modules. Random additional edge weights are uniform on $[0.25,1]$ before spectral normalization.

\begingroup
\AVTableSetup
\setlength{\tabcolsep}{3pt}
\setlength{\LTleft}{0pt}
\setlength{\LTright}{0pt}
\renewcommand{\arraystretch}{1.18}
\begin{longtable}{P{0.25\linewidth}P{0.27\linewidth}P{0.40\linewidth}}
\caption{Complete RC23 simulation configuration.}
\label{tab:simulation-parameters}\\
\toprule
Component & Prespecified value & Role \\
\midrule
\endfirsthead
\multicolumn{3}{c}{\tablename\ \thetable\ (continued)}\\
\toprule
Component & Prespecified value & Role \\
\midrule
\endhead
\midrule
\multicolumn{3}{r}{Continued on next page}\\
\endfoot
\bottomrule
\endlastfoot
Actors & 40 & Fixed active population within each run \\
Primary seed & 20260819 & Main network and figure \\
Topology robustness seeds & 20260819--20260823 & Five seeds on each of three topology families \\
Experiment 2 seeds & 20260819--20260828 & Ten networks for simulation intervals \\
Topologies & Core--periphery, sparse, modular & Structural robustness \\
Primary graph & 219 directed edges; density 0.14038 & Seed-20260819 core--periphery instance \\
Spectral normalization & $\rho(G)=1$ & Makes $\kappa=\beta$ \\
Stable feedback conditions & $0,0.25,0.50,0.75,0.95$ & Named theorem-consistent conditions \\
Boundary test & $\kappa=1.05$ & Assumption violation; reported separately \\
Convergence tolerance & $10^{-9}$ in Euclidean step norm & Prespecified stopping rule \\
Iteration horizons & 1,500 generally; 700 in Experiment 2 & Prespecified finite computational limits \\
Experiment 1 shock mass & 1.0 & Equal central, peripheral, or broad aggregate shock \\
Experiment 2 shock mass & 6.0 & Exactly equal across the three allocations \\
Experiment 3 chain shares & 60\% issuance; 40\% destination & Identical venue dispersion across route structures \\
Activation regimes & Simultaneous, fixed sequential, randomized asynchronous & Order robustness \\
\end{longtable}
\endgroup

\subsubsection{Process and convergence}

For simultaneous activation all agents read $x^{(\ell)}$ before updating. Fixed sequential and randomized asynchronous activation update the current state one actor at a time. For every stable run, the independently computed analytical equilibrium is the validation target. The executable procedure is:

\begin{AVPseudocode}{Theorem-consistent network-activity simulation}{alg:network-activity-simulation}
\AVAlgLine{\AVAlgKeyword{Inputs:} directed influence matrix $G$; direct benefits $b$; feedback $\beta$; activation schedule; seed; iteration horizon $L$; tolerance $\varepsilon=10^{-9}$.}
\AVAlgLine{\AVAlgKeyword{Outputs:} activity path $\{x^{(\ell)}\}$; total activity; HHI; step norm; equilibrium error; convergence or failure status.}
\AVAlgLine{Normalize $G$ so that its spectral radius satisfies $\rho(G)=1$; set $\kappa=\beta\rho(G)$.}
\AVAlgLine{If $\kappa<1$, compute the validation target $x^{\star}=(I-\beta G)^{-1}b$; otherwise record that no finite target is asserted.}
\AVAlgLine{Initialize nonnegative actor activities $x^{(0)}$ under the prespecified condition.}
\AVAlgLine{\AVAlgKeyword{For} iteration $\ell=0,\ldots,L-1$ \AVAlgKeyword{do}}
\AVAlgLineI{Activate actors simultaneously, sequentially, or asynchronously under the specified schedule.}
\AVAlgLineI{For each activated actor $i$, update $x_i\leftarrow\max\{0,b_i+\beta G_{i,:}x\}$ using the schedule-appropriate state.}
\AVAlgLineI{Record $\sum_i x_i$, activity HHI, $\lVert x^{(\ell+1)}-x^{(\ell)}\rVert_2$, and, when defined, $\lVert x^{(\ell+1)}-x^{\star}\rVert_2$.}
\AVAlgLineI{If any state is non-finite or exceeds the defensive stability guard, terminate and report failure.}
\AVAlgLineI{If $\lVert x^{(\ell+1)}-x^{(\ell)}\rVert_2<\varepsilon$, terminate and report convergence.}
\AVAlgLine{\AVAlgKeyword{End for}; if the horizon is reached without the stopping rule, report non-convergence.}
\AVAlgLine{Repeat the procedure across prespecified seeds, shocks, topologies, feedback conditions, and activation schedules; validate stable runs against $x^{\star}$.}
\end{AVPseudocode}

Convergence means satisfaction of the step-norm rule, not merely reaching the iteration horizon. Stable runs are additionally checked using $\lVert x^{(\ell)}-x^*\rVert_2$. Explosive or non-finite states terminate defensively. The $\kappa=1.05$ run has no asserted finite analytical target.

\subsubsection{Experiment-specific interventions}

Experiment 1 compares a zero-shock control, one influence-peripheral actor, one influence-central actor, and a broad shock distributed evenly across all actors. Centrality selection uses PageRank under the declared influence orientation $j\rightarrow i$. Outcomes include direct and indirect activity, directed-distance profiles, convergence iterations, and analytical error. For the three primary feedback values, the figure retains every actor with finite directed distance, groups actors by distance from the shocked node, and plots the group mean with plus or minus one standard error on an explicitly labeled logarithmic response axis.

Experiment 2 ranks actors by baseline equilibrium activity at $\kappa=0.75$. The peripheral and incumbent allocations divide mass 6 equally among the lowest and highest thirds. The proportional benchmark assigns $\Delta b_i=6b_i/\sum_jb_j$, which produces a proportional equilibrium scaling and therefore unchanged HHI up to numerical tolerance. Seed-level outcomes generate 5th--95th simulation intervals; they are not sampling confidence intervals.

Experiment 3 holds destination share at 0.40 and compares one route, two independent routes, four independent routes, and four nominal routes sharing one critical component. It reports route loss and component loss separately. The structures are synthetic and omit behavioral substitution, capacity, congestion, recovery, and rerouting.

\subsubsection{Software and artifacts}

The recorded run used Python 3.12.13, Mesa 3.5.1, NetworkX 3.6.1, NumPy 2.3.5, SciPy 1.17.0, pandas 2.2.3, Matplotlib 3.10.8, Pillow 12.3.0, and pytest 8.4.2. The release manifest records the environment, runtime, parameters, paper values, and release-relative hashes for every generated artifact. Nine automated tests cover analytical recovery, near-boundary convergence, directed influence orientation, equal shock mass and HHI signs, route/component separation, spectral normalization, retention of all three primary feedback values in the distance-response summary, portable external-output manifests, and fail-closed PDF/SVG/PNG validation. Figure exports are validated in temporary files and atomically published only after their format checks pass. Passing these tests establishes internal consistency and artifact integrity, not empirical validity.

%% file: appendix/simulation_sensitivity.tex
\subsection{Simulation validation, robustness, and evidence boundaries}\label{app:simulation-sensitivity}

\subsubsection{Analytical-equilibrium validation}

Every stable primary and robustness run is compared with $x^*=(I-\beta G)^{-1}b$. Across the released experiments, the largest final Euclidean error is $1.90\times10^{-8}$. In the primary central-shock simultaneous-update run, convergence requires 18 iterations at $\kappa=0.25$, 79 at $\kappa=0.75$, and 438 at $\kappa=0.95$. This monotone slowdown is consistent with feedback approaching the spectral boundary. The $\kappa=1.05$ run is an assumption-violation diagnostic and does not converge within the computational rule; it is never summarized as a stable equilibrium.

Simultaneous, fixed sequential, and randomized asynchronous schedules are run on the primary network. Five deterministic seeds are also evaluated across core--periphery, sparse, and modular topologies for each stable positive feedback condition. Every stable topology--seed run converges and matches the analytical benchmark. These checks validate the implementation across updating order and synthetic graph families; they do not establish external validity for observed Aave behavior.

\subsubsection{Directed-distance response profiles}

The directed-distance comparison retains $1$, $13$, $19$, and $7$ actors at distances $0$, $1$, $2$, and $3$ from the influence-central shocked actor. At those distances, mean responses equal $(1.014982,0.046682,0.003905,0.000106)$ for $\kappa=0.25$, $(1.326297,0.321124,0.097987,0.008869)$ for $\kappa=0.75$, and $(3.399475,1.822490,0.767513,0.091824)$ for $\kappa=0.95$. Figure~\ref{fig:three-theory-simulations} plots all three profiles on a logarithmic response scale. Error bars are plus or minus one standard error across actors within a directed-distance group; distance zero has one actor and therefore no cross-actor error bar. These quantities describe heterogeneity in one fixed synthetic network, not confidence intervals or estimated distance effects in Aave.

\subsubsection{Seed-level growth--concentration results}

Experiment 2 reports ten deterministic network seeds rather than a single selected draw. All three interventions have exact direct-benefit mass 6.0. Table~\ref{tab:theory-simulation-results} reports medians; the 5th--95th intervals are 13.01--18.63\% for peripheral-expansion activity and $-5.18$--$-3.61$\% for its HHI, 17.35--20.62\% for proportional activity with numerical HHI change below $1.1\times10^{-9}$ percentage points in absolute value, and 22.88--27.40\% for incumbent-amplification activity with 3.99--6.19\% HHI growth. These are simulation intervals over deterministic seeds, not confidence intervals for a historical population.

The proportional benchmark is exact by construction. Because $\Delta b$ is proportional to $b$, linearity of $(I-\beta G)^{-1}$ scales every actor's equilibrium activity by the same factor. Actor shares and HHI therefore remain unchanged up to numerical solver tolerance. Peripheral and incumbent allocations demonstrate the theorem's two opposing signs without changing aggregate shock mass.

\subsubsection{Route and shared-component sensitivity}

With chain shares fixed at $(0.60,0.40)$, chain HHI equals $0.52$ in every route configuration. Worst immediate route loss is 40\% with one route, 20\% with two independent routes, and 10\% with four routes. Four nominal routes sharing one critical component retain a 10\% worst single-route loss but a distinct 40\% shared-component loss. Reporting the two columns separately prevents nominal lane count from being interpreted as independent-stack resilience.

\subsubsection{Fixed-network and active-set boundary}

The experiment holds the actor set and $G$ fixed. It therefore represents the intensive-margin response
\[
x_i^0\rightarrow x_i^1
\]
conditional on a predetermined interaction structure, not the extensive and structural response
\[
\text{entry, exit, and }G_0\rightarrow G_1.
\]
The empirical temporal networks may turn over substantially; no claim is made that the fixed-network approximation is validated by those address graphs. Node retention, entrant and exit shares, edge retention, Jaccard similarity, weighted edge correlation, PageRank rank correlation, core persistence, and continuously active-actor restrictions are appropriate diagnostics for a future local-approximation test.

Negative interactions, strategic substitutes, active-set changes, adaptive rerouting, endogenous links, capacity constraints, and correlated component failures can alter the model's conclusions or require different solution methods. Simulated magnitudes are not fitted to event counts, chain shares, concentration levels, or protocol-event responses and cannot be cited as estimates or causal effects.

\subsubsection{From simulated feedback to empirically identified network complementarity}

The present data cannot recover $\beta$ from an ordinary least-squares regression such as
\[
x_{it}=\alpha_i+\tau_t+\beta\sum_jg_{ij}x_{jt}+\varepsilon_{it}.
\]
Mutual responses are simultaneous; actors face common shocks; observed links may form endogenously; the theoretical and empirical networks differ; addresses do not necessarily identify actors; and direct event exposure is heterogeneous and incompletely observed. Future work could use structural estimation matched to entry, activity, concentration, and propagation moments; defensible network instruments based on exogenous exposure $z,Gz,G^2z$; dynamic panels with predetermined network lags and explicit sorting controls; or simulation-based calibration validated on held-out periods. A resulting coefficient would need to be labeled according to its identification strategy---for example, a calibrated mechanism parameter rather than a causal estimate.

%% file: appendix/longitudinal_network_robustness.tex
\subsection{Longitudinal network robustness and superseded diagnostics}\label{app:longitudinal-robustness}

\subsubsection{Aggregation identities}
Two reported Ethereum concentration changes answer different questions. Pooling all pre-period events and all post-period events yields position-holder HHI values 0.005647 and 0.003555, a 37.0\% decline. Averaging the 16 weekly HHI values on each side yields 0.00956 and 0.00655, a 31.5\% decline. HHI is nonlinear, so the pooled-period HHI is not the mean weekly HHI. The main text reports the weekly comparison with longitudinal outcomes and the pooled comparison with the actor-bound calculation.

\subsubsection{Topology specification}
Participation and concentration retain self-directed events. Structural measures drop self-loops and use the non-self delegated/third-party projection. Directed out-strength and in-strength HHI preserve action-side and position-holder roles; PageRank is computed on the directed graph; maximum $k$-core, Borgatti--Everett, and Rombach methods use the documented undirected projection. All numerical statistics use the complete graph, while main-text Figure~\ref{fig:real-v5-core-periphery} uses a fixed 160-node display backbone. The display sample is never an estimation sample.

The binary Borgatti--Everett fits hit their two-node lower bound, and both nodes are contained in each maximum-$k$ core. The unequal-set Jaccards (0.143 and 0.019) therefore equal the small-to-large set-size ratios and do not establish conflicting membership. Rombach--$k$ rank correlations are 0.521 and 0.578. The backbone figure has been moved to the structural-results sequence in Section~\ref{sec:causal}; this appendix retains the construction and sensitivity audit without duplicating or renumbering the visual.

\subsubsection{Display-backbone community sensitivity}
The available supplementary data do not contain the full directed weighted edge table, so a population community partition, temporal persistence analysis, Leiden/Infomap comparison, or stochastic-block-model fit cannot be estimated. A narrower diagnostic can be computed on the fixed 160-node display backbone for each chain. Starting from singleton communities, the procedure repeatedly merges the adjacent pair with the largest positive weighted modularity increment, breaking ties by address order. It repeats the calculation with raw event weights, $\log(1+w)$ weights, and binary edges, and compares partitions using normalized mutual information (NMI).

With raw weights, the display diagnostic yields 11 communities on Ethereum ($Q=0.383$) and 14 on Arbitrum ($Q=0.532$). Under $\log(1+w)$ and binary weights, the counts are 11 and 10 on Ethereum and 16 and 16 on Arbitrum. Relative to the raw-weight partition, NMI is 0.740 and 0.412 on Ethereum and 0.663 and 0.509 on Arbitrum. Figure~\ref{fig:display-community-audit} therefore visualizes sensitivity rather than a stable substantive partition. The result is that displayed membership depends materially on the edge-weight definition, especially under binarization; no full-graph or economic-actor community claim follows.

\input{figures/figA_display_community_audit}

\subsubsection{Superseded non-common-calendar contrast}
An earlier table aligned Ethereum's July 2023 issuance window with Arbitrum's July 2024 cross-chain activation window. Both units were exposed to different events and their weeks did not share a calendar. The source file \texttt{tables/tab06\_pilot\_did.tex} is retained for provenance but is no longer rendered or numbered; its result status is \texttt{SUPERSEDED} by the audited Arbitrum--Gnosis common-calendar comparison and it must not be interpreted as an ATT.

\subsubsection{Finite decomposition boundary}
Action and reserve labels remain in the documented event rows, but the analysis dataset does not expose a symmetric row-level Ethereum--Arbitrum panel. Leave-one-action-out, each-action-only, GHO/non-GHO, and high-volume-reserve exclusions are therefore not reported. This absence is a hard interpretation limit: the aggregate patterns are Pool-event-frequency descriptions and are not attributed to stablecoin issuance, a particular reserve, or a policy mechanism.\par

%% file: figures/figA_display_community_audit.tex
\begin{figure}[t]
\centering
\suspendrefereelines
\includegraphics[width=0.98\textwidth]{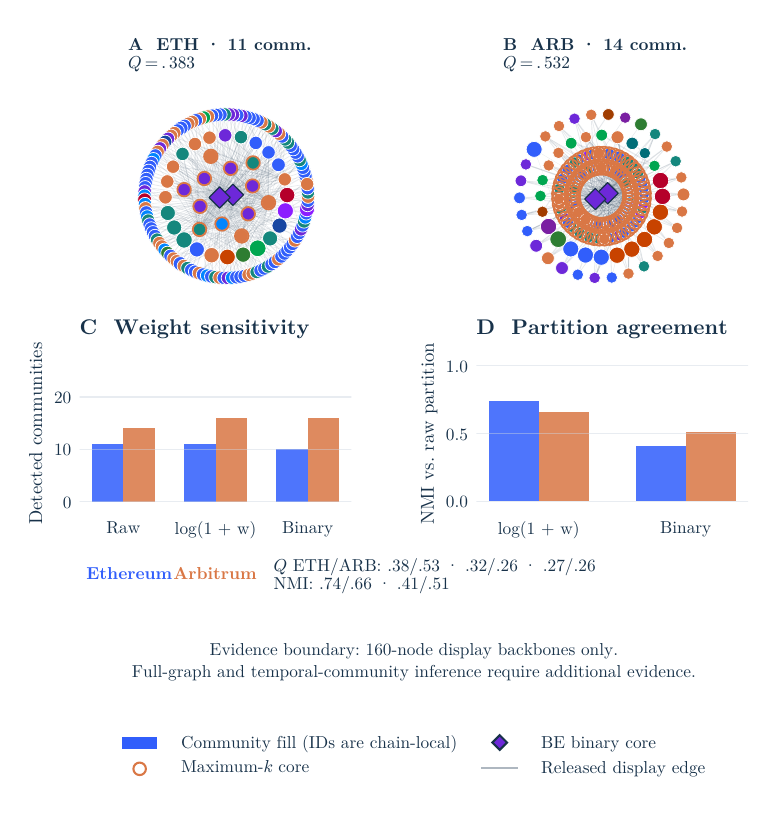}
\resumerefereelines
\caption{Community sensitivity analysis on the fixed display backbones. Panels A and B color the 160 displayed nodes per chain by a weighted greedy-modularity partition of the top-weight backbone; community identifiers are size-ranked and chain-local, orange rings mark maximum-$k$-core nodes, violet diamonds mark the two-node Borgatti--Everett core, node size encodes PageRank, and faint edges are the displayed relations. Panel C reports community counts and modularity $Q$ after applying raw, $\log(1+w)$, or binary edge weights. Panel D reports normalized mutual information (NMI) between the raw-weight partition and each alternative. The visible changes under weight definitions are the result: community assignments on this compact backbone are specification-sensitive. The figure is a display-subset diagnostic, not a partition of the complete directed weighted graphs, a temporal-community estimate, an economic-actor grouping, or an estimation sample. Full-graph community and persistence claims require the complete edge-event table and corresponding validation checks.}
\label{fig:display-community-audit}
\end{figure}

%% file: appendix/be_core_infrastructure_evidence.tex
\subsection{Borgatti--Everett infrastructure-core evidence}\label{app:be-core-infrastructure}

The Borgatti--Everett (BE) binary core in Figure~\ref{fig:real-v5-core-periphery} is interpreted as a limited infrastructure-decentralization diagnostic. The algorithm selects two nodes on each chain because the fitted binary solution reaches the imposed two-node lower bound. This prevents a strong core-size interpretation, but the identities of the selected address-role nodes still help diagnose whether the displayed core is composed of ordinary user addresses, protocol infrastructure, or unlabeled addresses. Table~\ref{tab:be-core-infrastructure-evidence} records the exact address-level evidence used for that interpretation.

\begin{table}[!htbp]
\centering
\caption{Address evidence for the two-node BE binary cores in Figure~\ref{fig:real-v5-core-periphery}.}
\label{tab:be-core-infrastructure-evidence}
\AVTableSetup
\begin{tabularx}{0.99\textwidth}{P{15mm}>{\raggedright\arraybackslash\hsize=1.15\hsize}X>{\raggedright\arraybackslash\hsize=.85\hsize}X}
\toprule
\AVTableHeader
\textbf{Chain} & \textbf{BE-core address, verification, and public trace} & \textbf{Interpretation boundary} \\
\midrule
Ethereum &
\texttt{0xb748952c7bc638f317}\newline\texttt{75245964707bcc5ddfabfc}\par
\href{https://etherscan.io/address/0xb748952c7bc638f31775245964707bcc5ddfabfc}{Explorer record}. Aave migration helper; verified contract name \emph{MigrationHelperMainnet}. &
Protocol-adjacent migration infrastructure; not evidence of a natural person or terminal economic controller. \\
\addlinespace[2pt]
Ethereum &
\texttt{0xd322a49006fc828f9b5}\newline\texttt{b37ab215f99b4e5cab19c}\par
\href{https://etherscan.io/address/0xd322a49006fc828f9b5b37ab215f99b4e5cab19c}{Explorer record}. Aave Wrapped Token Gateway V3 label. &
Protocol-adjacent gateway infrastructure in the address-role projection; not a user-level concentration claim. \\
\addlinespace[2pt]
Arbitrum &
\texttt{0xecd4bd3121f9fd604f}\newline\texttt{fac631bf6d41ec12f1fafb}\par
\href{https://etherscan.io/address/0xecd4bd3121f9fd604ffac631bf6d41ec12f1fafb}{Explorer record}. Aave stake/ABPT migrator label and address-book trace. &
Protocol-adjacent migration infrastructure; supports an infrastructure-core reading only. \\
\addlinespace[2pt]
Arbitrum &
\texttt{0xf3c3f14dd7bdb7e03e}\newline\texttt{6ebc3bc5ffc6d66de12251}\par
\href{https://www.arbiscan.io/address/0xf3c3f14dd7bdb7e03e6ebc3bc5ffc6d66de12251}{Explorer record}. Unlabeled in the audited package; public search surfaced token-holder context but no reliable entity label. &
Unlabeled address-role core node; retained as a verifiable address, not assigned to an actor or infrastructure class without primary-source evidence. \\
\bottomrule
\end{tabularx}
\begin{tablenotes}[flushleft]
\AVTableNoteFont
\item \emph{Notes:} The source figure marks BE-core nodes as diamonds in \path{figures/assets/real_v5_core_periphery.drawio}. The table records public labels only as address-level annotations. It does not infer ownership, governance power, beneficial control, infrastructure-removal loss, or bridge-route dependence. The appropriate main-text conclusion is that the BE core is shaped by protocol-adjacent infrastructure and address-role mechanics, while entity decentralization and functional infrastructure resilience require additional verified data.
\end{tablenotes}
\end{table}

%% file: appendix/arbitrum_gnosis_benchmark_robustness.tex
\subsection{Arbitrum--Gnosis comparative robustness and identification}\label{app:benchmark-did}

\subsubsection{Common-calendar construction and estimand boundary}
The comparison aligns Arbitrum and Gnosis to the 2 July 2024 Arbitrum GHO reserve activation. Both chains contain every event week from $-16$ through $+16$ in the half-open UTC windows documented with the released panel. The primary sample uses clean-pre weeks $-16$ through $-9$ and post weeks $+1$ through $+16$; weeks $-8$ through $-1$ and event week 0 are displayed but excluded. Gnosis is a contemporaneous Aave deployment without the focal 2024 Arbitrum expansion. It can absorb some common calendar movement, but no assumption is made that it reproduces Arbitrum's unobserved counterfactual path.

The two-chain interaction in Equation~\eqref{eq:did-style} equals the arithmetic difference in changes
\begin{equation}
\widehat{\Delta}_{h}=
\left(\overline{y}_{A,1:h}-\overline{y}_{A,-16:-9}\right)
-\left(\overline{y}_{G,1:h}-\overline{y}_{G,-16:-9}\right),
\qquad h\in\{8,12,16\},
\label{eq:benchmark-did}
\end{equation}
where $A$ denotes Arbitrum and $G$ denotes Gnosis. The released script estimates the equivalent weekly-gap regression and its Bartlett/Newey--West covariance matrix, then cross-checks that matrix against \texttt{statsmodels}. Table~\ref{tab:arbitrum-gnosis-benchmark} reports the primary mean-change results.

\input{tables/tab07_arbitrum_gnosis_benchmark}

\subsubsection{Controlled level and slope changes}
Table~\ref{tab:arbitrum-gnosis-its} reports every coefficient in Equation~\eqref{eq:controlled-its}. The clean-pre differential slopes are $-0.17259$ per week for log participation and +0.001902 for position-holder-event HHI. The fitted post segments require both a level and a slope change under the primary eight-week anticipation exclusion. Because the omitted interval contains substantial movement, the week-0 level coefficient is an extrapolated model quantity and should not be read as an observed instantaneous jump.

\input{tables/tabA_arbitrum_gnosis_its}

\subsubsection{Anticipation-window sensitivity}
The anticipation exclusion is consequential. With no excluded anticipation weeks, the comparative HHI mean-change coefficient is positive and both controlled-ITS slope shifts have the opposite sign from the primary specification. A four-week exclusion produces intermediate mean-change coefficients but does not make the controlled-ITS terms stable. Table~\ref{tab:arbitrum-gnosis-anticipation} therefore serves as a specification-sensitivity diagnostic rather than a search for a preferred significant result.

\input{tables/tabA_arbitrum_gnosis_anticipation}

\subsubsection{Alternative post horizons}
The $+8$, $+12$, and $+16$ arithmetic contrasts preserve the same clean pre-window. Log-participation contrasts are 1.9941, 1.9829, and 1.9833; HHI contrasts are $-0.01409$, $-0.01227$, and $-0.01842$. Directional stability across post horizons is informative, but it does not eliminate the differential trends or the sensitivity to the anticipation definition.

\input{tables/tabA_arbitrum_gnosis_horizons}

\subsubsection{Pre-period slopes and placebo dates}
Table~\ref{tab:arbitrum-gnosis-diagnostics} reports the clean-pre slopes and transparent two-week placebo calculations at pseudo-event weeks $-13$, $-12$, and $-11$. The model-based slope ranges exclude zero and the placebo contrasts are material relative to several observed weekly changes. These results do not function as a pass/fail parallel-trends test; pretests can have low power and can distort subsequent inference when used as a selection rule \cite{roth2022pretest}. They instead document that a causal interpretation would be sensitive to the assumed counterfactual trajectory.

\input{tables/tabA_arbitrum_gnosis_diagnostics}

\subsubsection{Comparison movement and unimplemented diagnostics}
For log participation, $1.9519/1.9833=98.4\%$ of the absolute primary contrast is the Gnosis decline. For HHI, $0.02343/(0.00501+0.02343)=82.4\%$ of combined absolute movement is the Gnosis increase. These are arithmetic decompositions of the observed changes, not causal shares.

The released common-calendar artifact supports weekly participation, HHI, alternative anticipation exclusions, post horizons, differential slopes, and placebo dates. Daily aggregation, longer comparable pre-periods, alternative execution clocks, action- or reserve-specific common-calendar outcomes, EOA-only or contract-excluded comparisons, and leave-one-comparison-chain analyses are \emph{planned diagnostics or future extensions}; no estimates for them are presented. They require the corresponding event-level or multi-chain release and validation gates. Future sensitivity to bounded trend departures should follow an explicit design such as Rambachan and Roth \cite{rambachan2023credible}, while staggered estimators require multiple valid adoption cohorts rather than relabeling the present centered gaps.\par

%% file: tables/tab07_arbitrum_gnosis_benchmark.tex
\begin{table*}[t]
\centering
\caption{Arbitrum--Gnosis DiD-style comparative change and model-based uncertainty.}
\label{tab:arbitrum-gnosis-benchmark}
\AVTableSetup
\begin{tabularx}{0.99\textwidth}{>{\raggedright\arraybackslash}X*{4}{>{\centering\arraybackslash}X}}
\toprule
\AVTableHeader
\textbf{Outcome} & \textbf{Arbitrum change} & \textbf{Gnosis change} & \textbf{DiD-style change} & \textbf{Descriptive HAC range} \\
\midrule
Log participation & +0.0314 & -1.9519 & \textbf{+1.9833} & [1.6077, 2.3589] \\
Position-holder-event HHI & +0.00501 & +0.02343 & \textbf{-0.01842} & [-0.02900, -0.00784] \\
\bottomrule
\end{tabularx}
\begin{minipage}{0.97\textwidth}
\AVTableNoteFont\emph{Notes:} Direct empirical output. The clean pre-period is weeks $-16$ to $-9$; weeks $-8$ to $-1$ and week 0 are excluded; the post-period is weeks $+1$ to $+16$. The last column is a normal approximation based on Bartlett/Newey--West HAC standard errors with lag 3 and no finite-sample correction. It describes modeled serial dependence in the aggregate weekly gap; it is not a confidence interval for a treatment effect. Position holders are protocol-observed addresses, not verified economic actors. The negative HHI contrast means that HHI rose less on Arbitrum than on Gnosis; it does not mean that Arbitrum HHI fell in absolute terms.
\end{minipage}
\end{table*}

%% file: tables/tabA_arbitrum_gnosis_its.tex
\begin{table}[!ht]
\centering
\caption{Controlled interrupted-time-series coefficients for the Arbitrum--Gnosis gap.}
\label{tab:arbitrum-gnosis-its}
\AVTableSetup
\begin{tabularx}{0.99\linewidth}{P{0.21\linewidth}Yrrrr}
\toprule
\AVTableHeader
\textbf{Outcome} & \textbf{Term} & \textbf{Estimate} & \textbf{HAC SE} & \textbf{Lower} & \textbf{Upper} \\
\midrule
Log participation & Intercept & $-0.91883$ & 0.26750 & $-1.44313$ & $-0.39453$ \\
 & Pre-gap slope/week & $-0.17259$ & 0.02429 & $-0.22020$ & $-0.12498$ \\
 & Level shift at week 0 & +4.03961 & 0.41905 & 3.21827 & 4.86094 \\
 & Slope shift/week & +0.18448 & 0.03016 & 0.12537 & 0.24359 \\
\midrule
Position-holder-event HHI & Intercept & +0.017107 & 0.011197 & $-0.004839$ & 0.039053 \\
 & Pre-gap slope/week & +0.001902 & 0.000798 & 0.000338 & 0.003466 \\
 & Level shift at week 0 & $-0.028487$ & 0.012673 & $-0.053327$ & $-0.003647$ \\
 & Slope shift/week & $-0.003515$ & 0.001078 & $-0.005628$ & $-0.001403$ \\
\bottomrule
\end{tabularx}
\begin{minipage}{0.96\linewidth}
\AVTableNoteFont\emph{Notes:} Direct empirical output. The dependent variable is the weekly Arbitrum-minus-Gnosis outcome gap. The model uses clean-pre weeks $-16$ to $-9$ and post weeks $+1$ to $+16$ ($n=24$); weeks $-8$ through 0 are omitted. Bartlett/Newey--West HAC standard errors use lag 2 and no finite-sample correction. The week-0 level is extrapolated because week 0 is excluded. Ranges are normal approximations describing model-based time-series uncertainty, not treatment-effect confidence intervals.
\end{minipage}
\end{table}

%% file: tables/tabA_arbitrum_gnosis_anticipation.tex
\begin{table}[!ht]
\centering
\caption{Sensitivity to the excluded anticipation window.}
\label{tab:arbitrum-gnosis-anticipation}
\AVTableSetup
\begin{tabularx}{0.98\linewidth}{P{0.24\linewidth}*{4}{>{\centering\arraybackslash}X}}
\toprule
\AVTableHeader
\textbf{Outcome} & \textbf{Excluded pre-event weeks} & \textbf{DiD-style change} & \textbf{cITS level at week 0} & \textbf{cITS slope change/week} \\
\midrule
Log participation & None & +1.3414 & +0.1551 & $-0.1158$ \\
 & 4 weeks & +1.6935 & +0.9104 & $-0.0531$ \\
 & 8 weeks (primary) & +1.9833 & +4.0396 & +0.1845 \\
\midrule
Position-holder-event HHI & None & +0.00595 & +0.06308 & +0.00350 \\
 & 4 weeks & $-0.00747$ & +0.04635 & +0.00221 \\
 & 8 weeks (primary) & $-0.01842$ & $-0.02849$ & $-0.00352$ \\
\bottomrule
\end{tabularx}
\begin{minipage}{0.96\linewidth}
\AVTableNoteFont\emph{Notes:} Direct empirical robustness output. Event week 0 is excluded in every specification; the post-period ends at week $+16$. Expanding the anticipation exclusion shortens the clean pre-period from weeks $-16{:}-1$ to $-16{:}-9$. The sensitivity, including sign changes in the controlled-ITS terms and the HHI mean-change contrast, reinforces the need to treat the estimates as comparative longitudinal evidence rather than as identified causal effects.
\end{minipage}
\end{table}

%% file: tables/tabA_arbitrum_gnosis_horizons.tex
\begin{table}[!ht]
\centering
\caption{Descriptive contrasts across alternative post-window horizons.}
\label{tab:arbitrum-gnosis-horizons}
\AVTableSetup
\begin{tabularx}{0.98\linewidth}{>{\raggedright\arraybackslash}X*{5}{>{\centering\arraybackslash}X}}
\toprule
\AVTableHeader
\textbf{Outcome} & \textbf{Through week} & \textbf{Arb.} & \textbf{Gnosis} & \textbf{Diff. in changes} & \textbf{NW diagnostic} \\
\midrule
Log participation & $+8$ & +0.1967 & -1.7974 & +1.9941 & [1.5700, 2.4183] \\
Log participation & $+12$ & +0.0824 & -1.9004 & +1.9829 & [1.5968, 2.3690] \\
Log participation & $+16$ & +0.0314 & -1.9519 & +1.9833 & [1.6077, 2.3589] \\
Position-holder-event HHI & $+8$ & +0.00680 & +0.02089 & -0.01409 & [-0.02034, -0.00785] \\
Position-holder-event HHI & $+12$ & +0.00713 & +0.01940 & -0.01227 & [-0.01927, -0.00527] \\
Position-holder-event HHI & $+16$ & +0.00501 & +0.02343 & -0.01842 & [-0.02900, -0.00784] \\
\bottomrule
\end{tabularx}
\begin{minipage}{0.96\linewidth}
\AVTableNoteFont\emph{Notes:} Direct empirical robustness output. The clean pre-period remains weeks $-16$ to $-9$ for every row. Bartlett/Newey--West HAC standard errors use lag 3 and no finite-sample correction. The ranges are descriptive normal approximations, not treatment-effect confidence intervals.
\end{minipage}
\end{table}

%% file: tables/tabA_arbitrum_gnosis_diagnostics.tex
\begin{table}[!ht]
\centering
\caption{Differential pre-trends and pre-period placebo contrasts.}
\label{tab:arbitrum-gnosis-diagnostics}
\AVTableSetup
\begin{tabular}{llrrr}
\toprule
\AVTableHeader
\textbf{Outcome} & \textbf{Diagnostic} & \textbf{Estimate} & \textbf{Lower} & \textbf{Upper} \\
\midrule
Log participation & gap slope/week & -0.17259 & -0.22020 & -0.12498 \\
Position-holder-event HHI & gap slope/week & +0.00190 & 0.00034 & 0.00347 \\
\midrule
Log participation & placebo week -13 & -0.2975 & -- & -- \\
Log participation & placebo week -12 & -0.2298 & -- & -- \\
Log participation & placebo week -11 & -0.4950 & -- & -- \\
Position-holder-event HHI & placebo week -13 & -0.00720 & -- & -- \\
Position-holder-event HHI & placebo week -12 & +0.00327 & -- & -- \\
Position-holder-event HHI & placebo week -11 & +0.02075 & -- & -- \\
\bottomrule
\end{tabular}
\begin{minipage}{0.96\linewidth}
\AVTableNoteFont\emph{Notes:} Direct empirical diagnostics. Slope ranges use Bartlett/Newey--West HAC lag 2 without a finite-sample correction. Placebos compare two weeks before and after each pseudo-event within the clean pre-period. The differential slopes and material placebo movements show that causal interpretation would require additional comparison support and trend-robust designs.
\end{minipage}
\end{table}

%% file: appendix/evidence_status_summary.tex
\subsection{Evidence-status summary}\label{app:evidence-status}

\input{tables/tabA_sample_coverage_audit}

\begin{table}[!htbp]
\centering
\caption{Evidence categories and supported claims.}
\label{tab:evidence-status}
\AVTableSetup
\begin{tabularx}{0.99\textwidth}{P{0.14\textwidth}P{0.22\textwidth}YP{0.18\textwidth}}
\toprule
\AVTableHeader
\textbf{Evidence category} & \textbf{Verified object} & \textbf{Supported use} & \textbf{Current limitation} \\
\midrule
Source audit & Ethereum source and recovery boundary & Method assessment & Not the longitudinal analysis panel \\
Ethereum events & 118,806 Pool events & Address-level measurement and action composition & No actor or causal result \\
Contract roles & Historical code and 10 curated labels & Contract/address-type sensitivity & Economic-actor coverage 0\% \\
Actor bounds & Position-holder panel and period HHI bounds & Pooled address HHI and period-specific actor bounds & Change envelope is not a sharp joint set \\
Longitudinal\newline panels & Ethereum--Arbitrum weekly panels, full-node topology, and display backbones & Longitudinal levels and membership transitions, structural distributions, and display-only community sensitivity & Clocks differ; no full edge-event or route table; no ATT \\
Comparative analysis & Audited Arbitrum--Gnosis common-calendar panel, DiD-style coefficient, controlled ITS, and dynamic contrasts & Comparative longitudinal evidence and specification diagnostics & One comparison chain; differential pre-path; no treatment effect \\
Causal extension & Corrected multichain collection procedure and checks & Future data collection & Multichain ATT not produced \\
Route extension & Route schema and proposed diagnostics & Future infrastructure analysis & Verified route-event panel absent \\
Subgraph comparison & Specified reconciliation procedure & Future independent validation & Not conducted; supports no current result \\
\bottomrule
\end{tabularx}
\end{table}

The structured evidence table records the documented collection, reported results, analysis procedures, and evidentiary status. A category may contain conventional comparative coefficients while leaving the causal estimand unidentified. This distinction is intentional: completing a regression does not establish its counterfactual assumptions.

\begin{table}[!htbp]
\centering
\caption{Orphan-result audit and unique manuscript status.}
\label{tab:orphan-result-audit}
\AVTableSetup
\renewcommand{\arraystretch}{1.20}
\begin{tabularx}{0.99\textwidth}{P{0.23\textwidth}P{0.13\textwidth}YP{0.22\textwidth}}
\toprule
\AVTableHeader
Empirical asset family & Status & Authoritative use & Conflicting or unavailable use \\
\midrule
Ethereum--Arbitrum weekly trajectories and sample documentation & \texttt{MAIN} & Chain-relative participation, weekly HHI, turnover, and typed topology & Not a common-calendar treatment comparison \\
Action-layer topology profile & \texttt{MAIN} & Role- and action-specific structural heterogeneity & Not economic ownership or full multilayer routing \\
Full-graph core diagnostics and 160-node display backbone & \texttt{MAIN} & Full-graph counts/coreness plus display-only inspection & Display is not an estimation sample or actor-control test \\
Centrality, core membership, and role composition & \texttt{MAIN} & Scale-aware full-node structural distributions & Not causal removal dependence \\
Arbitrum--Gnosis six-panel comparison and main coefficient table & \texttt{MAIN} & DiD-style comparative change, controlled ITS, and dynamic contrasts & Not an identified GHO effect or event-study effect \\
Entity audit and pooled-period actor bounds & \texttt{APPENDIX} & Address/contract sensitivity and period-specific partial-identification bounds & No verified economic-actor partition \\
Display-backbone community sensitivity & \texttt{APPENDIX} & Weight-definition sensitivity on released 160-node displays & No full-graph community population claim \\
Anticipation, horizon, slope, placebo, and ITS robustness tables & \texttt{APPENDIX} & Specification and comparison-suitability diagnostics & Not treatment-assignment inference \\
Verified route panel, multi-cohort estimators, actor resolution, and dynamic multilayer models & \texttt{FUTURE} & Prespecified evidence and design requirements & No current numerical result \\
Ethereum--Arbitrum non-common-calendar pilot table & \texttt{SUPERSEDED} & Source retained for provenance only; not rendered & Replaced by the audited common-calendar comparison \\
Fixed 18-work literature Sankey & \texttt{SUPERSEDED} & Coding CSVs retained for provenance only; not rendered & Replaced by the chronological literature landscape \\
\bottomrule
\end{tabularx}
\AVTableNoteFont\emph{Notes:} Each listed asset family has exactly one release status. \texttt{MAIN} and \texttt{APPENDIX} identify rendered evidence; \texttt{FUTURE} identifies unimplemented requirements; \texttt{SUPERSEDED} identifies retained source or provenance that is excluded from manuscript numbering and claims.
\end{table}

%% file: tables/tabA_sample_coverage_audit.tex
\begin{table}[!htbp]
\centering
\caption{Extended sample-coverage audit and unavailable fields.}
\label{tab:extended-sample-coverage-audit}
\AVTableSetup
\begin{tabularx}{0.98\textwidth}{P{0.18\textwidth}Y Y Y}
\toprule
\AVTableHeader
\textbf{Audit item} & \textbf{Ethereum} & \textbf{Arbitrum} & \textbf{Gnosis comparison} \\
\midrule
Calendar role & Chain-relative panel & Focal/common calendar & Contemporaneous comparison series \\
Reference clock & 15 Jul 2023 & 2 Jul 2024 & 2 Jul 2024 \\
Event weeks & $-16{:}+16$ & $-16{:}+16$ & $-16{:}+16$ \\
Aave V3 Pool events & 118,806 & 1,837,410 & 81,814 \\
Transactions & 103,171 & unavailable: aggregate RC26 inputs omit transaction identifiers & 78,353 \\
Position-holder addresses & 15,351 & unavailable: aggregate RC26 inputs omit row-level address identifiers & 26,020 \\
Delegated topology events & 27,579 & 422,834 & not estimated: comparison inputs contain no delegated-edge records \\
Provider cross-check & pass & pass & pass \\
Permitted main use & descriptive & descriptive + comparative longitudinal analysis & comparative longitudinal analysis \\
Causal interpretation & not identified & not identified & not identified \\
\bottomrule
\end{tabularx}
\begin{minipage}{0.96\textwidth}
\AVTableNoteFont\emph{Notes:} This appendix table preserves the complete coverage audit underlying Table~\ref{tab:aggregate-sample}. ``Unavailable'' and ``not estimated'' are evidence-status labels, not zeros, and the affected quantities are not used as main comparative outcomes. The committed RC26 evidence contains weekly aggregates rather than the governed row-level transaction and address identifiers needed to deduplicate Arbitrum transactions or position holders across weeks; those totals therefore cannot be reconstructed without evidence outside the anonymous release. Gnosis is aligned to the verified Arbitrum expansion date and is evaluated only as a contemporaneous comparison series. Address and address-role counts do not identify people, organizations, or common economic control.
\end{minipage}
\end{table}

%% file: appendix/subgraph_validation_protocol.tex
\subsection{Pre-specified Subgraph validation protocol}\label{app:subgraph-validation}

This protocol describes a future validation exercise that has not yet been conducted. Consensus RPC logs remain authoritative for all reported results. The purpose of a Subgraph comparison is to detect indexing, mapping, schema, or local-decoder discrepancies, not to replace block-level evidence.

\begin{enumerate}
\item \textbf{Source documentation.} Record the Aave protocol-subgraphs source files and retrieval date, network specification, Graph deployment ID, endpoint class, and retrieval timestamp. A changing endpoint alone is insufficient for historical validation.
\item \textbf{Indexing criterion.} Query \texttt{\_meta} and require the indexed block to be at or beyond every requested boundary, with no indexing error. Retain the query and response with the source record.
\item \textbf{Canonical key.} Reconcile on chain ID, block number, transaction hash, and log index. Event names, Pool emitter, reserve, action-side address, position-holder address, and raw integer amount are compared field by field.
\item \textbf{Role semantics.} Supply and Borrow must map \texttt{onBehalfOf} from indexed topic 2; Withdraw, Repay, and LiquidationCall must map the canonical \texttt{user}. Zero-address rates and action-specific disagreement tables are mandatory.
\item \textbf{Aggregate reconciliation.} Publish unmatched keys, field-level differences, weekly action counts, active position-holder addresses, HHI differences, and the maximum absolute and relative discrepancy for every chain--week.
\item \textbf{Pass/fail rule.} Exact key coverage and role equality are required for headline validation. Any tolerated metadata-only difference must be pre-specified, non-outcome-bearing, and listed. A failure leaves RPC results intact but prevents a ``cross-source validated'' label.
\end{enumerate}

Even a perfect reconciliation cannot validate policy exogeneity, parallel trends, address-to-actor mappings, or bridge-route dependence. Those are separate identification and measurement problems. Deployment IDs currently recorded in the study materials are advertised Subgraph identifiers; the exact deployment endpoint, mapping files, and complete retrieval record must be documented before the comparison is conducted.

%% file: appendix/future_causal_infrastructure_extensions.tex
\subsection{Future systemic and agent-mediated extensions}\label{app:future-extensions}\label{app:future-systemic}

This appendix separates implementable research designs from results established in the present study. Every item below is prospective. None changes the evidence status of the four measured dimensions, identifies an average treatment effect, converts addresses into economic actors, proves route resilience, or establishes autonomous-agent prevalence in the analyzed Aave events.

\subsubsection{Evidence and design gates}

Future multichain analysis must first recollect any compact deployment created before the corrected Supply/Borrow \texttt{onBehalfOf} decoding and zero-address rejection. Each panel must pass ABI-semantic, window-continuity, duplicate-key, provider, and reproducibility checks. Staggered designs additionally require verified reserve activations, common outcome semantics, overlapping calendars, sufficient pre-treatment support, explicit anticipation windows, suitable not-yet-treated deployments, and chain-level inference. Group-time, interaction-weighted, or imputation estimators become relevant only after those gates are met \cite{callaway2021difference,sun2021estimating,goodmanbacon2021difference,borusyak2024revisiting}; synthetic comparisons require a larger verified donor pool, stable pre-fit, predictor balance, and placebo support.

Table~\ref{tab:future-research-roadmap} preserves the six information fields that appeared in the former main-text roadmap while presenting them in a two-column, interdisciplinary format. It consolidates the prior identification, causal-requirements, and staged-extension tables so that each unresolved mechanism has one authoritative future status.

\input{tables/tab08_future_research_roadmap}

\subsubsection{Dynamic heterogeneous multilayer representation}

A future protocol system at time \(t\) can be represented as
\[
\mathcal{G}_t=\left(V_t,E_t^{\mathrm{lending}},E_t^{\mathrm{GHO}},E_t^{\mathrm{transfer}},E_t^{\mathrm{bridge}},E_t^{\mathrm{governance}},E_t^{\mathrm{infrastructure}}\right).
\]
The layers encode functional, contractual, or institutional relations. Chains may define partitions within those layers, but they are not geographical regions. Nodes may represent addresses or evidence-supported entities, contracts, assets, facilitators, bridges, or infrastructure providers. Identity links across layers must retain provenance and uncertainty rather than silently treating heuristic clusters as known actors.

Memory-aware community analysis can identify persistent groups and role changes \cite{clemente2024memory}. Higher-order events can retain joint relations among users, contracts, assets, and infrastructure rather than reducing every event to pairwise projections \cite{iacopini2024temporal}; the choice between hypergraph and simplicial representations must be justified because it can change implied dynamics \cite{zhang2023higherorder}. Temporal aggregation can likewise be selected according to its dynamical consequences \cite{allen2024compressing}. These methods motivate future analysis but do not retroactively alter the weekly pairwise networks used in the paper.

\input{figures/fig21_future_systemic_decentralization}

\subsubsection{Actor, route, and decision provenance}

Economic-actor resolution should accept only dated primary-source controller evidence with validity intervals and a correction history. Bytecode, co-timing, shared counterparties, or behavioral similarity alone do not create must-links; where coverage remains incomplete, actor concentration remains an identified set. Route analysis separately requires occurrence records linking chain, source and destination contracts, facilitator, messaging provider, asset, transaction, amount, and time. A governance configuration transaction or Pool event does not prove that a route carried the measured activity.

The prospective fifth dimension, \(\mathcal D_t\), concerns concentration in effective decision making across principals, decision systems, executors, and beneficiaries. Candidate summaries include principal HHI across agent-operated wallets, model- or strategy-provider HHI, the number of principals depending on one execution service, synchronized actions after shocks, and the share of autonomous versus human-confirmed transactions. The qualitative typology in \citet{ante2026autonomous} motivates hypotheses about possible roles but does not label this study's transactions.

\subsubsection{Validation before complexity}

The present theorem-based simulation remains the transparent baseline. Behavioral or language-model agents could add heterogeneous objectives, risk preferences, memory, information, tools, entry, exit, and adaptive strategies, but they require controlled increments, repeated seeds, architecture and provider ablations, and validation against held-out protocol behavior. EconAgent illustrates one route to heterogeneous multi-period simulation \cite{li2024econagent}; it does not validate a DeFi calibration. Likewise, failure simulations need observed outages, rerouting, recovery, capacity, and substitution before removal loss can be interpreted as resilience. The higher-order multilayer framework of \citet{nicoletti2024information} helps formulate propagation, while empirical calibration remains a separate evidence gate.

The governing progression is therefore
\[
\begin{aligned}
\text{observed measurement}
&\rightarrow\text{validated multilayer construction}\\
&\rightarrow\text{failure and formation tests}\\
&\rightarrow\text{verified decision concentration},
\end{aligned}
\]
with each arrow conditional on new data, technology, application scenarios, and validation rather than on complexity alone.

%% file: tables/tab08_future_research_roadmap.tex
\suspendrefereelines
\begingroup
\setstretch{1.02}
\AVTableSetup
\setlength{\tabcolsep}{4pt}
\setlength{\LTleft}{0pt}
\setlength{\LTright}{0pt}
\renewcommand{\arraystretch}{1.18}
\begin{longtable}{P{0.205\linewidth}P{0.735\linewidth}}
\caption{Detailed six-field roadmap from the current evidence boundary to falsifiable extensions.}
\label{tab:future-research-roadmap}\\
\toprule
\AVTableHeader
\textbf{Research domain} & \textbf{Unresolved mechanism, required data, candidate method, falsifiable outcome, and claim enabled} \\
\midrule
\endfirsthead
\multicolumn{2}{c}{\tablename\ \thetable\ (continued)}\\
\toprule
\AVTableHeader
\textbf{Research domain} & \textbf{Five prospective fields} \\
\midrule
\endhead
\midrule
\multicolumn{2}{r}{Continued on next page}\\
\endfoot
\bottomrule
\endlastfoot
\textbf{Dynamic network formation} & \textbf{Unresolved mechanism:} Do entry, exit, and edge rewiring mediate the observed sign regimes? \textbf{Required data:} Time-stamped nodes and edges around repeated, comparable protocol changes. \textbf{Candidate method:} Event-history or actor-oriented formation model linked to the activity equation. \textbf{Falsifiable outcome:} The formation response is positive, null, or sign-reversing for concentration growth. \textbf{Claim enabled:} Whether $dG/dp$ reinforces or offsets peripheral expansion. \\
\textbf{Economic actors} & \textbf{Unresolved mechanism:} Does address growth survive verified entity linkage and common-control bounds? \textbf{Required data:} Audited must-links, control and beneficiary evidence, and time-varying entity provenance. \textbf{Candidate method:} Partial identification plus entity-resolved sensitivity analysis. \textbf{Falsifiable outcome:} Actor-level breadth or HHI remains inside, or departs from, address-level bounds. \textbf{Claim enabled:} Economic-actor participation and concentration. \\
\textbf{Route dependencies} & \textbf{Unresolved mechanism:} Does multichain access diversify bridge, facilitator, sequencer, oracle, and execution routes? \textbf{Required data:} Verified calls, routes, authorizations, shared components, and dependency maps. \textbf{Candidate method:} Provenance-aware multilayer dependency reconstruction. \textbf{Falsifiable outcome:} Independent routes increase, remain unchanged, or collapse to common bottlenecks. \textbf{Claim enabled:} Route redundancy and shared-component concentration. \\
\textbf{Comparative identification} & \textbf{Unresolved mechanism:} Which changes follow rollout rather than anticipation, spillovers, or common shocks? \textbf{Required data:} Multiple credible comparison chains, repeated rollout cohorts, and longer prespecified pre-periods. \textbf{Candidate method:} Cohort event studies with appropriate not-yet-treated comparisons. \textbf{Falsifiable outcome:} Cohort estimates agree, attenuate, or reverse under common support. \textbf{Claim enabled:} A bounded average rollout effect for supported cohorts. \\
\textbf{Robust comparison design} & \textbf{Unresolved mechanism:} How sensitive are conclusions to nonparallel trends and donor choice? \textbf{Required data:} Verified donor pool, predictor balance, pre-fit, placebo, and outcome support. \textbf{Candidate method:} Bounded-trend sensitivity and synthetic comparison designs. \textbf{Falsifiable outcome:} Signs persist or fail at stated trend deviations and placebo ranks. \textbf{Claim enabled:} Design-conditional comparative effects, not from the current one-chain contrast. \\
\textbf{Multilayer failures} & \textbf{Unresolved mechanism:} Can activity and liquidity substitute after component or provider failure? \textbf{Required data:} Observed outages, rerouting, recovery, liquidity migration, and cross-layer exposure. \textbf{Candidate method:} Component-removal, cascading-failure, and substitution experiments. \textbf{Falsifiable outcome:} Service recovers through independent paths or exhibits measurable cascade loss. \textbf{Claim enabled:} Failure resilience and substitution capacity. \\
\textbf{Autonomous agents and delegation} & \textbf{Unresolved mechanism:} Is effective decision making concentrated behind expanding wallet counts? \textbf{Required data:} Verified principals, wallet permissions, delegation records, providers, models, and execution traces. \textbf{Candidate method:} Entity-resolved delegation graph with provider and strategy concentration measures. \textbf{Falsifiable outcome:} Principal or provider HHI diverges from address HHI, or does not. \textbf{Claim enabled:} Independently evidenced autonomous-agent and delegation concentration. \\
\end{longtable}
\noindent\begin{minipage}{0.98\linewidth}
\AVTableNoteFont\emph{Notes:} All six original fields are preserved, but the five detailed fields are stacked within each disciplinary row to avoid sub-8-point type. These are future evidence requirements and candidate tests, not analyses completed in the present study. Any actor-, route-, or delegation-level claim requires evidence beyond Pool-event addresses.
\end{minipage}
\endgroup
\resumerefereelines

%% file: figures/fig21_future_systemic_decentralization.tex
\begin{figure}[!htbp]
\centering
\begin{tikzpicture}[x=1mm,y=1mm,
  column/.style={av/card,minimum width=40mm,minimum height=70mm,text width=34mm,inner sep=2.0mm},
  boundary flow/.style={av/pending-flow,draw=AVViolet!76},
  observed flow/.style={av/flow,draw=AVTeal!84},
  status/.style={av/tag,minimum width=24mm,align=center}
]
\node[column,draw=AVTeal!78,fill=AVTeal!5] (observed) at (22,42) {%
  {\AVFigureTitleFont\textcolor{AVTeal}{Observed execution}}\\[3mm]
  {\AVFigureSmallFont\textbf{Verified objects}\\Pool events, contract roles, and weekly address-role networks}\\[2.5mm]
  {\AVFigureSmallFont\textbf{Current measures}\\participation; activity distribution; structure; infrastructure proxies}\\[2.5mm]
  {\AVFigureTinyFont Descriptive address evidence, not actor-level or causal.}
};
\node[column,draw=AVViolet!72,dashed,fill=AVLilac] (systemic) at (67,42) {%
  {\AVFigureTitleFont\textcolor{AVViolet}{Future systemic\\ evidence}}\\[3mm]
  {\AVFigureSmallFont\textbf{Required objects}\\typed multilayer events; entity provenance; routes and shared components; outages and substitution}\\[2.5mm]
  {\AVFigureSmallFont\textbf{Candidate tests}\\role transitions; cross-layer concentration; independent paths; failure cascades}\\[2.5mm]
  {\AVFigureTinyFont Requires new collection and validation.}
};
\node[column,draw=AVOrange!82,dashed,fill=AVWarm] (decision) at (112,42) {%
  {\AVFigureTitleFont\textcolor{AVOrange}{Future decision\\ \& delegation}}\\[3mm]
  {\AVFigureSmallFont\textbf{Required provenance}\\principal $\dashrightarrow$ agent or strategy $\dashrightarrow$ wallet or contract $\rightarrow$ protocol action $\dashrightarrow$ beneficiary}\\[2.5mm]
  {\AVFigureSmallFont\textbf{Candidate measures}\\principal, provider, strategy, and shared-dependency HHI; autonomous-execution share}\\[2.5mm]
  {\AVFigureTinyFont No agent prevalence is inferred from current addresses.}
};

\node[status,fill=AVTeal] at (22,80) {Observed};
\node[status,fill=AVViolet] at (67,80) {Future / required};
\node[status,fill=AVOrange] at (112,80) {Future / required};
\draw[boundary flow] (observed.east)--(systemic.west);
\draw[boundary flow] (systemic.east)--(decision.west);
\end{tikzpicture}
\caption{\textbf{Evidence architecture from observed protocol execution to systemic and decision-level decentralization.} The first column summarizes the present address and address-role construction. The second and third columns define distinct future evidence layers for systemic dependencies and effective decision making. Dashed outlines and arrows mark prospective objects; they are not current empirical results.}
\label{fig:future-systemic-decentralization}
\end{figure}